\documentclass[pdflatex,sn-mathphys,iicol,amsmath,amssymb,aps,a4paper]{sn-jnl}

\usepackage{graphicx}%
\usepackage{multirow}%
\usepackage{amsmath,amssymb,amsfonts}%
\usepackage{amsthm}%
\usepackage{mathrsfs}%
\usepackage[title]{appendix}%
\usepackage{textcomp}%
\usepackage{manyfoot}%
\usepackage{booktabs}%
\usepackage{algorithm}%
\usepackage{algorithmicx}%
\usepackage{algpseudocode}%
\usepackage{listings}%

\usepackage{graphicx}
\usepackage{dcolumn}
\usepackage{bm}
\usepackage{multirow}
\usepackage{orcidlink}
\usepackage[normalem]{ulem}
\usepackage{subcaption}
\usepackage{xspace}
\usepackage{mathtools}
\usepackage{hyperref}
\hypersetup{
    colorlinks=true,
    linkcolor=blue,
    filecolor=magenta,      
    urlcolor=cyan,
    }
\usepackage{listings}
\usepackage[utf8]{inputenc}
\DeclarePairedDelimiter{\ceil}{\lceil}{\rceil}

\begin{document}

\newcommand{\ccnumu}{CC~$\nu_\mu$}
\newcommand{\ccnue}{CC~$\nu_e$}

\newcommand{\ntrainepochs}{25\xspace}
\newcommand{\nuimagestrainone}{93218\xspace}
\newcommand{\nvimagestrainone}{92806\xspace}
\newcommand{\nwimagestrainone}{91252\xspace}
\newcommand{\nuimagestraintwo}{85981\xspace}
\newcommand{\nvimagestraintwo}{85555\xspace}
\newcommand{\nwimagestraintwo}{84658\xspace}

\newcommand{\nuimagesvalone}{31069\xspace}
\newcommand{\nvimagesvalone}{30935\xspace}
\newcommand{\nwimagesvalone}{30417\xspace}
\newcommand{\nuimagesvaltwo}{28660\xspace}
\newcommand{\nvimagesvaltwo}{28518\xspace}
\newcommand{\nwimagesvaltwo}{28219\xspace}

\newcommand{\vdnuimagestrainone}{89044\xspace}
\newcommand{\vdnvimagestrainone}{88966\xspace}
\newcommand{\vdnwimagestrainone}{85785\xspace}
\newcommand{\vdnuimagestraintwo}{83589\xspace}
\newcommand{\vdnvimagestraintwo}{83631\xspace}
\newcommand{\vdnwimagestraintwo}{78595\xspace}

\newcommand{\vdnuimagesvalone}{29681\xspace}
\newcommand{\vdnvimagesvalone}{29655\xspace}
\newcommand{\vdnwimagesvalone}{28595\xspace}
\newcommand{\vdnuimagesvaltwo}{27863\xspace}
\newcommand{\vdnvimagesvaltwo}{27876\xspace}
\newcommand{\vdnwimagesvaltwo}{26198\xspace}

\newcommand{\nvalidationevents}{179558\xspace}
\newcommand{\ntotalevents}{300000\xspace}

\newcommand{\parenbar}[1]{\overset{\scriptscriptstyle(-)}{#1}}

\title{Neutrino Interaction Vertex Reconstruction in DUNE with Pandora Deep Learning}


%
\author[35]{A.~Abed Abud}
\author[66]{R.~Acciarri}
\author[12]{M.~A.~Acero}
\author[192]{M.~R.~Adames}
\author[72]{G.~Adamov}
\author[66]{M.~Adamowski}
\author[20]{D.~Adams}
\author[19]{M.~Adinolfi}
\author[30]{C.~Adriano}
\author[81]{A.~Aduszkiewicz}

% ...
\affil[12]{Universidad del Atl\'antico, Barranquilla, Atl\'antico, Colombia}
% ...
\affil[19]{University of Bristol, Bristol BS8 1TL, United Kingdom}
\affil[20]{Brookhaven National Laboratory, Upton, NY 11973, USA}
% ...
\affil[30]{Universidade Estadual de Campinas, Campinas - SP, 13083-970, Brazil}
% ...
\affil[35]{CERN, The European Organization for Nuclear Research, 1211 Meyrin, Switzerland}
% ...
\affil[66]{Fermi National Accelerator Laboratory, Batavia, IL 60510, USA}
% ...
\affil[72]{Georgian Technical University, Tbilisi, Georgia}
% ...
\affil[81]{University of Houston, Houston, TX 77204, USA}
% ...
\affil[192]{Universidade Tecnol\'ogica Federal do Paran\'a, Curitiba, Brazil}
% ...


\abstract{The Pandora Software Development Kit and algorithm libraries perform reconstruction of neutrino interactions in liquid argon time projection chamber detectors. Pandora is the primary event reconstruction software used at the Deep Underground Neutrino Experiment, which will operate four large-scale liquid argon time projection chambers at the far detector site in South Dakota, producing high-resolution images of charged particles emerging from neutrino interactions. While these high-resolution images provide excellent opportunities for physics, the complex topologies require sophisticated pattern recognition capabilities to interpret signals from the detectors as physically meaningful objects that form the inputs to physics analyses. A critical component is the identification of the neutrino interaction vertex. Subsequent reconstruction algorithms use this location to identify the individual primary particles and ensure they each result in a separate reconstructed particle. A new vertex-finding procedure described in this article integrates a U-ResNet neural network performing hit-level classification into the multi-algorithm approach used by Pandora to identify the neutrino interaction vertex. The machine learning solution is seamlessly integrated into a chain of pattern-recognition algorithms. The technique substantially outperforms the previous BDT-based solution, with a more than 20\% increase in the efficiency of sub-1\,cm vertex reconstruction across all neutrino flavours.}

\keywords{Deep Learning, Neutrino Physics, Pandora}



\maketitle
\section{Introduction}\label{sec:introduction}
The Deep Underground Neutrino Experiment (DUNE) \cite{DUNE:2020jqi}, currently under construction, will be a world-leading observatory for the study of neutrinos and nucleon decay. The DUNE far detector modules will be hosted approximately 1.5\,km underground at the Sanford Underground Research Facility (SURF), in South Dakota, USA. The liquid argon time projection chambers (LArTPCs) comprising the far detector will contain 70\,kt of liquid argon, with a fiducial mass of at least 40\,kt.

Among DUNE's many physics goals are the measurement of the charge-parity violation phase in the lepton sector, determination of the neutrino mass ordering and the octant in which the $\theta_{23}$ mixing angle lies, along with a search for supernova neutrino bursts and to test the three-flavour paradigm itself. Physics analyses depend upon determination of event properties such as the flavour of the neutrino interacting or an estimation of the incident neutrino energy. The determination of such quantities depends upon high quality reconstruction of the interactions that will take place inside DUNE's far detectors. The Pandora Software Development Kit (SDK) acts as one of the main reconstruction tools used by DUNE, providing pattern recognition algorithms to build up a picture of the interactions. This article presents details of the integration of deep learning into the set of algorithms previously described in \cite{pandora_uboone} and \cite{pandora_protodune}.

The Pandora SDK was originally developed to identify the energy deposits of particles in fine-granularity detectors, in particular guiding the design and optimisation of future $e^{+}e^{-}$ linear colliders \cite{Thomson2009, Marshall2013}. The multi-algorithm approach to pattern recognition seeks to apply focused, decoupled algorithms to input building blocks. Input is provided into Pandora in the form of a sparse list of hits (localised charge deposits), determined by a low-level hit-finding procedure developed for MicroBooNE \cite{wirecell_microboone}. Complex topologies are deferred to later algorithms, when more is understood about the context, in an effort to avoid making errors that will be difficult to correct later. As a result of the multi-algorithm approach, it is not necessary to choose between hand-engineered or machine-learned algorithms, but one can combine the approaches, leveraging the power of modern machine learning techniques where appropriate, while taking advantage of physics and detector knowledge in the form of `hand-tuned' algorithms where appropriate. Algorithms are designed to be generic and tunable, such that multiple experiments can be supported.

Identifying the neutrino interaction vertex is a critical aspect of reconstructing a neutrino interaction within a LArTPC. All charge deposition proceeds from this location, so its accurate and precise determination can exert a strong influence on the quality of the subsequent reconstruction. Misidentification of the interaction vertex location can result in splitting of the trajectory of a single particle and merging of multiple trajectories, as well as incorrect parent-child relationships (e.g. the two photons from $\pi^0$ decay are children of the parent pion) between particles. Depending on the magnitude of the error, these problems can affect estimates of reconstructed energy and incident neutrino direction, or could lead to mischaracterisation of the interaction type and thereby alter the interpretation of an interaction. In this article we describe an approach to determine the interaction vertex location using a deep neural network, which represents the first integration of support for deep learning algorithms into Pandora.

\begin{figure*}[tbh]
  \centering
  \includegraphics[clip, trim=0.0cm 4.5cm 0.0cm 4.0cm, width=1\textwidth]{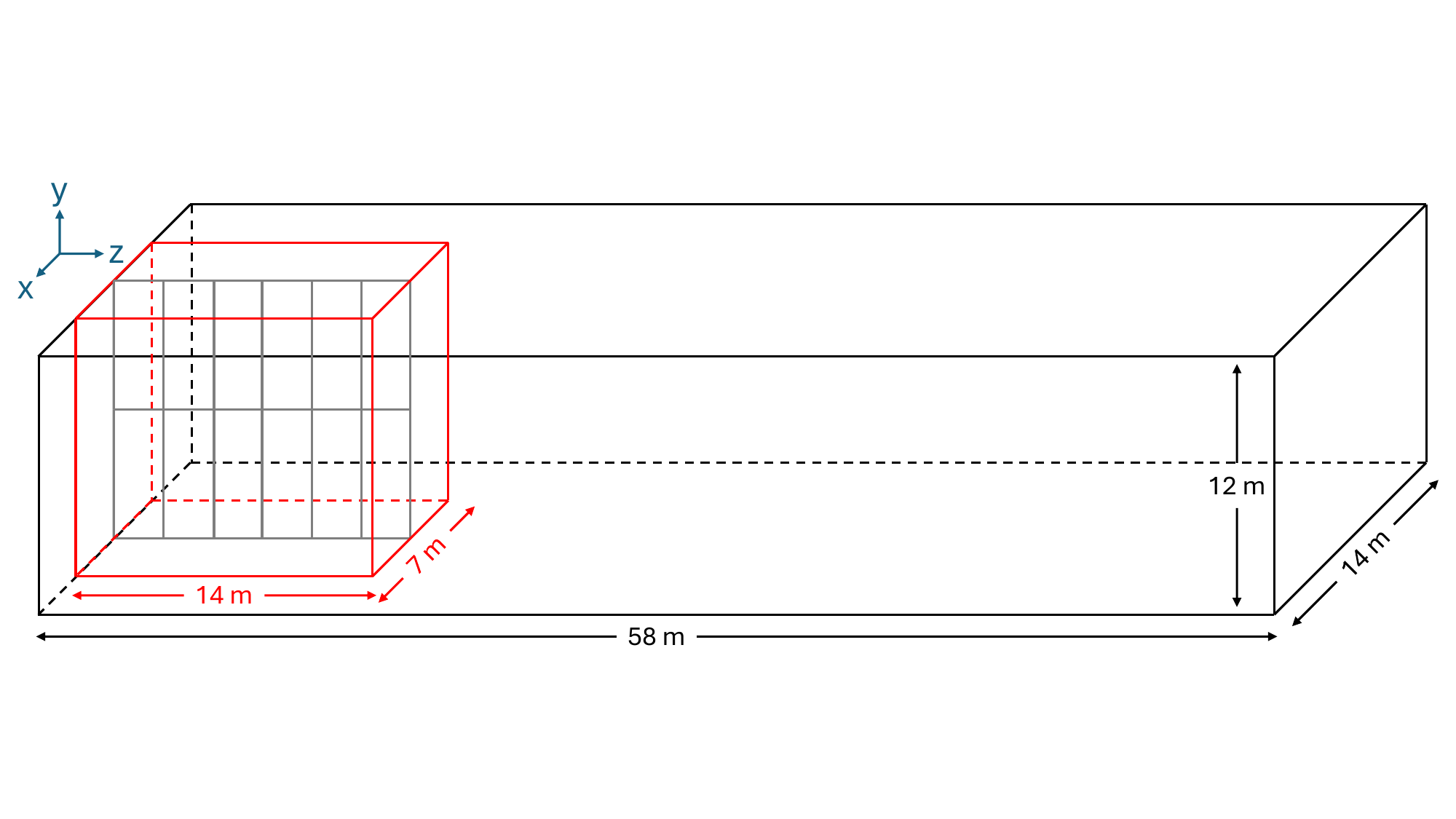}
  \caption{Schematic of the workspace geometry (red) within the context of the full detector geometry (black). APAs are indicated by gray boxes in the YZ plane of the workspace volume, while CPA walls are indicated in red in the YZ plane. APA walls in the full geometry are indicated by the black boxes in the YZ plane.}
  \label{fig:workspace_geo}
\end{figure*}

Section~\ref{sec:lartpcs} briefly introduces LArTPCs, while Section~\ref{sec:simulated_data} describes the simulated data used for this work. The conceptual approach to finding the vertex and the corresponding truth definition are presented in Section~\ref{sec:concept}. Section~\ref{sec:training} summarises the network architecture, provides details of the training metrics and assesses the behaviour of the loss function and network weights. Results in a simulated experimental environment are presented in Section~\ref{sec:results:accel}. The performance characteristics, sources of bias and robustness considerations are discussed in Section~\ref{sec:model_dep} and a number of extensions/alternative approaches planned for future work are described in Section~\ref{sec:extensions}.

\section{Liquid argon time projection chambers}\label{sec:lartpcs}
\begin{figure*}[tbh]
  \centering
  \includegraphics[clip, trim=0.0cm 1.0cm 0.0cm 1.0cm, width=0.75\textwidth]{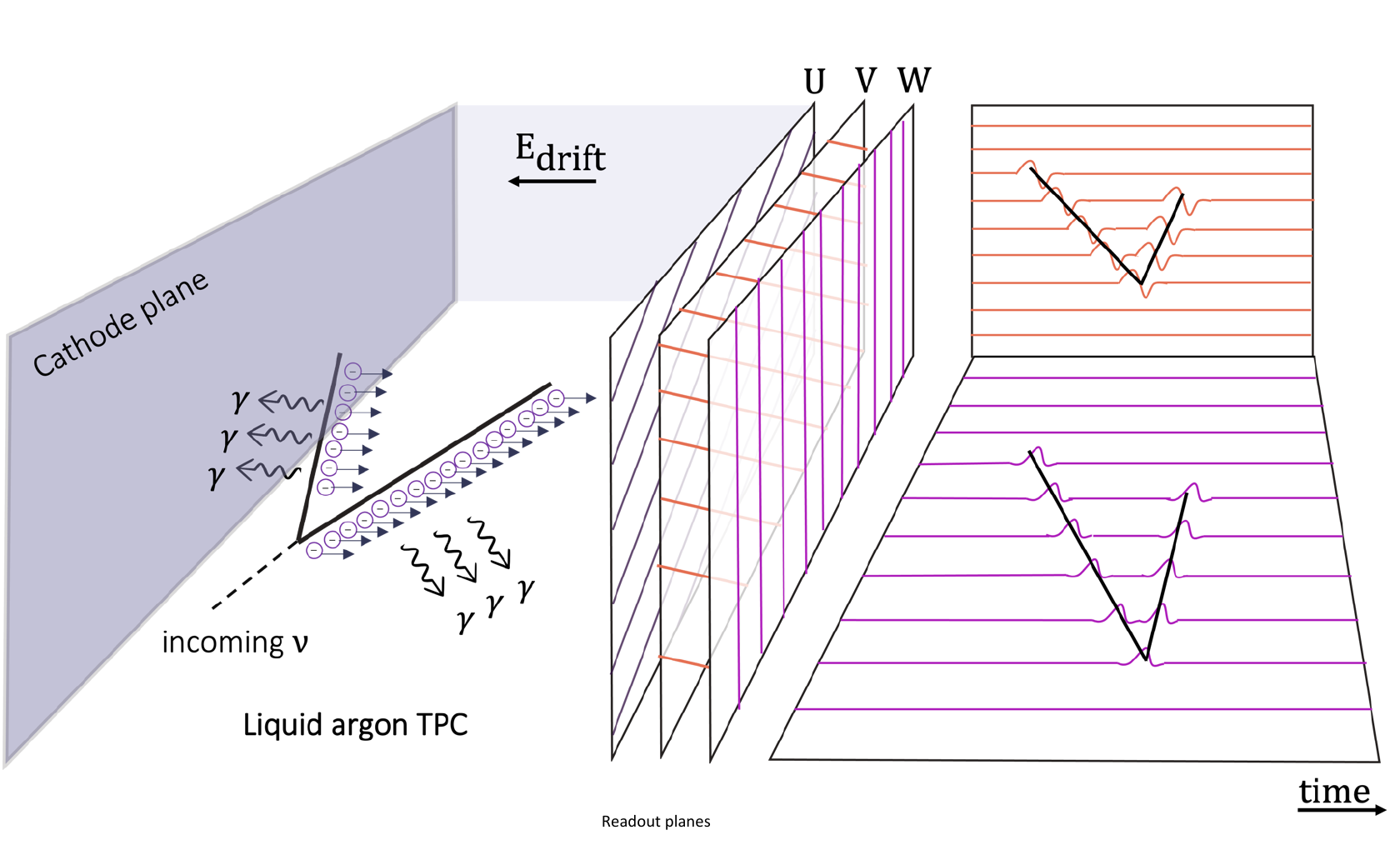}
  \caption{Operating principle of a LArTPC \cite{dunehd}. Charged particles produced by the incident neutrino interaction ionise the argon medium, producing ionization electrons that drift towards anode induction (U and V) and collection (W) planes, along with scintillation photons.}
  \label{fig:lartpc}
\end{figure*}

Two far detector designs currently exist for DUNE, the horizontal-drift \cite{dunehd} and vertical-drift \cite{dunevd} detectors. These two designs utilise different geometries, orientations, drift lengths and readout technologies, the details of which are left to the aforementioned references. Conceptually the operation is the same, and Pandora is agnostic to the detector differences. A LArTPC volume is a fully active liquid argon target, with a uniform electric field applied between a cathode and anode. Charged particles propagating through this medium ionise the argon (see Fig.~\ref{fig:lartpc}), producing drift electrons and scintillation photons. The drift electrons are carried by the electric field to the anode where they are read out on three readout planes: Two planes (U and V) with voltage biased to be transparent to the electrons and thereby allow current to be induced in readout channels as the electrons pass by, and a readout plane (W) with voltage set to permit collection of the drift electrons on the readout channels. The scintillation light is detected by photon detectors embedded in anode and cathode planes \cite{xarapuca}, though Pandora currently only reads charge information coming from the TPC waveforms, and does not yet read in scintillation light information and therefore this will not be discussed further in this article.

\section{Simulated data}\label{sec:simulated_data}
Given Pandora's agnosticism to the detector design, this article will focus on samples in the horizontal-drift detector and that come from beam neutrinos. Each simulated event represents a single readout window (a span of time allowing the ionisation electrons from a triggered event to drift the width of the detector), with the triggered neutrino beam placed within the window such that the time of the neutrino interaction corresponds to $t_0 = 0$. Given the rock overburden the rate of cosmic ray  muons is expected to be approximately $1\times10^{-4}$ per typical readout window (5.4\,ms) in the full volume of a horizontal-drift far detector and therefore no cosmics are simulated.

\begin{figure*}[tbh]
  \centering
  \includegraphics[width=0.5\textwidth]{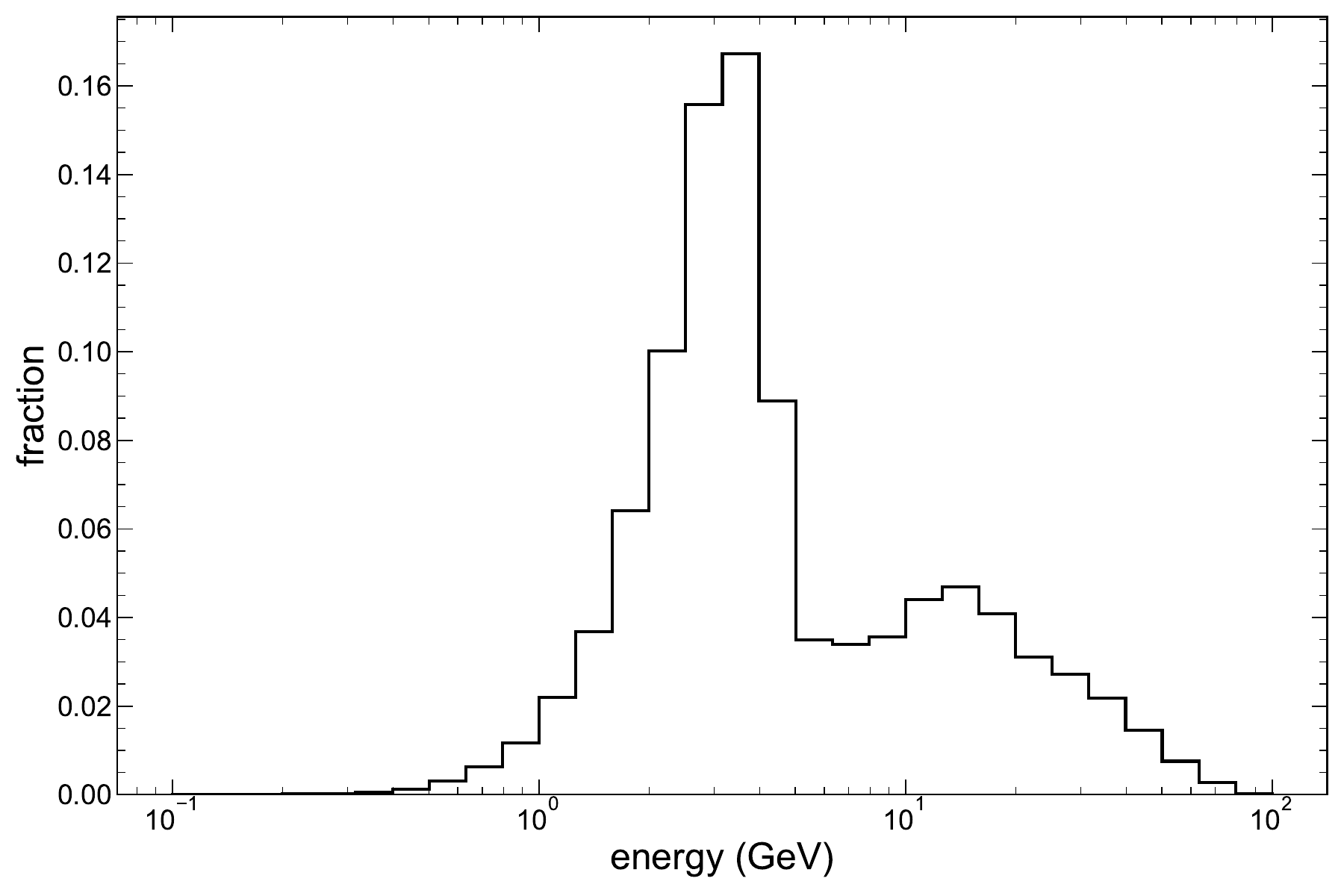}
  \caption{Distribution of true neutrino energy for the combined (FHC + RHC) simulated, unoscillated far detector interactions.}
  \label{fig:nu_energy}
\end{figure*}

Three samples of simulated data were generated for each of forward horn current (FHC) and reverse horn current (RHC) beam running, producing beams composed predominantly of neutrinos and antineutrinos, respectively. The expected unoscillated long-baseline neutrino facility (LBNF) flux \cite{dunehd_physics}, folded with cross-section, yields a $\nu_\mu$-dominated sample with a small contribution from intrinsic $\nu_e$ in FHC operation. The other two samples in FHC are produced by applying flavour swapping to this sample, producing a $\nu_e$-dominated sample via the swaps $\nu_\mu\rightarrow\nu_e$ and $\nu_e\rightarrow\nu_\tau$, and a $\nu_\tau$-dominated sample via the swaps $\nu_\mu\rightarrow\nu_\tau$ and $\nu_e\rightarrow\nu_\mu$. Equivalent swaps are applied for the RHC samples. The true neutrino energy distribution of the combined sample (FHC + RHC) is shown in Fig.~\ref{fig:nu_energy}.

Samples are produced using a suite of software comprising DUNESW v09\_81\_00d02. In particular, neutrino events are generated using GENIE v3.04.00 \cite{genie, genie_toolkit} with the Liquid Argon Experiment tune AR23\_20i\_00\_000, the propagation of particles and their interaction within the detector is simulated by Geant4 v10.6.p01f \cite{geant, geant_update1, geant_update2} with the QGSP\_BERT physics list. The electronics and field response is simulated using Wire-Cell, which also processes the signal to recover the original wire waveforms, as described in \cite{wirecell_microboone}. Hit finding is then performed by LArSoft v09\_81\_00 \cite{larsoft,larsoft_web}. The resultant hits are passed to the Pandora pattern recognition, where reconstruction of events is performed using LArPandora v9\_21\_12, which depends upon PandoraSDK v03-04-01 \cite{pandora_sdk}.  In the case of the horizontal-drift far detector \cite{dunehd}, the detector is represented by a subset of the full far detector geometry, known as the 1x2x6 workspace geometry (see Fig~\ref{fig:workspace_geo}), with an active volume of approximately $719\text{\,cm}\times{}1208\text{\,cm}\times{}1394\text{\,cm}$ (X$\times{}$Y$\times{}$Z). This is intended to provide good containment across a broad range of events. The detector volume is defined by twenty four drift volumes, arranged about twelve anode plane assemblies (APAs --- the wire planes that sense the ionisation electrons), such that the workspace geometry is two APAs high and six deep.

\section{Conceptual overview of vertex finding}\label{sec:concept}

\begin{figure*}[tbh]
    \begin{subfigure}{0.45\textwidth}
          \centering
          \includegraphics[clip, trim=3cm 0.1cm 3cm 3cm, width=1\textwidth]{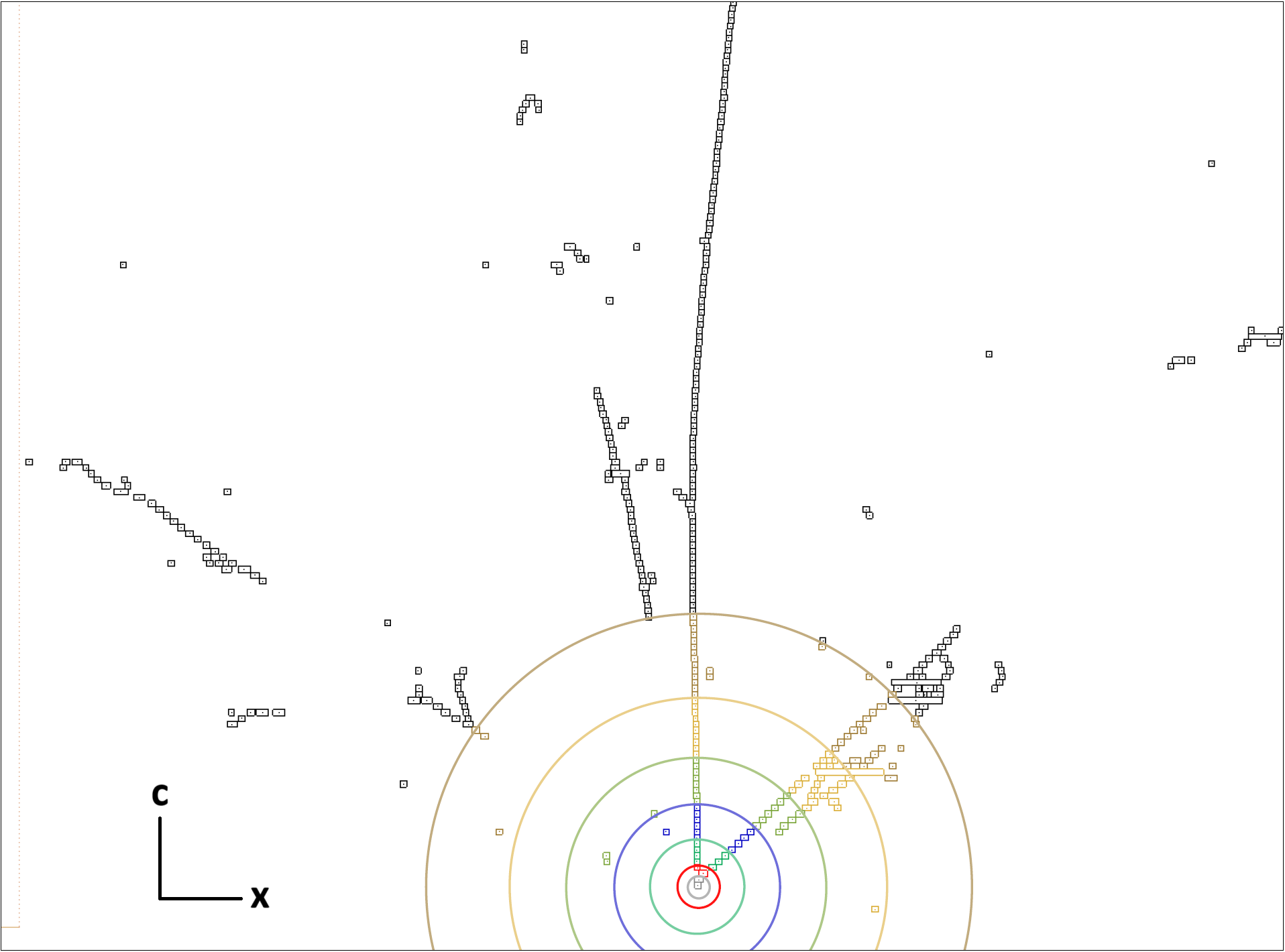}
          \caption{}
          \label{fig:hit_classes}          
    \end{subfigure}
    \hspace*{\fill}
    \begin{subfigure}{0.45\textwidth}
          \centering
          \includegraphics[clip, trim=3cm 0.1cm 3cm 3cm, width=1\textwidth]{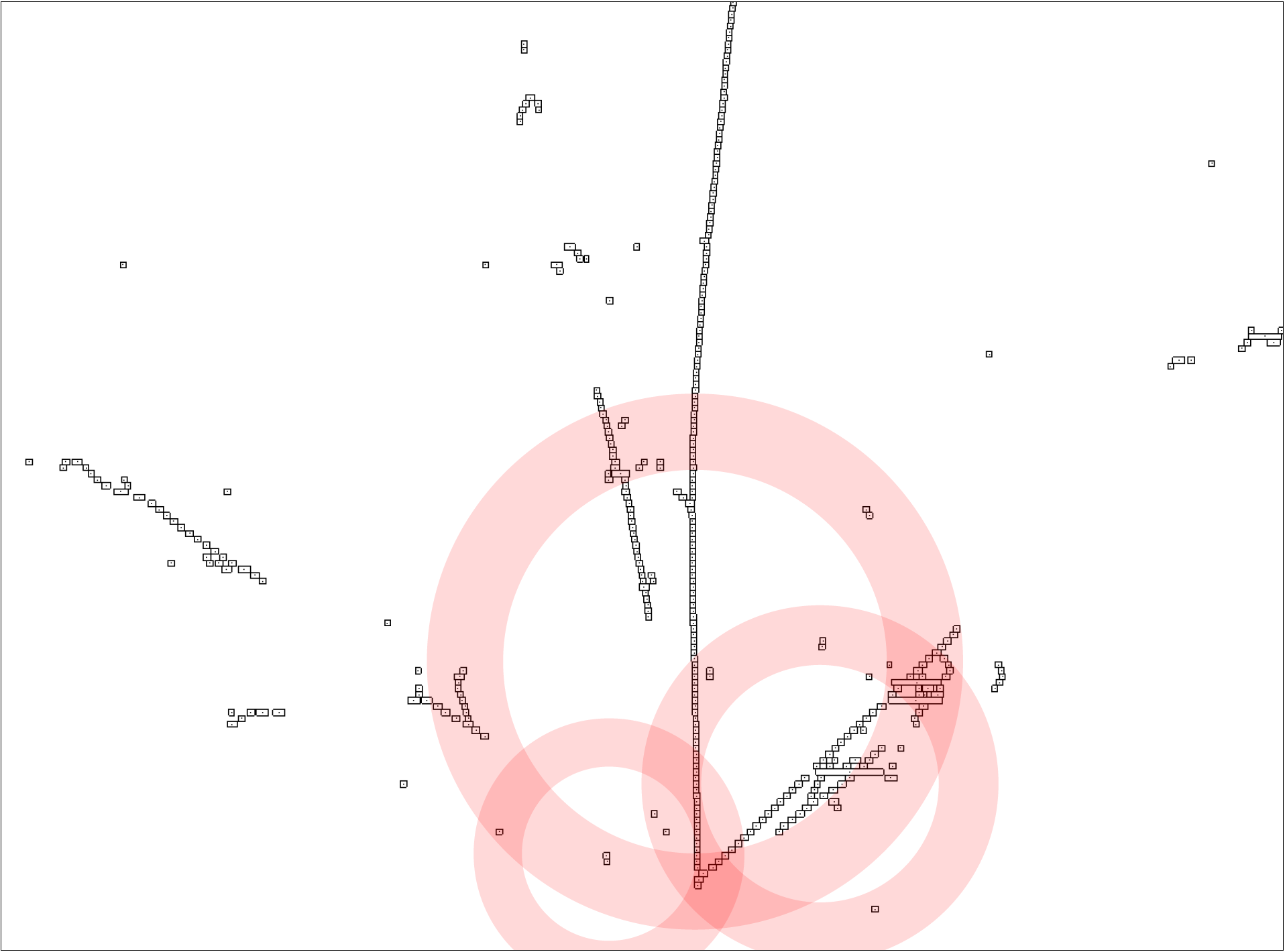}
          \caption{}
          \label{fig:hit_infer}
    \end{subfigure}
    \caption{An example of (a) the input hits and assignment of the first seven of the nineteen true distance classes for those hits and (b) a schematic of the heat map produced by three arbitrary hits during inference, for one view (W) of an event.}
\end{figure*}

The outputs of LArTPC detectors can be represented as two-dimensional images (see Fig~\ref{fig:hit_classes}) where charge deposition is represented as a hit. For each readout plane, a hit is composed of a triplet (x, c, q), where x is a drift coordinate indicating the distance traveled by drift electrons from the point of ionisation to the relevant readout plane, c is a coordinate representing the readout channel number, and q is a scalar quantity proportional to the amount of charge induced or deposited on the readout channel. In addition to the position and charge deposition for the hit, the width of the Gaussian fitted to the signal waveform is also retained, and is referred to as the hit width.

Upon visual inspection by a human expert, the interaction vertex is often, though not always, easily identified as a clear feature of the interaction from which all activity emanates. It therefore seems reasonable to suppose that modern machine learning techniques, such as deep convolutional neural networks, would be well-suited to automate this inherently visual task.

The concept for the network design adopted here is to relate each hit to the interaction vertex in terms of its distance from the vertex (see Fig.~\ref{fig:hit_classes}). This avoids the need to define coarse signal and background regions for the events, or target coordinates for regression. Each hit can contribute in a relatively precise way to the loss function that quantifies the network's accuracy, and spatial correlations between hits can help provide context. In particular, a hit can be allocated a pixel coordinate $(h_x, h_c)$, while the vertex can be similarly allocated to a pixel coordinate $(v_x, v_c)$ ($x$ indicating drift coordinate and $c$ readout channel in U, V, or W), with the distance between those pixels computed as

\begin{equation}\label{eq:distance}
    D = \frac{\sqrt{(v_x - h_x)^2 + (v_c - h_c)^2}}{\ceil[\Big]{\sqrt{2(L - 1)^2}}}
\end{equation}

where $L$ is the width (and height) of the image in pixels and acts to provide scale invariance for the network inputs. This distance measure is then allocated one of 19 discrete class labels for the network to learn (Fig.~\ref{fig:hit_classes}), covering a range of distances which can then be used as a target for pixel-level semantic segmentation~\cite{Ronneberger2015}. Thus, the network does not extract a vertex location directly, rather it infers a class that describes a range of distances between the pixel under consideration and the true vertex location. This information can then be used to project rings (Fig.~\ref{fig:hit_infer}) of appropriate inner and outer radii, centred on an active pixel of interest to form a heat map where the pixel with the most ring intersections would be considered the candidate vertex. The width of the rings depends upon the distance of the corresponding class to the interaction vertex. Rings describing pixels that are in close proximity to the candidate interaction vertex have narrow rings, while those rings broaden for classes at greater distances to the true/estimated interaction vertex. This reflects the expectation that it is more challenging to estimate the distance to the interaction vertex at large distances than nearby.

Neutrino interactions at DUNE can have a very large physical extent. Muons, for example, can travel many metres through the liquid argon medium. To avoid an excessive computational burden in processing, the images passed to the network are no more than $256 \times 256$ pixels. This can yield a resolution as coarse as $\sim{}5$\,cm per pixel for an event spanning the length of the detector, so an additional step is needed to improve the resolution to the scale of the channel spacing. As a result, a choice must be made about how to present the event to the network such that an acceptable vertex location precision can be obtained. Given that the interaction vertex location is unknown at this stage, it is desirable to retain all hit information from the image, rather than attempt to crop the image and risk losing the region around the interaction vertex. However, as noted, the large physical extent of many events can lead to individual pixels covering many readout channels, and therefore a two pass approach is adopted. The first pass uses as input all hits from the event, fitting this to the input image dimensions (the dimensions are scaled independently to fit the image size), allowing a low-resolution identification of the likely vertex location. Given this location, a second pass of the network is undertaken, zooming in on the region around this provisional location, to identify the vertex location to high-precision (the framing of the event in this second pass will be described in section~\ref{sec:training}). Each network produces a vertex position within its respective 2D readout plane, and these therefore need to be consolidated into a single 3D position. This is achieved by considering each combination of readout plane pairs to identify the Y and Z coordinate intersection of the two planar points based on the known position and orientation of the channels in each view, with X being directly available for each plane. The three sets of X, Y and Z coordinates are then averaged to produce a single 3D location, which is then projected back into each of three readout planes based on the known detector geometry, and a $\chi^2$ is computed between the projected and original vertex positions as a consistency check. If the $\chi^2$ value is less than 1, the 3D vertex is accepted, otherwise the 3D vertex corresponding to the two plane combination with the lowest difference between X coordinates is accepted.

To leverage the machine learning techniques within the standard Pandora reconstruction workflow, an interface to LibTorch, the binary distribution of PyTorch \cite{pytorch}, was implemented in Pandora. The interface is lightweight, wrapping the torch::jit::script::Module and the at::Tensor and torch::Tensor types, along with provision of a small number of helper functions to load the model, initialise tensors and run the inference step. This interface allows Pandora to run any TorchScript compatible network on a CPU as part of its chain of algorithms, requiring only suitably structured input tensors and code to process the network output upon its return. The computational costs associated with the vertex reconstruction step are modest: Running over 1182 events, approximately evenly split between the neutrino flavours, the maximum resident set size averaged $207\pm{}12$\,MB, while the average computation time for both passes on all three readout planes was $0.96\pm{}0.02$\,s running on an Intel Core i3-9100 CPU @ 3.60GHz.

\section{Training and understanding the network}\label{sec:training}
Given the previously described formulation of the problem in terms of semantic segmentation, the architecture used has a dense U-ResNet structure~\cite{Ronneberger2015, He-et-al-2015-deep}. The network has four blocks in the down-sampling path (extracting features) and in the up-sampling path (recovering spatial detail), with all convolutional filters of size 3x3. The convolutional blocks implement residual skip connections between the input to those blocks and the final ReLU activation function of the respective block (adding the input to the block to the intermediate network output prior to the activation function). Due to the limited training batch size of 32 images per batch required to fit within GPU memory (16~GB) during training, group, rather than batch, normalisation is applied within convolutional and transpose convolutional blocks to ensure the inputs to network layers don't drift as training proceeds.

%
Six networks were trained for the vertex finding task, one network for each wire plane for the low resolution pass and then three equivalent networks for the resolution refining second pass.

The network architecture and distance class definitions were identical in each case, as was the choice for the loss function, categorical cross-entropy loss, which measures the distance between the network's estimated probability distribution and the underlying true probability distribution for the classes, and is given by

\begin{equation}\label{eq:loss}
    L_n = -w_{y_n}\log{\frac{\exp{(x_{n,y_n})}}{\sum_{c=1}^{C}\exp{(x_{n,c})}}}
\end{equation}
(PyTorch's default weighted cross-entropy loss) where $n$ is the batch number, $C$ the number of classes, $x$ the input and $y$ the target, with class weightings $w$. The optimiser was Adam \cite{Adam2014}, with default PyTorch parameters. The two passes used different image sizes; $256\times{}256$ pixels in pass 1, with the entire event scaled down to fill the image canvas, while the second pass used a $128\times{}128$ pixel input image, where images were cropped, as needed, to ensure a single pixel per channel pitch for the channel coordinate and 0.5\,cm per pixel for the drift coordinate. While there are small variations between collection and induction readout plane channel pitches, for the horizontal-drift far detector, this corresponds to a region covering approximately 0.61\,m in channel coordinates and 0.64\,m in drift coordinates. The framing of the second pass images attempts to retain as much useful information around the vertex as possible. Centring the image on the provisional vertex location is likely wasteful, as activity around the vertex is highly likely to be downstream of the incident neutrino direction. As such, the fraction of hits on the left and right, and upstream and downstream of the provisional vertex location is determined, and the image centre is then shifted to better cover the region with greatest activity. That is, if 80\% of all hits are to the left of the provisional vertex, 80\% of the image area is also to the left of the provisional vertex.

For an image to be considered for training, the true neutrino interaction vertex was required to reside within a defined fiducial volume with  respect to the workspace geometry, 50\,cm inside the left, right, top, bottom and upstream (with respect to the beam direction) faces of the detector, and 150\,cm inside the downstream face. Furthermore, there must be at least 10 hits in the image and the true vertex must be no more than 1\,cm outside of the region containing the hits in either readout channel or drift coordinate, which excludes $\sim{}5\%$ of $\nu_\mu$ and $\nu_e$ images and $\sim{}6\%$ of $\nu_\tau$ images from consideration. This allows the network to learn the location of vertices that are slightly offset from the associated visible charge deposition, even when the closest hit is at an extreme edge of the image. Under these conditions the total number of images available for the first pass training were \nuimagestrainone, \nvimagestrainone, and \nwimagestrainone for the U, V and W views, respectively and for validation there were \nuimagesvalone, \nvimagesvalone, and \nwimagesvalone for the U, V and W views, respectively.

\begin{figure*}[tbh]
    \begin{subfigure}{0.49\textwidth}
        \centering
        \includegraphics[width=\textwidth]{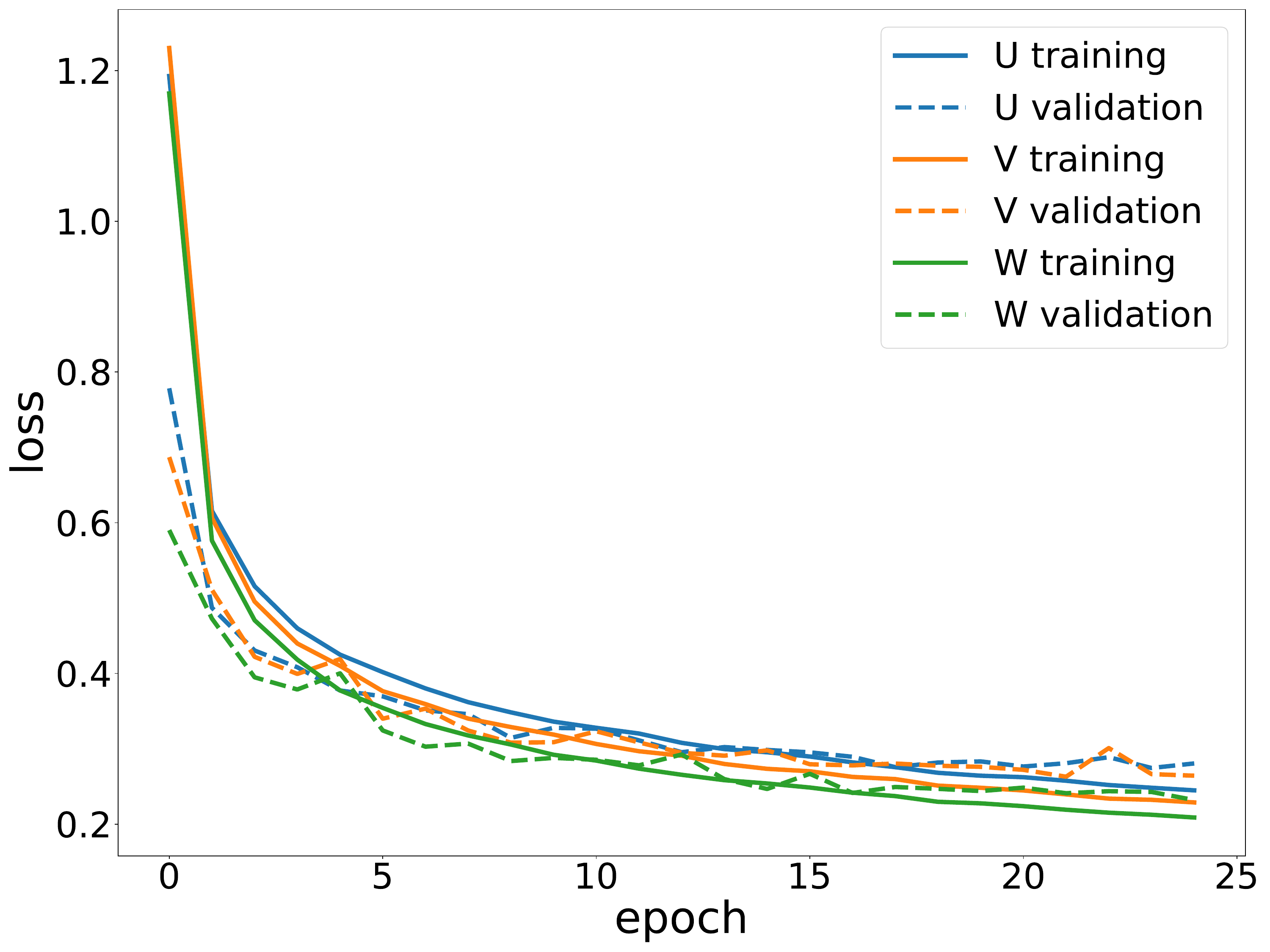}
        \caption{}
        \label{fig:stats_loss_1}
    \end{subfigure}\hfill
    \begin{subfigure}{0.49\textwidth}
        \centering
        \includegraphics[width=\textwidth]{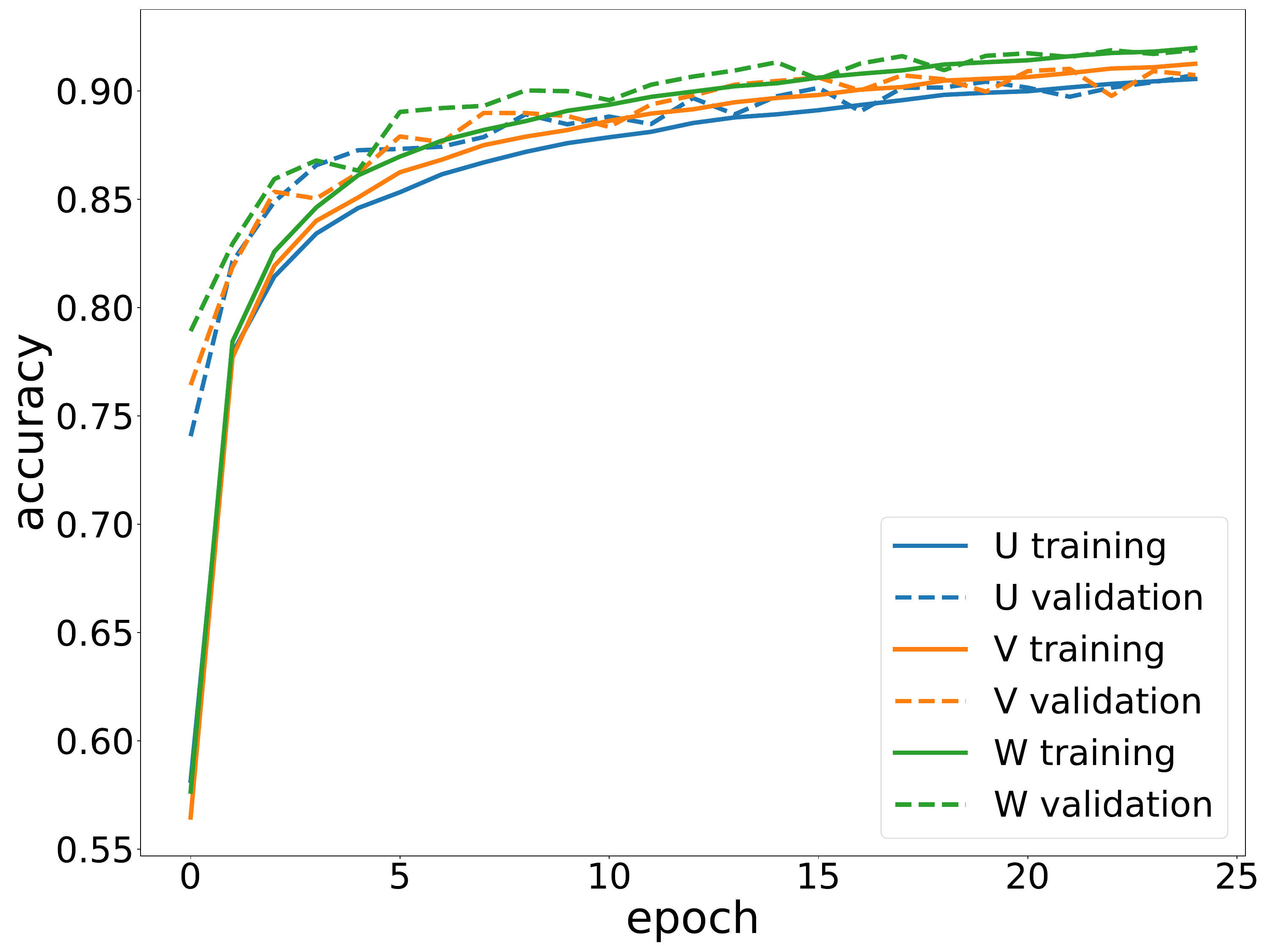}
        \caption{}
        \label{fig:stats_acc_1}
    \end{subfigure}
    \begin{subfigure}{0.49\textwidth}
        \centering
        \includegraphics[width=\textwidth]{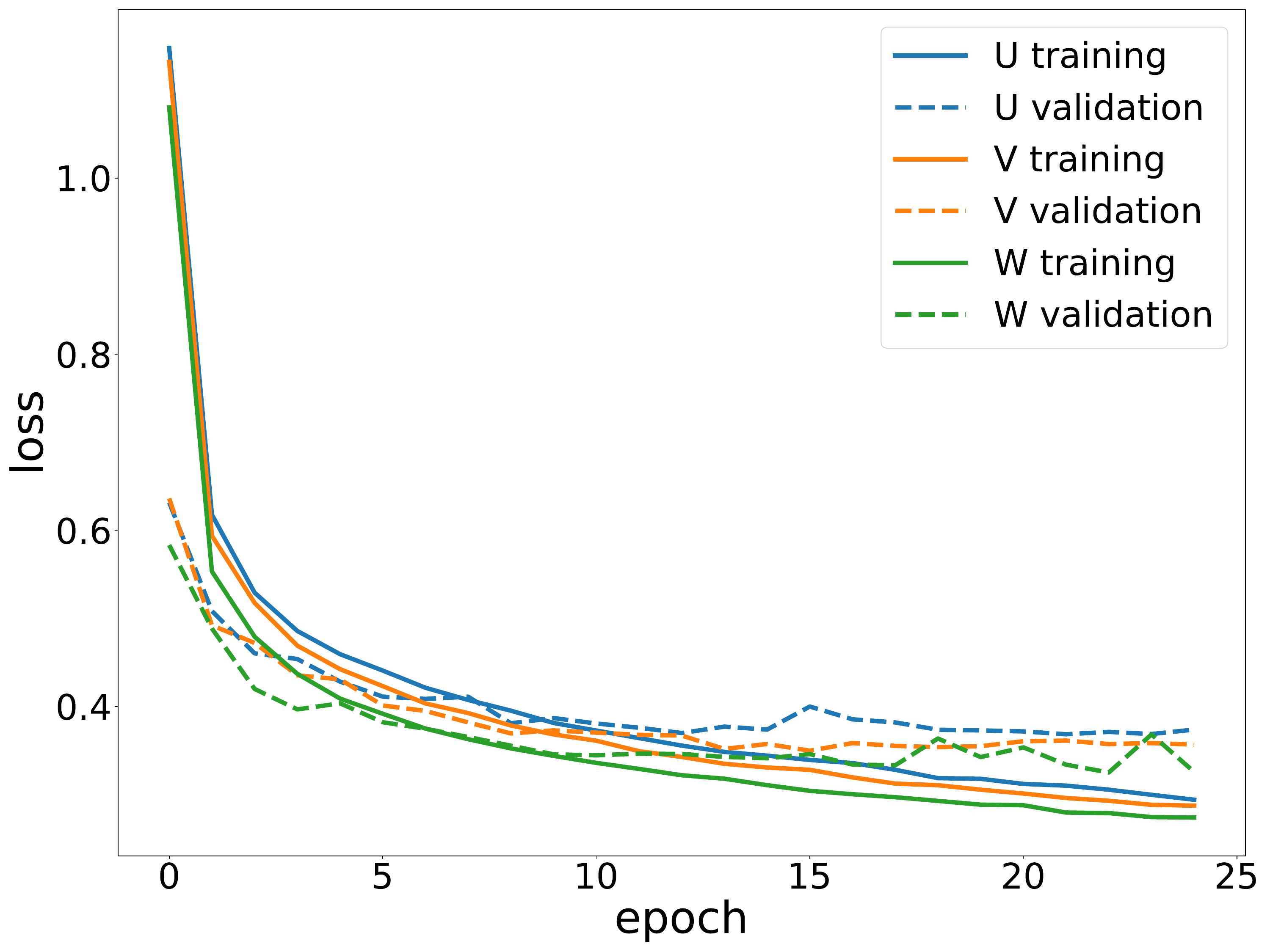}
        \caption{}
        \label{fig:stats_loss_2}
    \end{subfigure}\hfill
    \begin{subfigure}{0.49\textwidth}
        \centering
        \includegraphics[width=\textwidth]{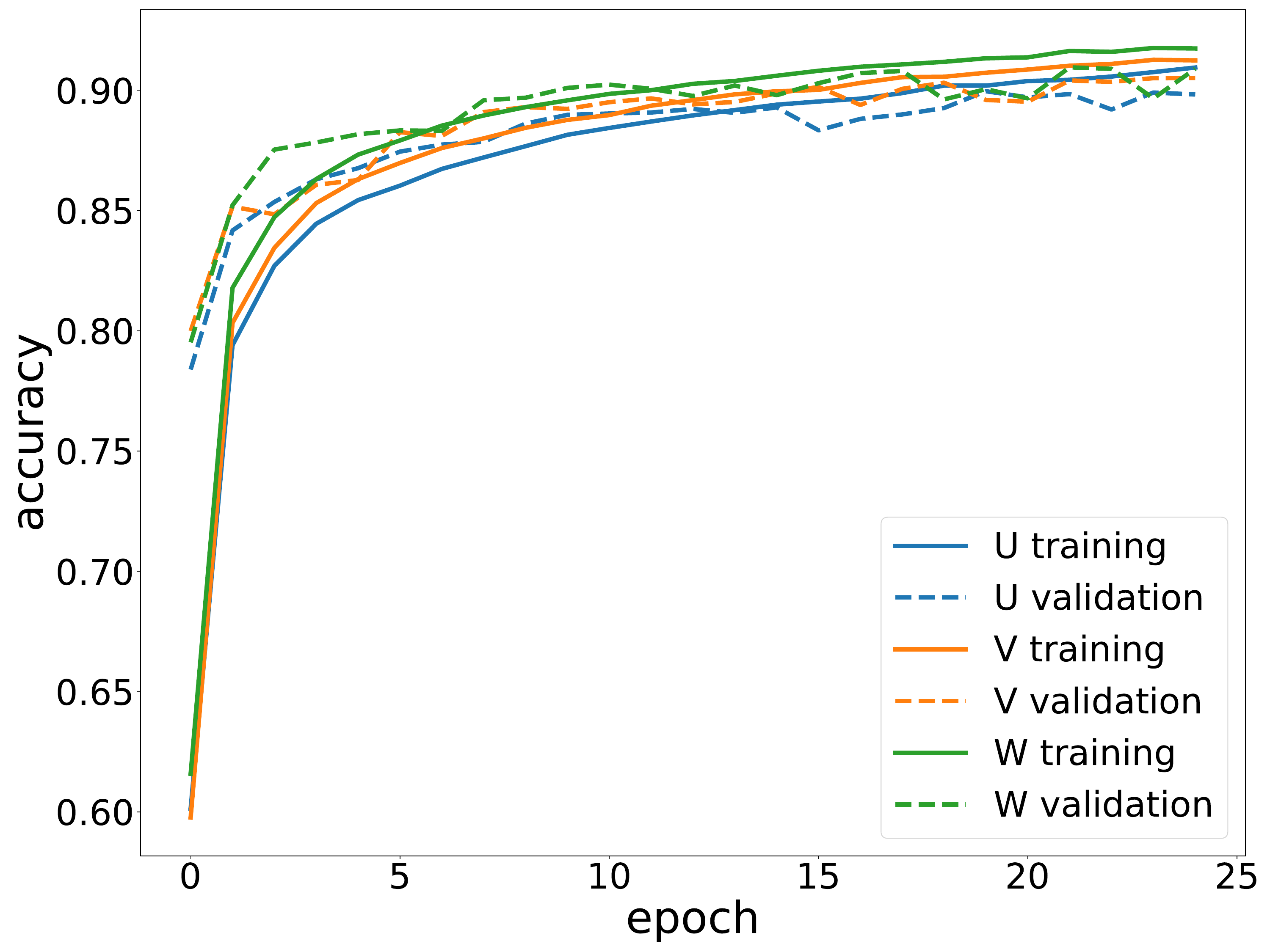}
        \caption{}
        \label{fig:stats_acc_2}
    \end{subfigure}
    \caption{Training (a) loss and (b) accuracy for all views in pass 1. Training (c) loss and (d) accuracy for all views in pass 2.}
\end{figure*}

For the second pass training, with equivalent inclusion cuts applied, there were \nuimagestraintwo, \nvimagestraintwo, and \nwimagestraintwo for the U, V and W views, respectively. Note that images are excluded if errors in the first pass produce framing of the candidate vertex region that does not contain the true vertex. For validation there were \nuimagesvaltwo, \nvimagesvaltwo, and \nwimagesvaltwo for the U, V and W views, respectively. Each network was trained for \ntrainepochs epochs and the evolution of the network performance across each of those training iterations is shown in Fig.~\ref{fig:stats_loss_1}-\ref{fig:stats_acc_2}.

In all first pass cases the validation loss and accuracy (the fraction of correctly classified hits) appear to plateau beyond the 20th epoch, with no evidence of divergence between the training and validation loss. In second pass training there is modest evidence of such divergence beyond epoch 12 and thus the chosen model is that of epoch 12. Across all views and both passes, approximately 90\% of hits in the validation set generate rings containing the true interaction vertex. As a result, a very large majority of hits will contribute weights to the heat map that contain the true vertex location, overwhelming any noise from the errant hits, and so one would expect effective reconstruction of the interaction vertex.

\begin{figure*}[tbh]
    \begin{subfigure}{0.32\textwidth}
        \centering
        \includegraphics[width=0.9\textwidth]{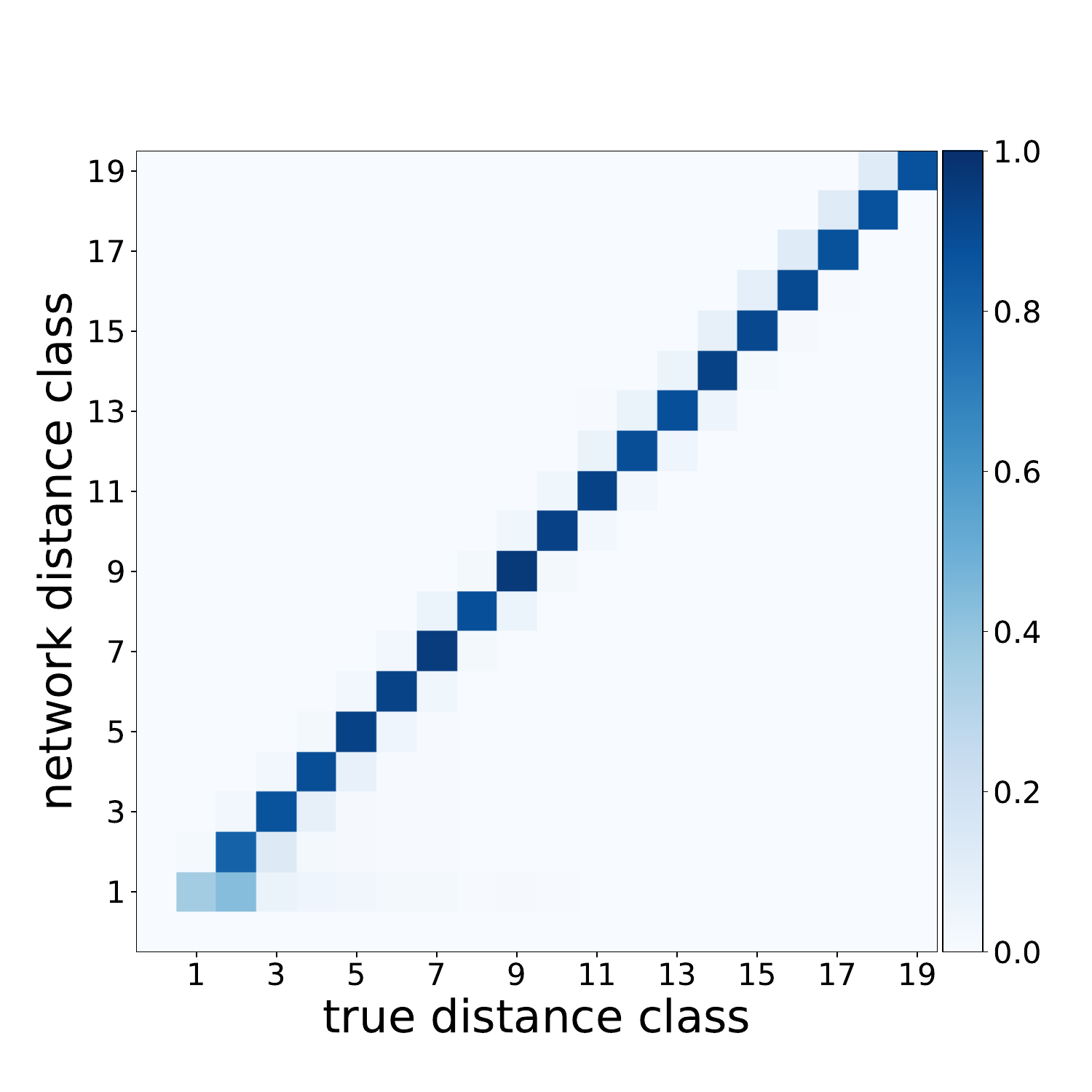}
        \caption{}\label{fig:confusion_1}
    \end{subfigure}\hfill
    \begin{subfigure}{0.32\textwidth}
        \centering
        \includegraphics[width=0.9\textwidth]{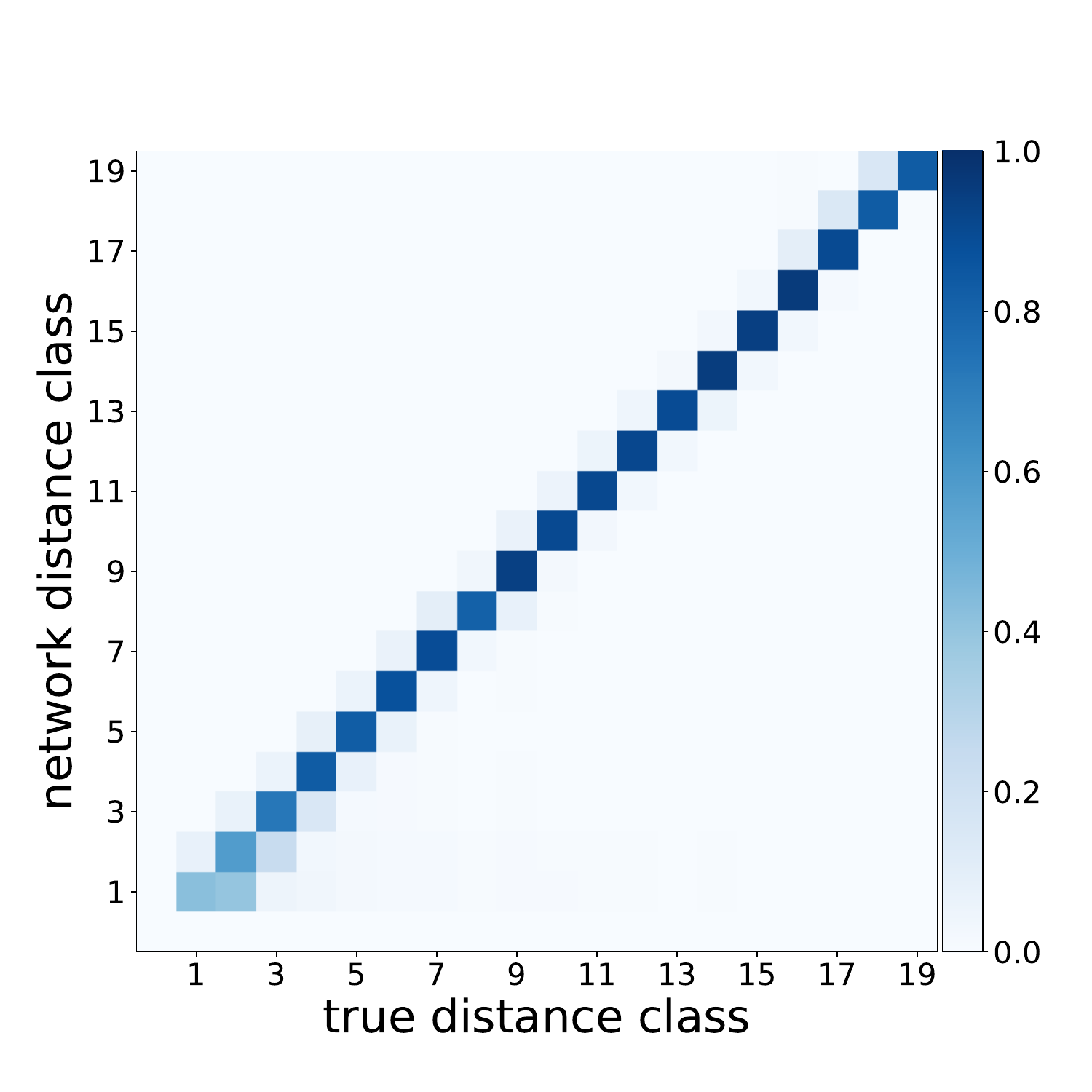}
        \caption{}\label{fig:confusion_2}
    \end{subfigure}\hfill
    \begin{subfigure}{0.32\textwidth}
        \centering
        \includegraphics[clip, trim=0cm 0cm 0cm 0cm, width=0.99\textwidth]{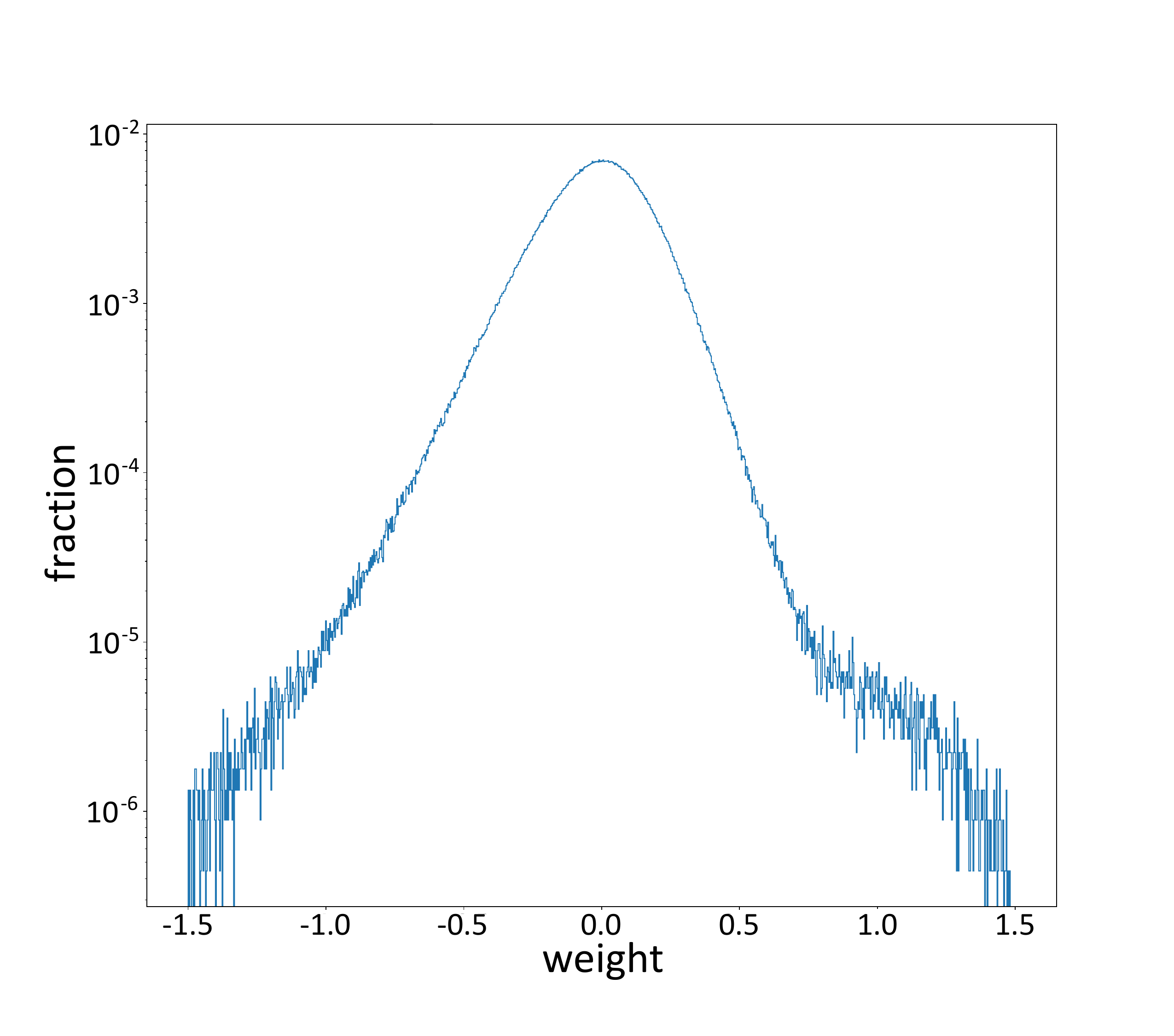}
        \caption{}\label{fig:weight_distribution}
    \end{subfigure}
    \caption{Confusion matrix for the W view in (a) pass 1 and (b) pass 2 (the null class (0) is zero-suppressed) and (c) the distribution of weights for the trained W view, pass 1 network.}
\end{figure*}

Also of interest are the confusion matrices for the various classes of the different networks. As already noted, per class accuracy is high, but one can also observe the distribution of classification errors in Fig.~\ref{fig:confusion_1}-\ref{fig:confusion_2} for the W view, with similar distributions for the U and V views, which are not shown. The off-diagonal contribution where errors are made is typically found in a class adjacent to the true one. The region in which most errors are made is that in the immediate vicinity of the true interaction vertex, which is unsurprising given the relative infrequency of hits in this region (geometric considerations result in fewer pixels belonging to the classes closest to the vertex).

\begin{figure*}[tbh]
    \begin{subfigure}{0.32\textwidth}
        \centering
        \includegraphics[clip, trim=15cm 4cm 9.5cm 5cm, width=\textwidth]{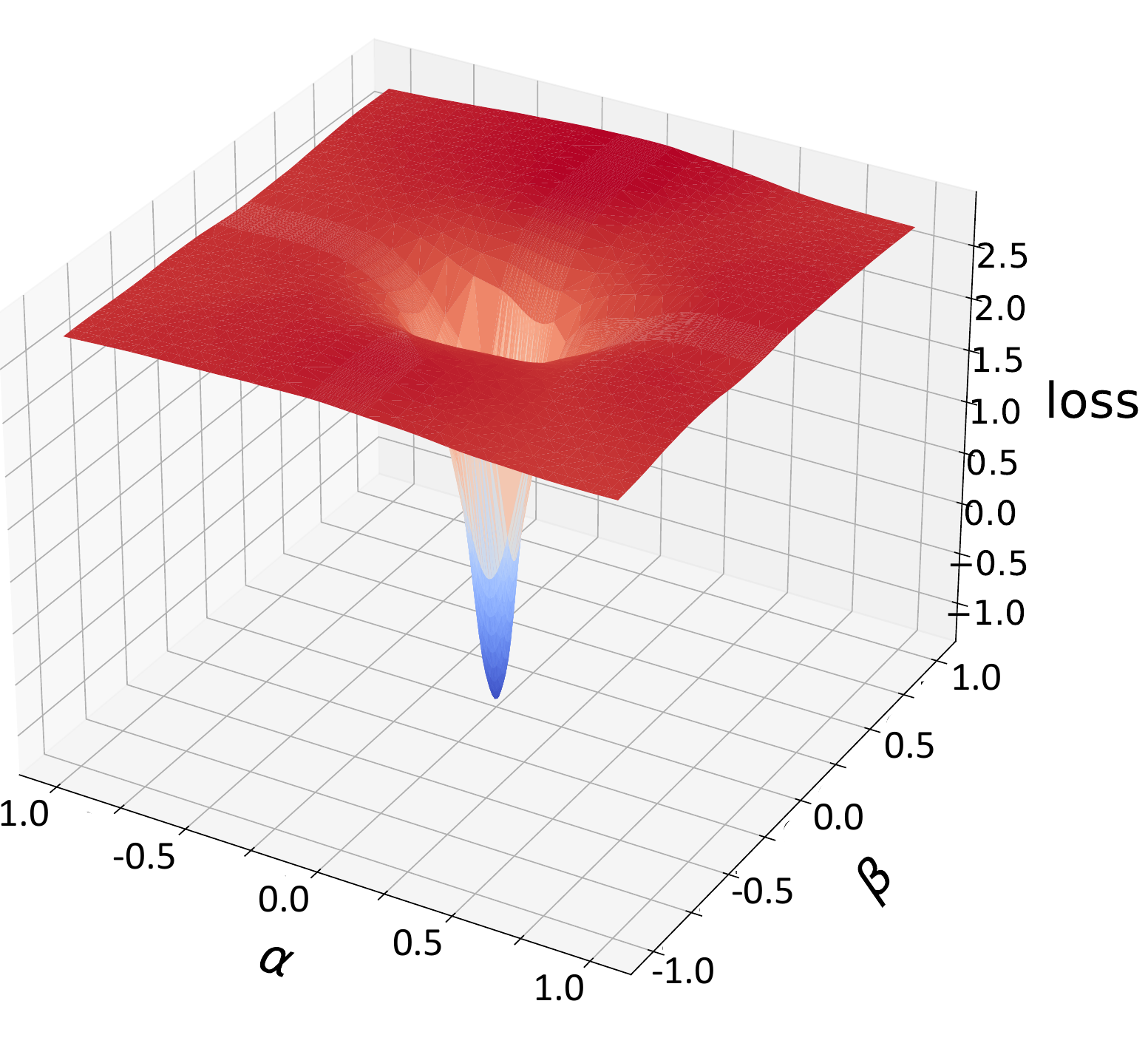}
        \caption{}\label{fig:loss_landscape_1}
    \end{subfigure}\hfill
    \begin{subfigure}{0.32\textwidth}
        \centering
        \includegraphics[width=\textwidth]{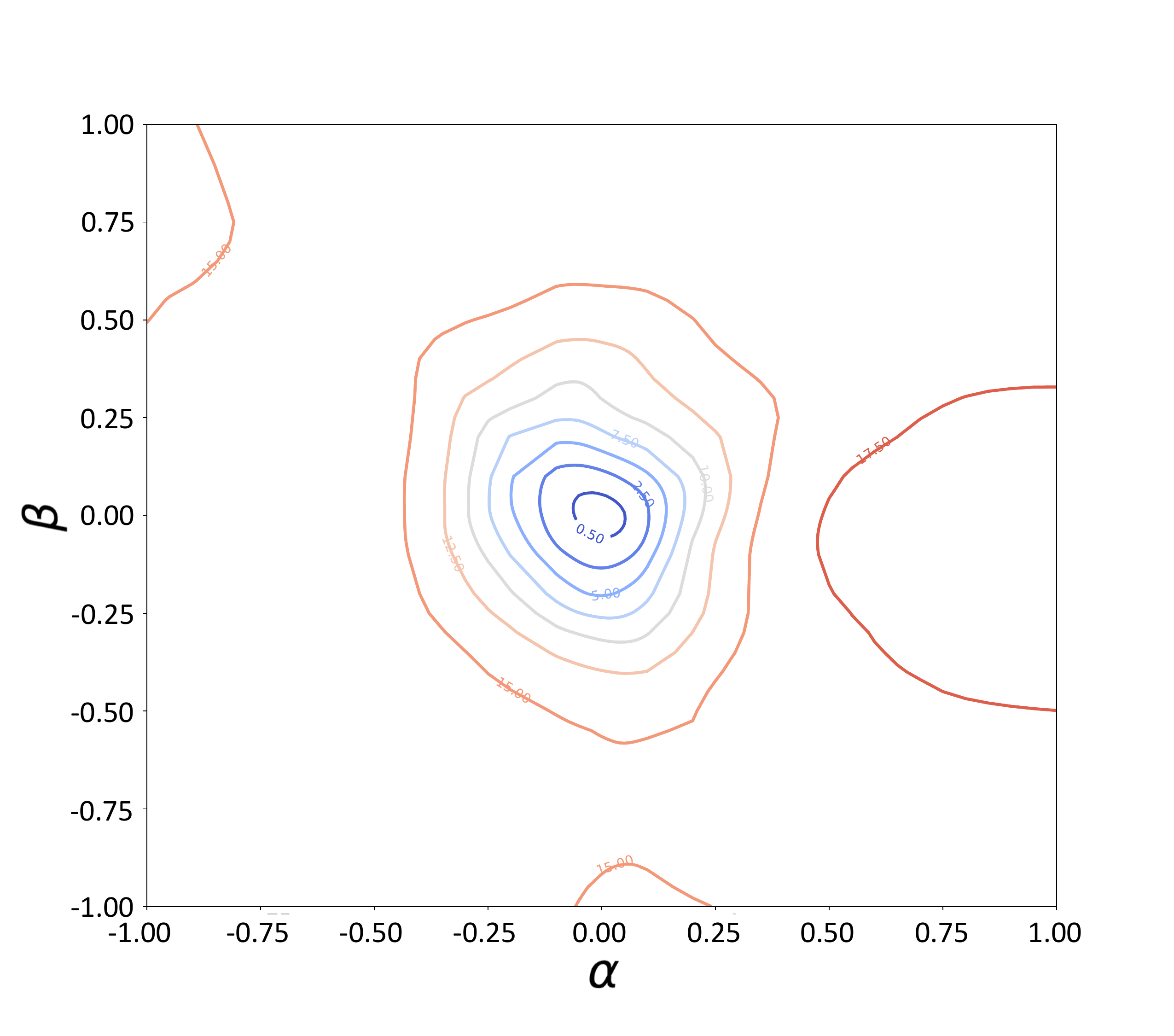}
        \caption{}\label{fig:loss_landscape_2}
    \end{subfigure}\hfill
    \begin{subfigure}{0.32\textwidth}
        \centering
        \includegraphics[clip, trim=15cm 4cm 9.5cm 5cm, width=\textwidth]{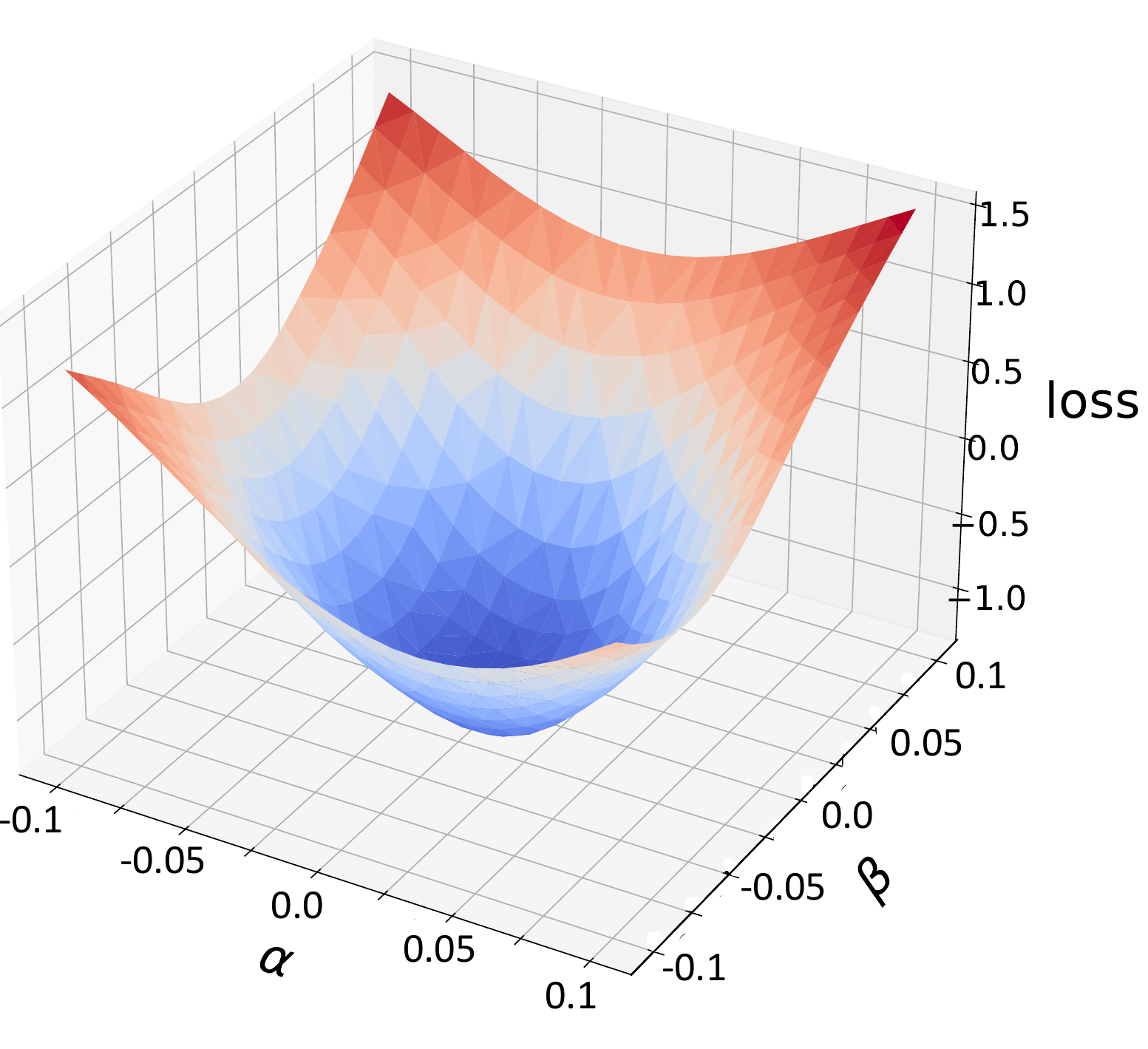}
        \caption{}\label{fig:loss_landscape_3}
    \end{subfigure}\hfill
    \caption{Loss landscapes showing in (a) the full 3D landscape, (b) the equivalent loss contour covering the full range of $\alpha$ and $\beta$ (coefficients of two random Gaussian direction vectors with dimensions compatible with the network weights) for the trained W view, pass 1 network, and (c) the 3D landscape for the region $\alpha, \beta \in [-0.1, 0.1]$.}\label{fig:loss_landscape}
\end{figure*}

The behaviour of the network also considered the distribution of weights in the trained model, the structure of the loss landscape in the vicinity of the minimum and we also explored the evolution of the learned weights in various convolutional layers to assess the stability of the training process. Fig.~\ref{fig:weight_distribution} shows the distribution of weights in the trained pass 1, W view network --  small weights reduce the sensitivity of the network to small perturbations~\cite{weight_tuning}, helping the network to generalise to previously unseen events. Fig.~\ref{fig:loss_landscape} shows how the average loss over 1024 validation events varies as we take steps away from the minimum using the method described by Li et al~\cite{Li2017}. The loss landscape (note the 3D surface depicts the logarithm of the average loss) appears smooth over the full grid of assessed weights and zooming into the region close to the minimum the landscape continues to exhibit features that lead to effective training.

\section{Vertex reconstruction performance}\label{sec:results:accel}
To assess the performance on any given event, the true neutrino interaction vertex was required to reside within the fiducial volume. A total of \nvalidationevents out of the \ntotalevents available events (independent of the training and validation samples) across all flavours and horn currents passed this cut.

\begin{figure*}[tbh]
    \begin{subfigure}{0.45\textwidth}
        \centering
        \includegraphics[width=\textwidth]{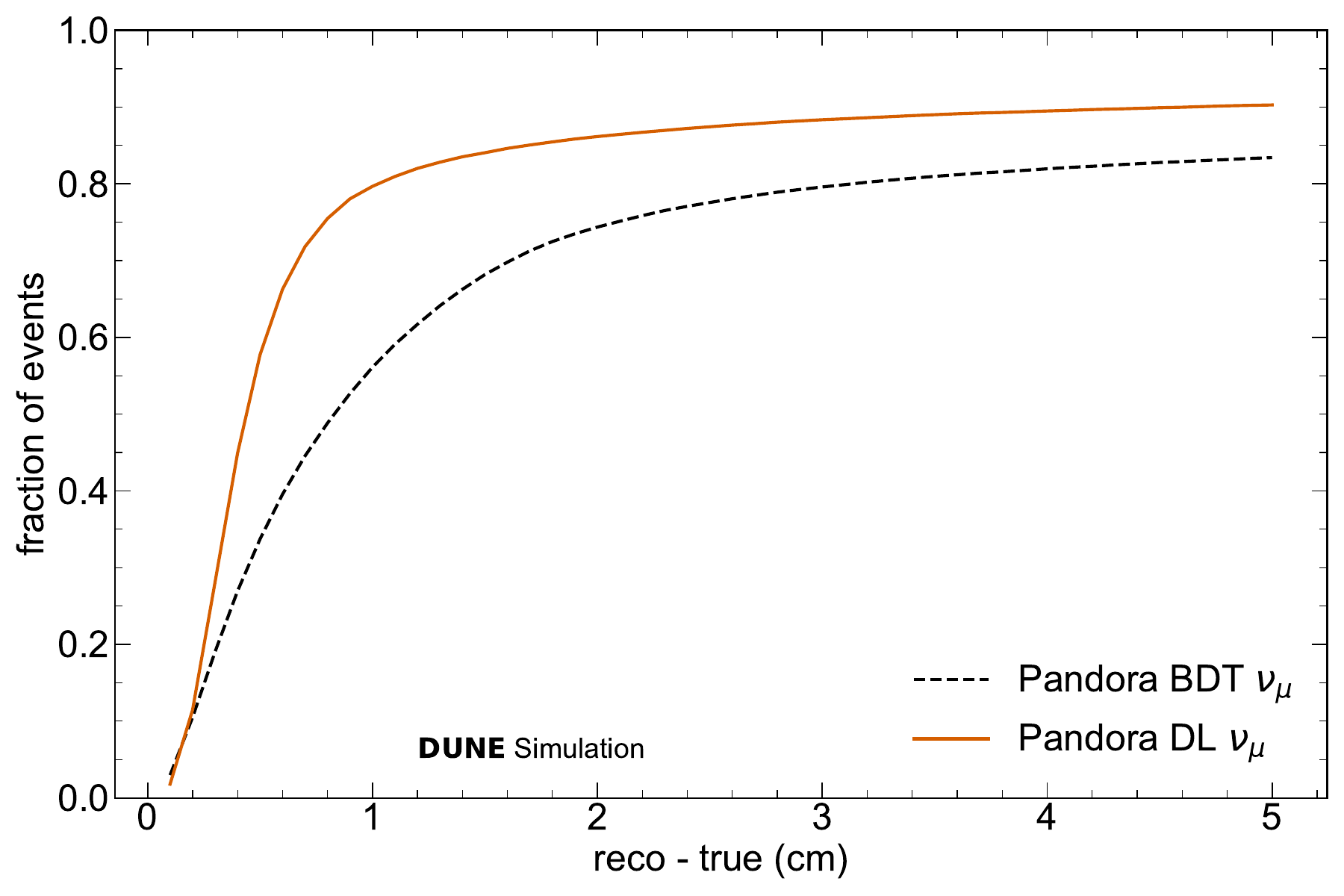}
        \caption{}\label{fig:accel_hd_eff_bdt_dl_1}
    \end{subfigure}\hfill
    \begin{subfigure}{0.45\textwidth}
        \centering
        \includegraphics[width=\textwidth]{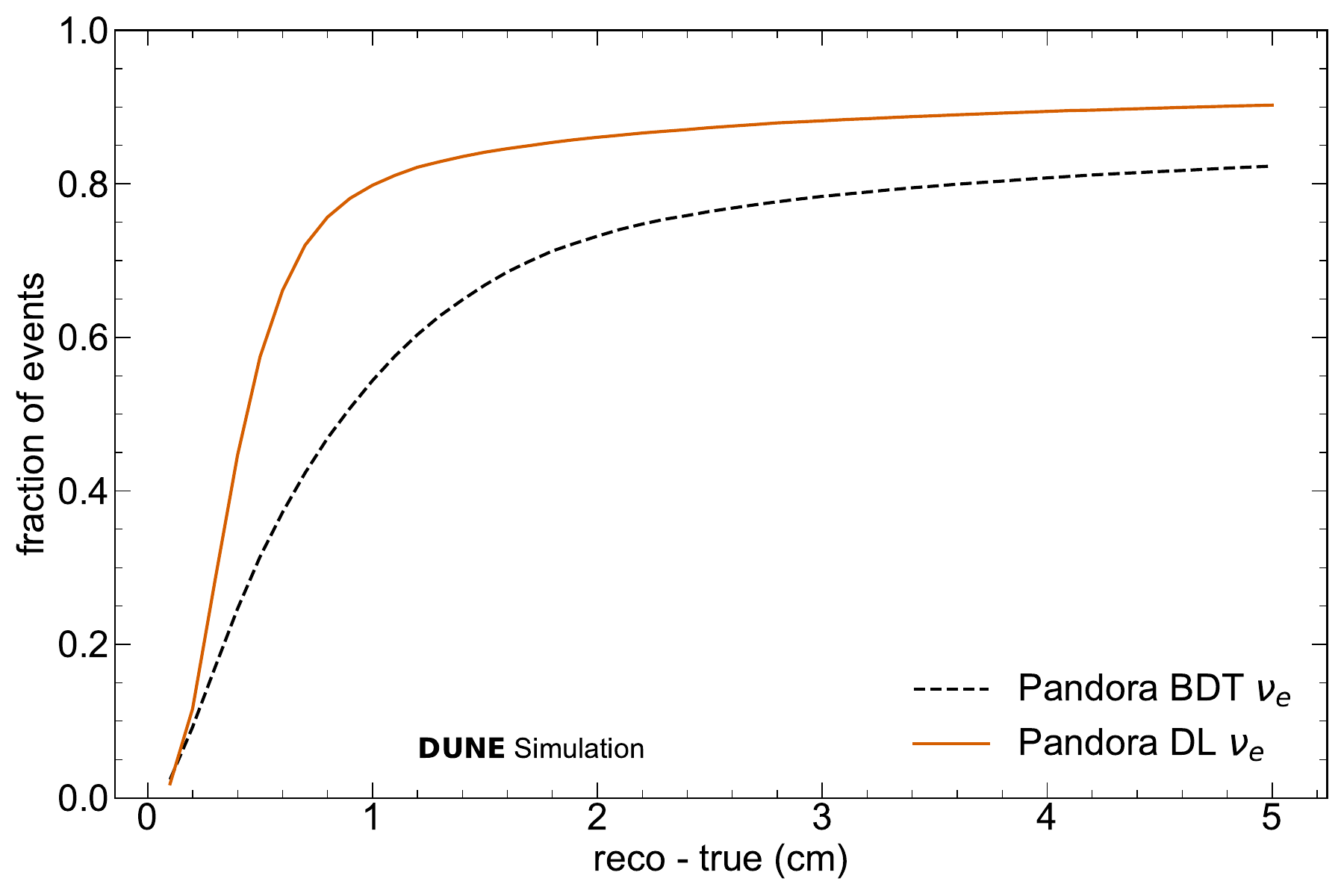}
        \caption{}\label{fig:accel_hd_eff_bdt_dl_2}
    \end{subfigure}
    \begin{subfigure}{0.45\textwidth}
        \centering
        \includegraphics[width=\textwidth]{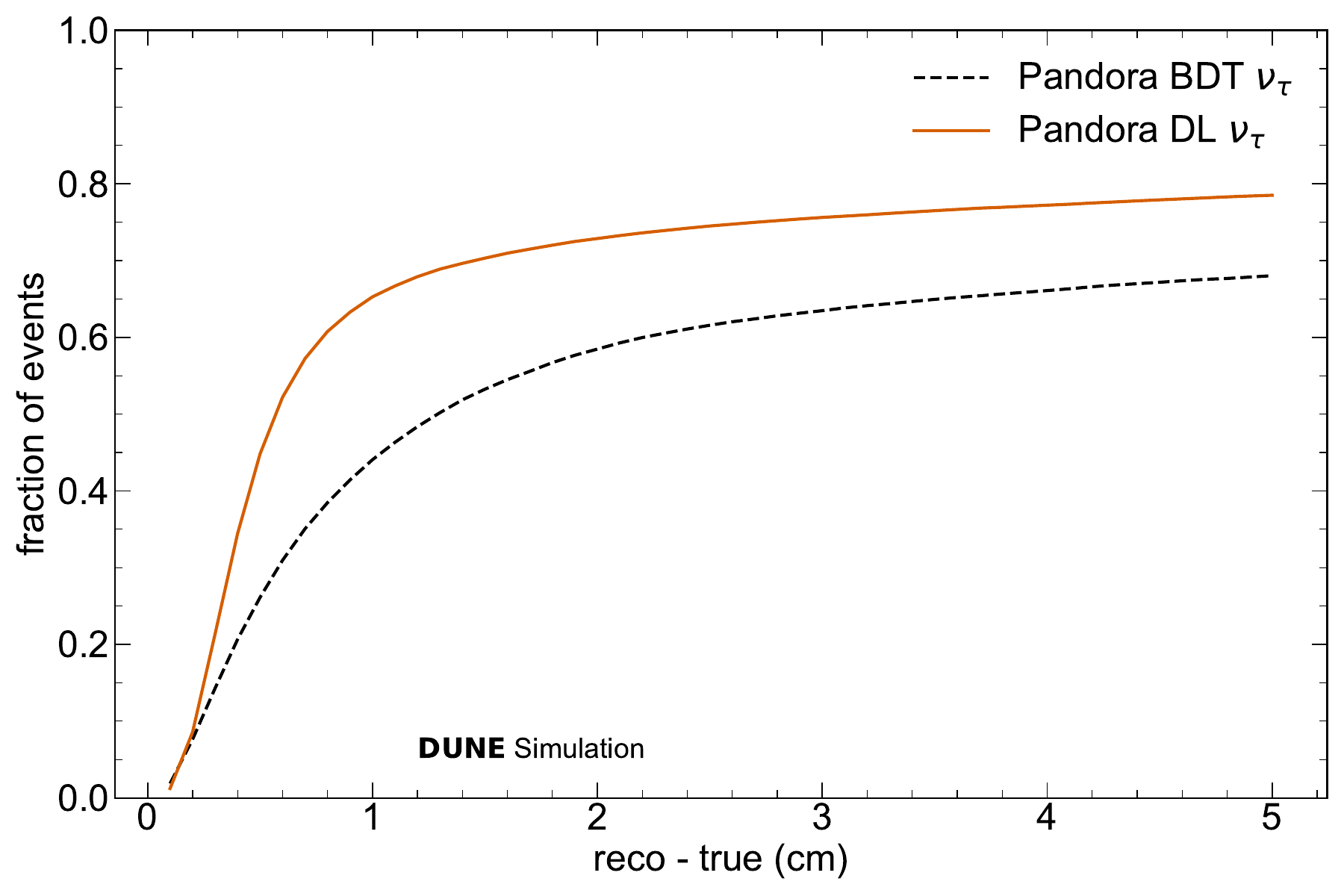}
        \caption{}\label{fig:accel_hd_eff_bdt_dl_3}
    \end{subfigure}\hfill
    \begin{subfigure}{0.45\textwidth}
        \centering
        \includegraphics[width=\textwidth]{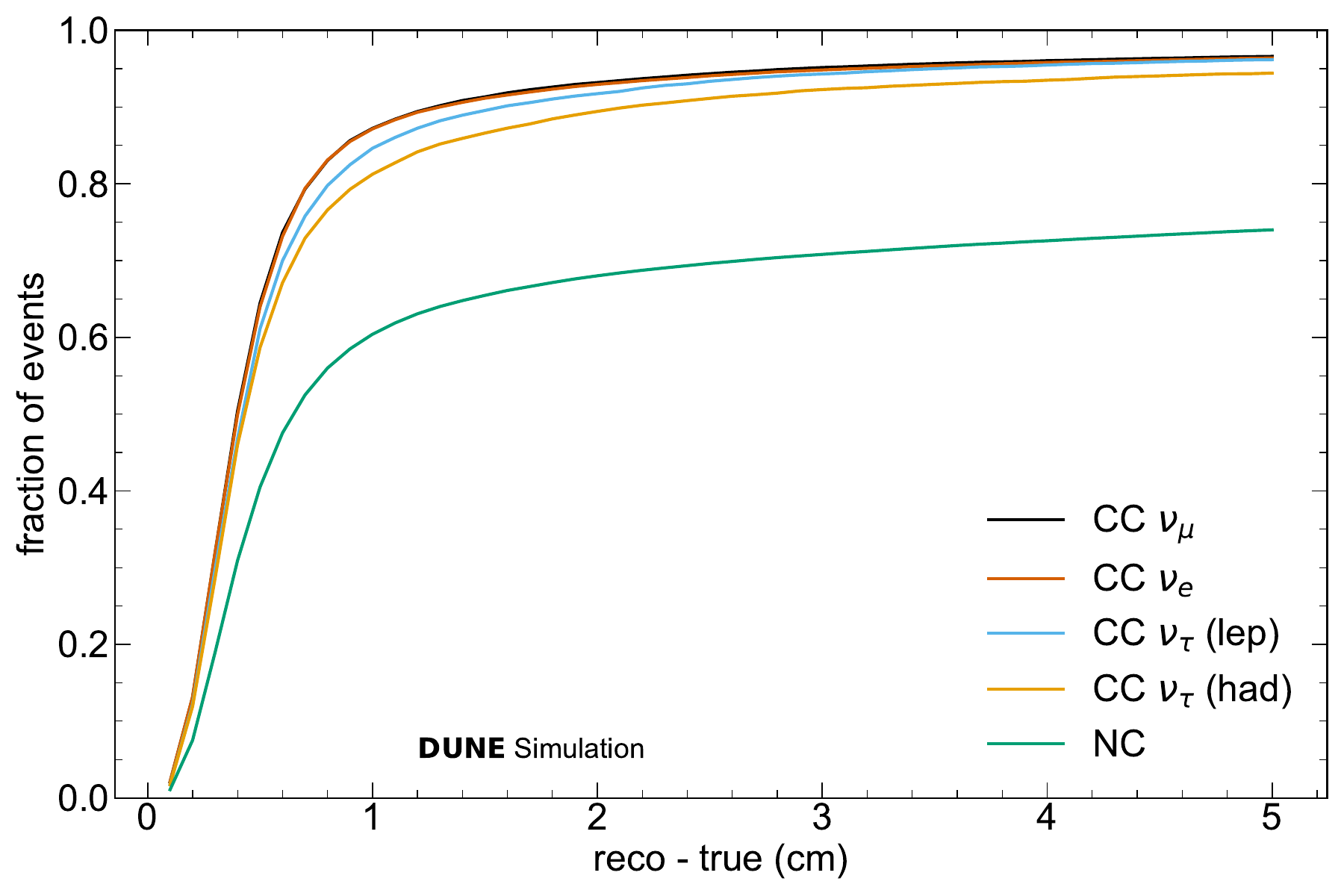}
        \caption{}\label{fig:accel_hd_eff_dl_charge}
    \end{subfigure}
    \caption{Fraction of vertices reconstructed within a given distance of the true neutrino interaction vertex for the previously used BDT and new network for (a) $\parenbar{\nu}_{\mu}$, (b) $\parenbar{\nu}_{e}$ and (c) $\parenbar{\nu}_{\tau}$. (d) shows the performance of the new deep learning network for different currents, flavours and in the case of CC $\nu_\tau$ interactions, for the leptonic and hadronic decays of the $\tau$. The network outperforms the BDT in all cases.}\label{fig:accel_hd_eff_bdt_dl}
\end{figure*}

Fig.~\ref{fig:accel_hd_eff_bdt_dl} compares the vertex reconstruction performance of our vertex finding network with the previous state-of-the-art boosted decision tree (BDT) for each neutrino flavour (inclusive of CC and NC interactions) for the horizontal-drift detector. The new method substantially outperforms the previous state-of-the-art across all samples. As the BDT depends upon hit clustering for its inputs, any reconstruction inefficiencies at this stage may hinder its performance, whereas the vertex finding network operates on hits and therefore is not subject to such inefficiencies. The vertex is reconstructed with high precision in a large fraction of events for the $\nu_\mu$ and $\nu_e$ samples, with approximately 80\% of all vertices reconstructed within 1\,cm of the true interaction vertex. The reconstruction performance is notably lower for the $\nu_\tau$ sample, where approximately 66\% of vertices are reconstructed within 1\,cm of the true interaction vertex. The reason for this difference is discussed further below. A notable minority of events reconstruct the vertex at $>5$\,cm from the true interaction vertex location. Such failures will be referred to as `catastrophic failures' and the nature of these failures will be discussed below, though we note here that one weakness of the two pass approach is that a sufficiently large error in pass 1 cannot be recovered in pass 2, because the true vertex will not be present in the image considered in pass 2.

Fig.~\ref{fig:accel_hd_eff_dl_charge} shows the vertex reconstruction performance for the network, broken down by flavour and weak current. It is clear in this figure that the presence of a leading lepton in the final state yields a highly performant vertex reconstruction. For the $\nu_\mu$ and $\nu_e$ samples approximately 87\% of all vertices are reconstructed within 1\,cm of the true interaction vertex, with approximately 83\% for the $\nu_\tau$ sample, and almost all vertices (95-97\%) are reconstructed within 5\,cm of the true interaction vertex. In conjunction with any hadronic activity, a leading lepton provides clear pointing information back to an interaction vertex, and so high quality reconstruction is expected. In contrast, the neutral current performance is lower, with 60-61\% of vertices reconstructed within 1\,cm of the true interaction vertex, and a catastrophic failure rate of 21-22\%. The absence of a leading lepton in such events reduces the available pointing information, and the topology becomes especially challenging for events dominated by diffuse neutron-induced activity, where the interaction location cannot be reasonably identified even by a human expert. This difference between charged current (CC) and neutral current (NC) events also provides an explanation for the inclusive $\nu_\tau$ vertex reconstruction performance. With the $\tau$ production threshold suppressing CC events below neutrino energies of $\sim{}3.5$\,GeV the fraction of CC interactions in the $\nu_\tau$ sample is only $\sim{}25$\%, as compared to $\sim{}70$\% for the $\nu_e$ and $\nu_\mu$ samples, and therefore the inclusive sample more closely tracks the NC performance than the CC performance.

\begin{figure*}[tbh]
    \begin{subfigure}{0.45\textwidth}
        \centering
        \includegraphics[width=\textwidth]{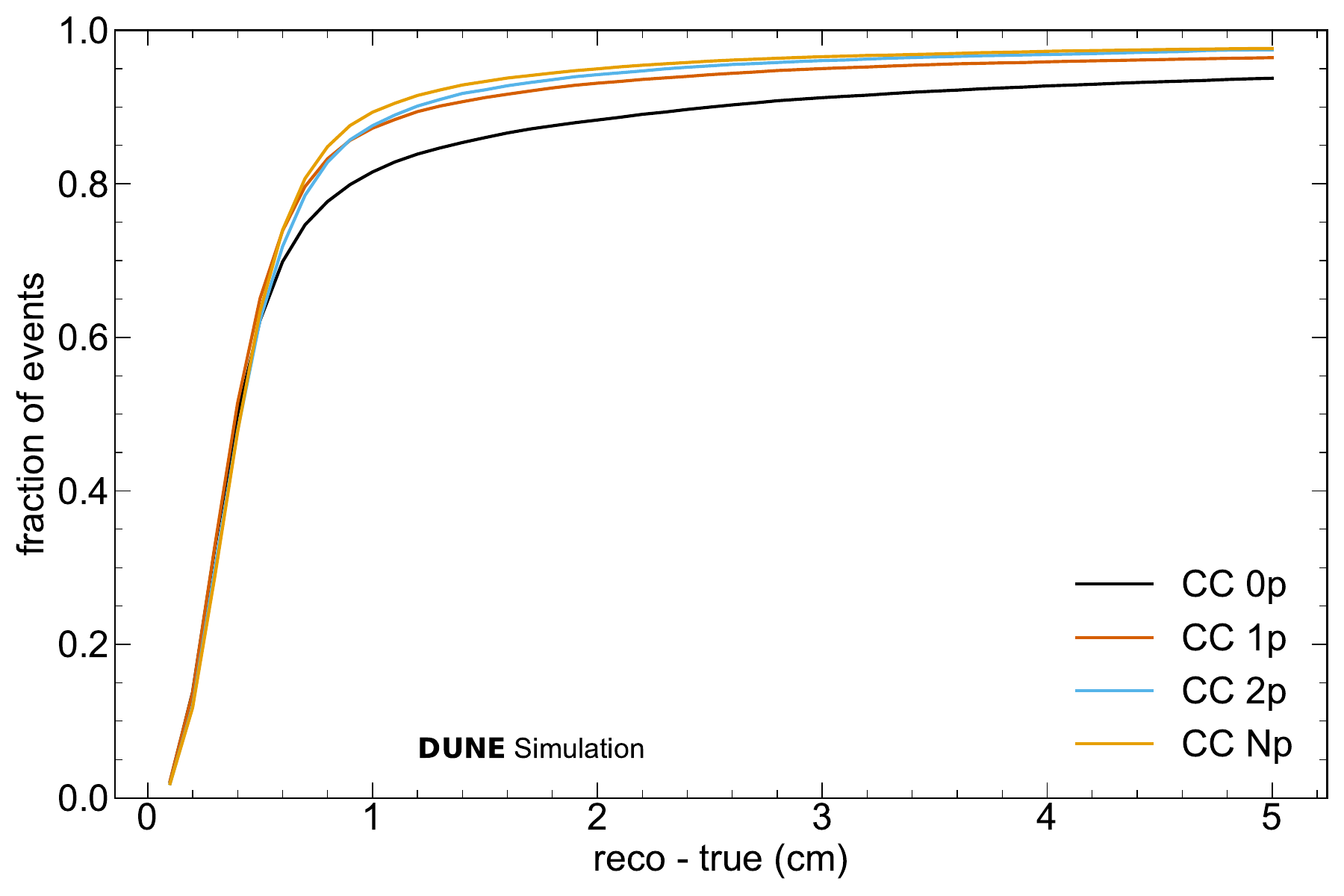}
    \end{subfigure}\hfill
    \begin{subfigure}{0.45\textwidth}
        \centering
        \includegraphics[width=\textwidth]{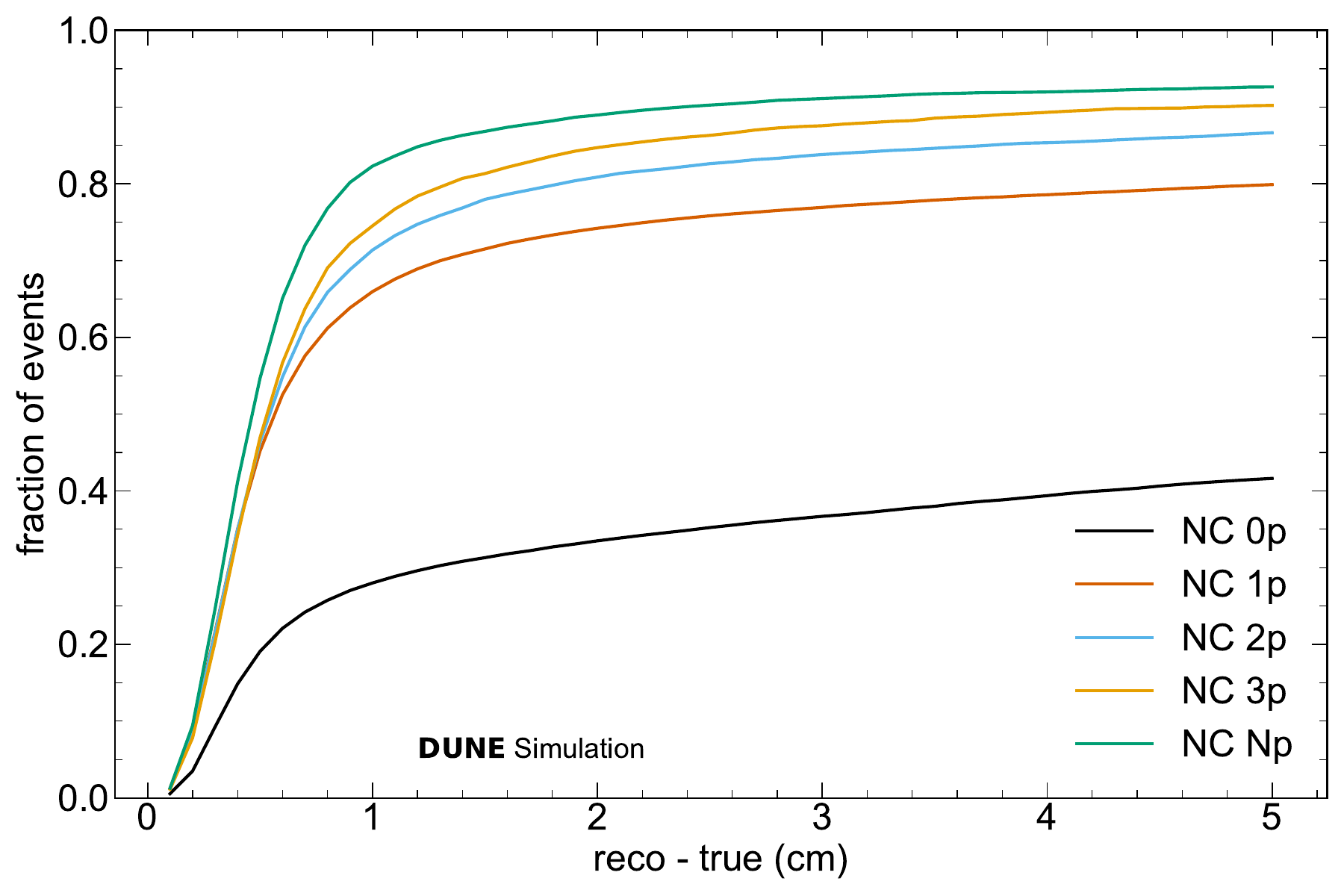}
    \end{subfigure}
    \begin{subfigure}{0.45\textwidth}
        \centering
        \includegraphics[width=\textwidth]{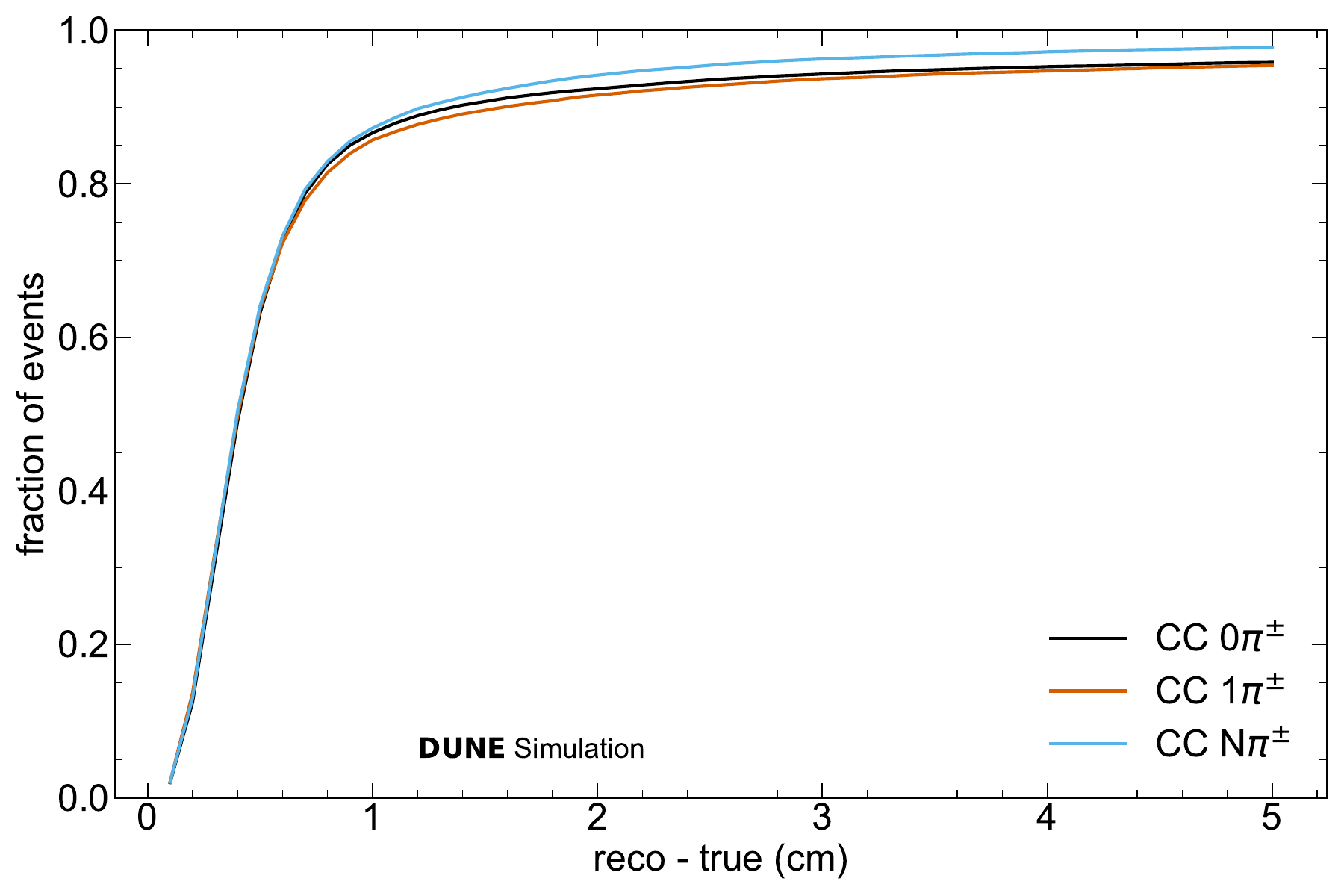}
    \end{subfigure}\hfill
    \begin{subfigure}{0.45\textwidth}
        \centering
        \includegraphics[width=\textwidth]{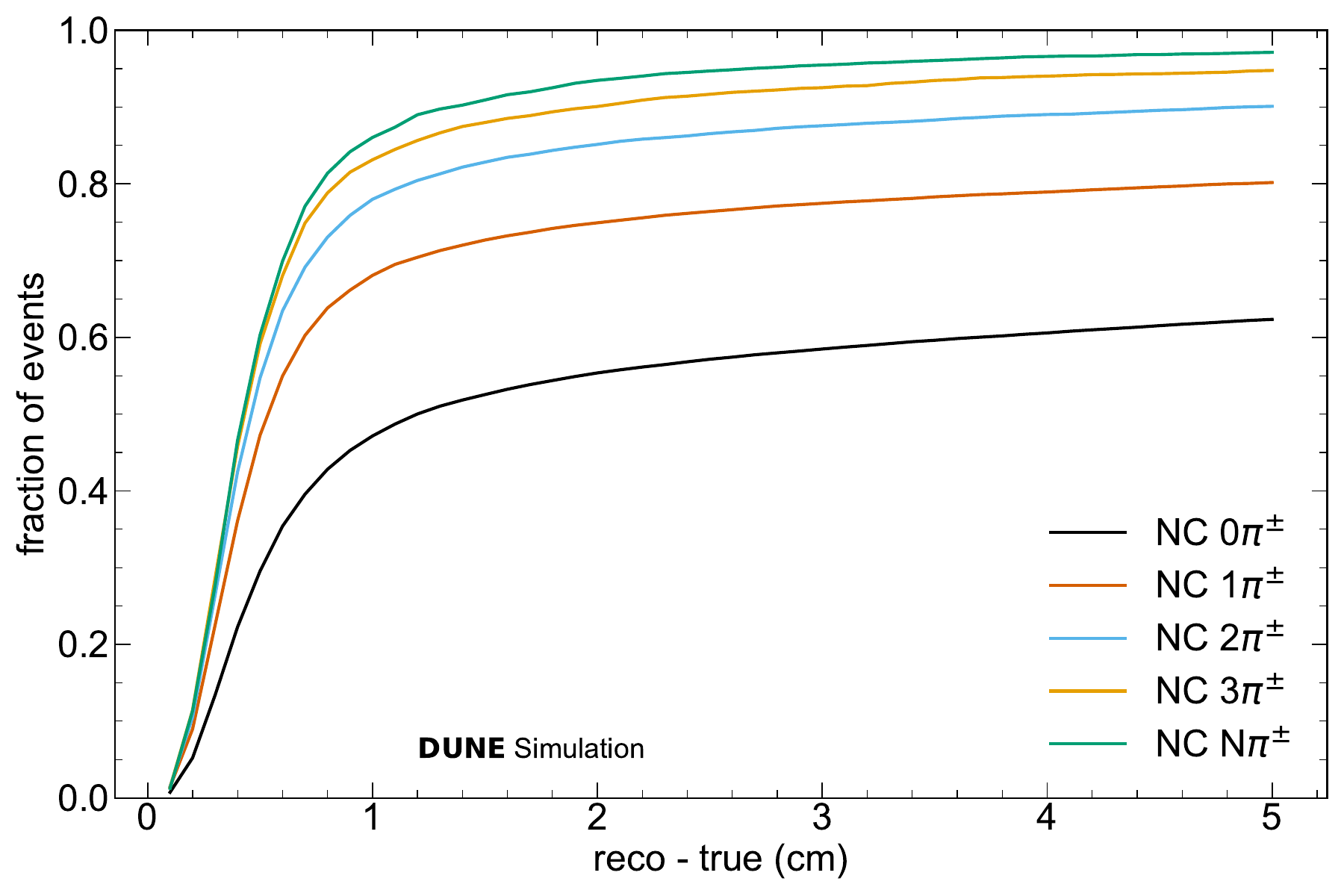}
    \end{subfigure}
    \begin{subfigure}{0.45\textwidth}
        \centering
        \includegraphics[width=\textwidth]{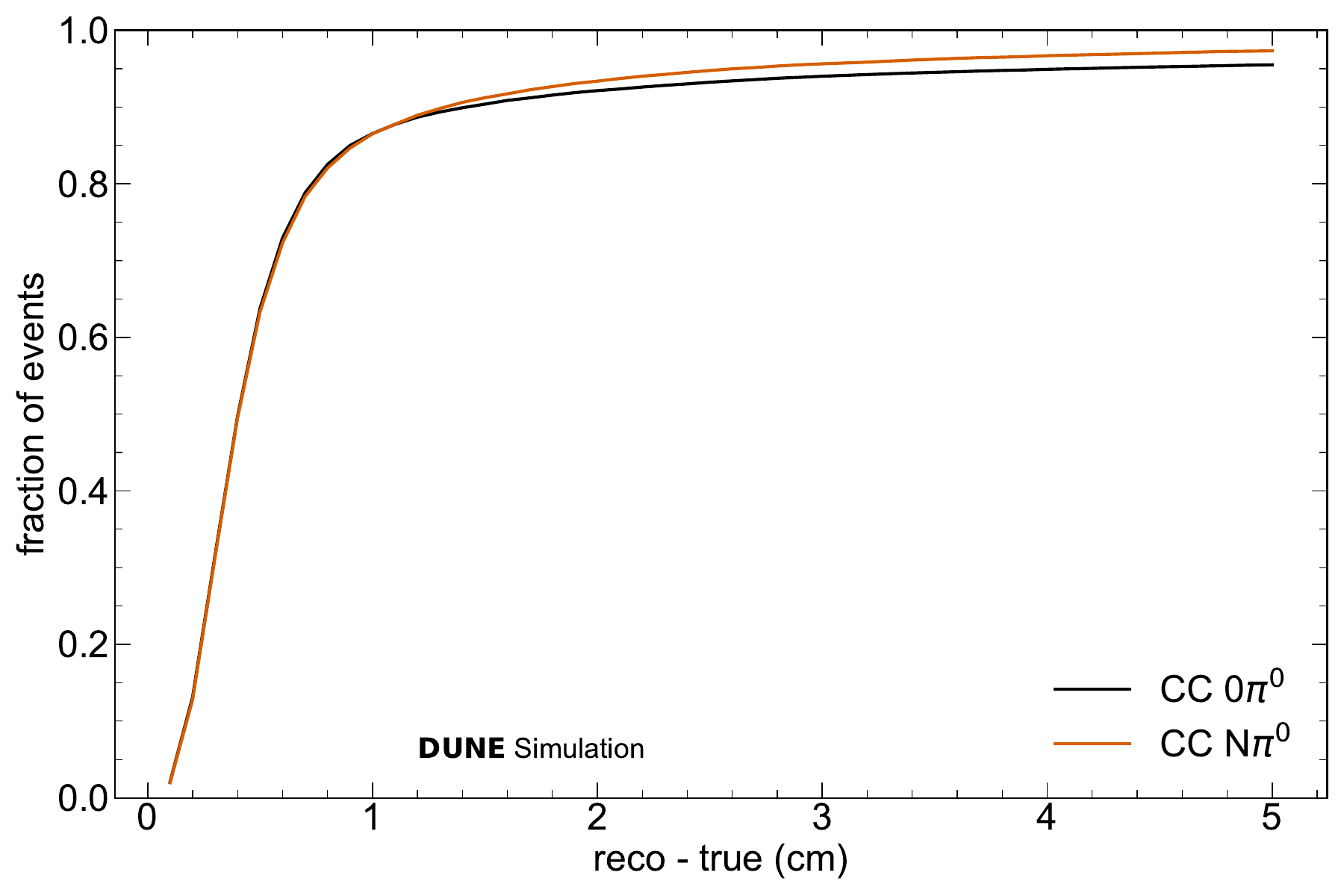}
    \end{subfigure}\hfill
    \begin{subfigure}{0.45\textwidth}
        \centering
        \includegraphics[width=\textwidth]{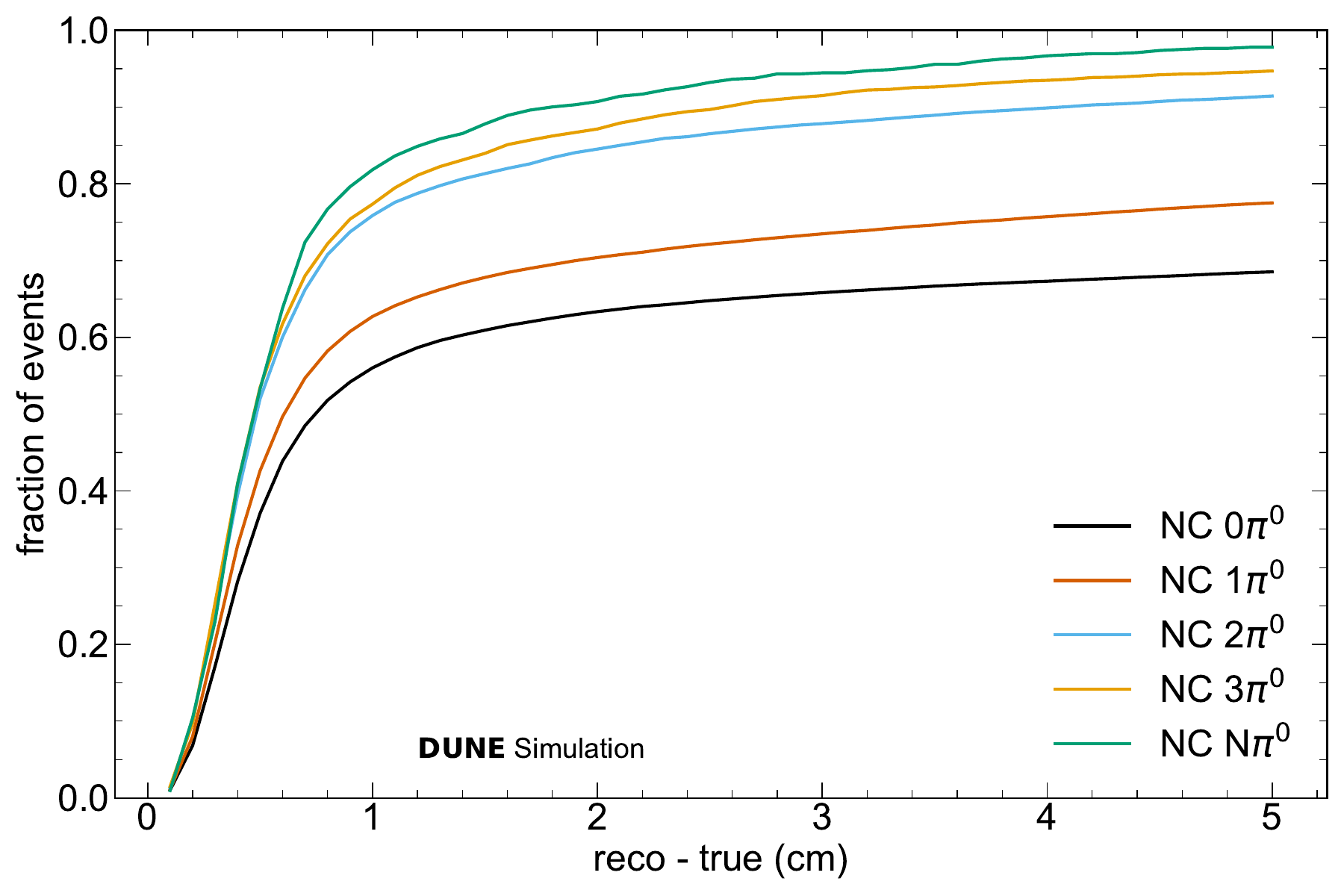}
    \end{subfigure}
    \caption{Fraction of vertices reconstructed within a given distance of the true neutrino (all flavours) interaction vertex for (left) CC and (right) NC events, for a given number of final state (top) protons, (middle) charged pions and (bottom) neutral pions. The CC interactions show little sensitivity to final state multiplicity, while NC interactions approach CC performance with increasing final state multiplicity.}\label{fig:accel_hd_eff_dl_final_states}
\end{figure*}

\begin{figure*}[tbh]
  \centering
  \includegraphics[width=0.65\textwidth]{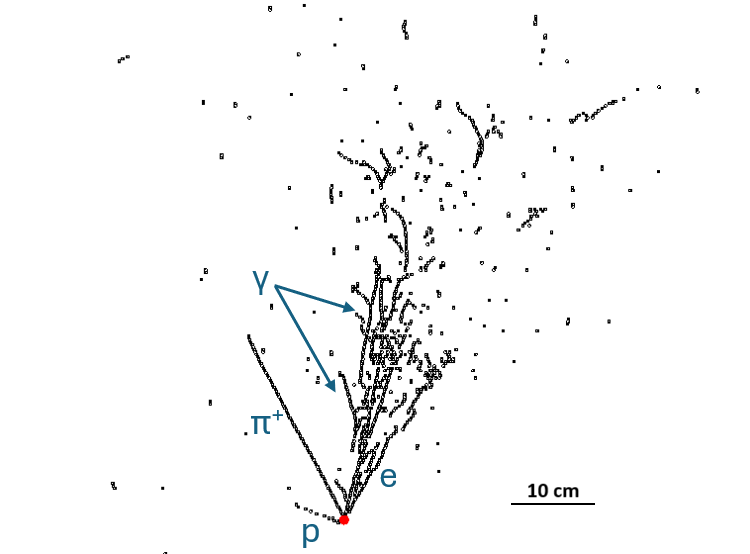}
  \caption{A 3\,GeV $\nu_e$ CC resonant pion production interaction with an electron, charged pion, neutral pion (decays to two photons) and proton in the final state. Hits from the W view in black, reconstructed vertex in red.}
  \label{fig:nue_cc_res_pointing}
\end{figure*}

Performance is further summarised according to selected final states in Fig.~\ref{fig:accel_hd_eff_dl_final_states}. Performance is assessed for the number of final state protons, charged pions and neutral pions, and CC and NC interactions. Immediately evident in the figures is the significance of the leading lepton. In the presence of the leading lepton, vertex reconstruction efficiency is very high at baseline, and moderately increases as the number of final state particles increases for each of the selected final state particles. This behaviour is expected, with each additional final state particle augmenting the existing pointing information to more clearly identify the vertex location. For example, Fig.~\ref{fig:nue_cc_res_pointing} depicts a well-reconstructed vertex given ample, consistent pointing information from the electron, charged pion and proton in a resonant pion production event. The effect of additional final state particles on NC events is much more substantial. In the absence of a leading lepton, fewer selected final state particles lead to large reductions in performance, particularly for protons, where little more than 40\% of all vertices are reconstructed within 5\,cm of the true vertex location. This is unsurprising, given that the failure of even a single proton to emerge from a nucleus will often imply no, or few, other final state particles that ionise the medium, and thus yield little guidance for even a human expert. As the number of these final state particles increases, the performance improves rapidly at first, and then slowly approaches CC-like performance. High pion multiplicity yields CC-like performance, with improved pointing information provided by the longer minimally ionising charged pion tracks relative to the shorter, highly ionising protons, and also by decay photons from final state neutral pions pointing to a common vertex.

\begin{figure*}[tbh]
    \begin{subfigure}{0.45\textwidth}
        \centering
        \includegraphics[width=\textwidth]{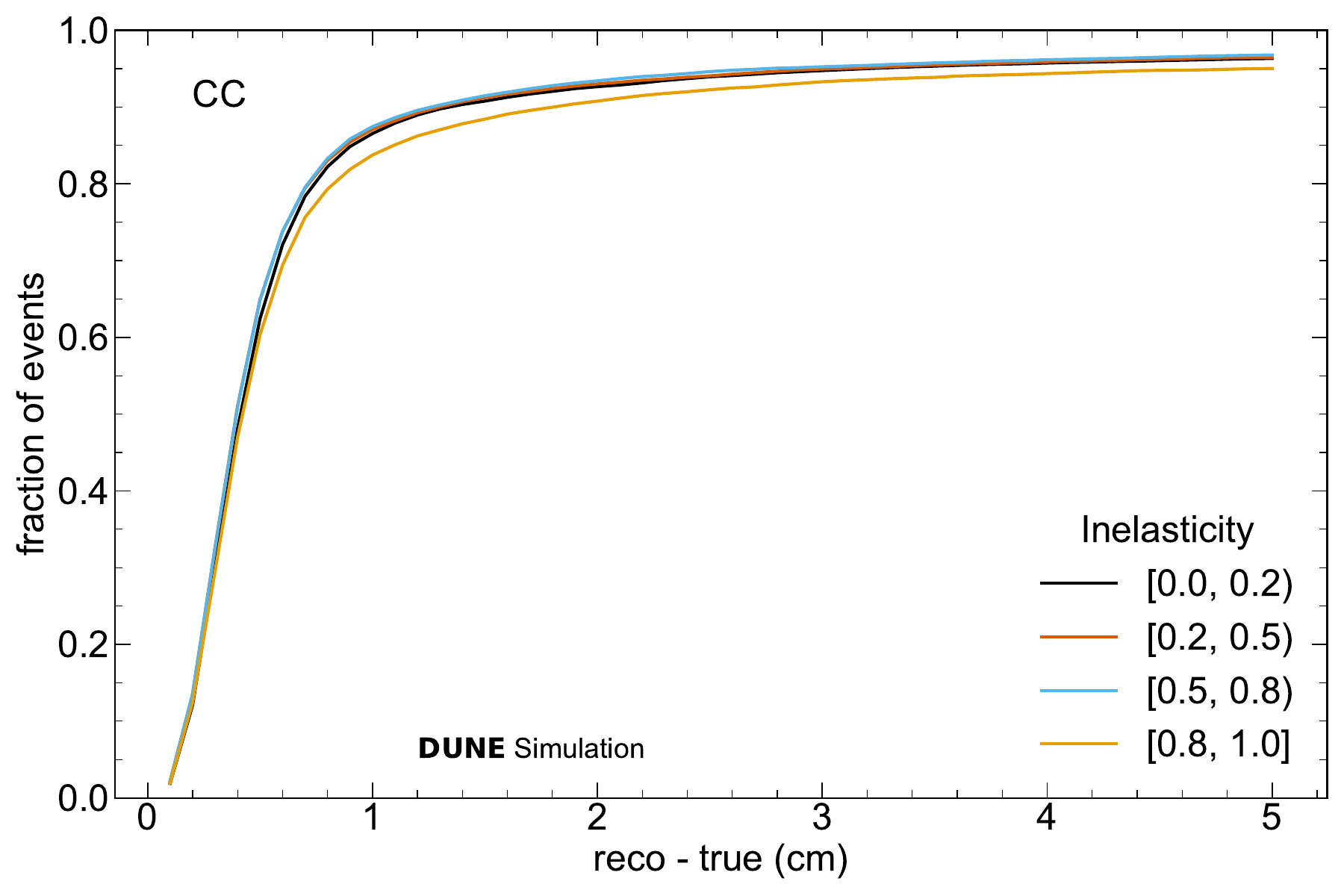}
        \caption{}
    \end{subfigure}\hfill
    \begin{subfigure}{0.45\textwidth}
        \centering
        \includegraphics[width=\textwidth]{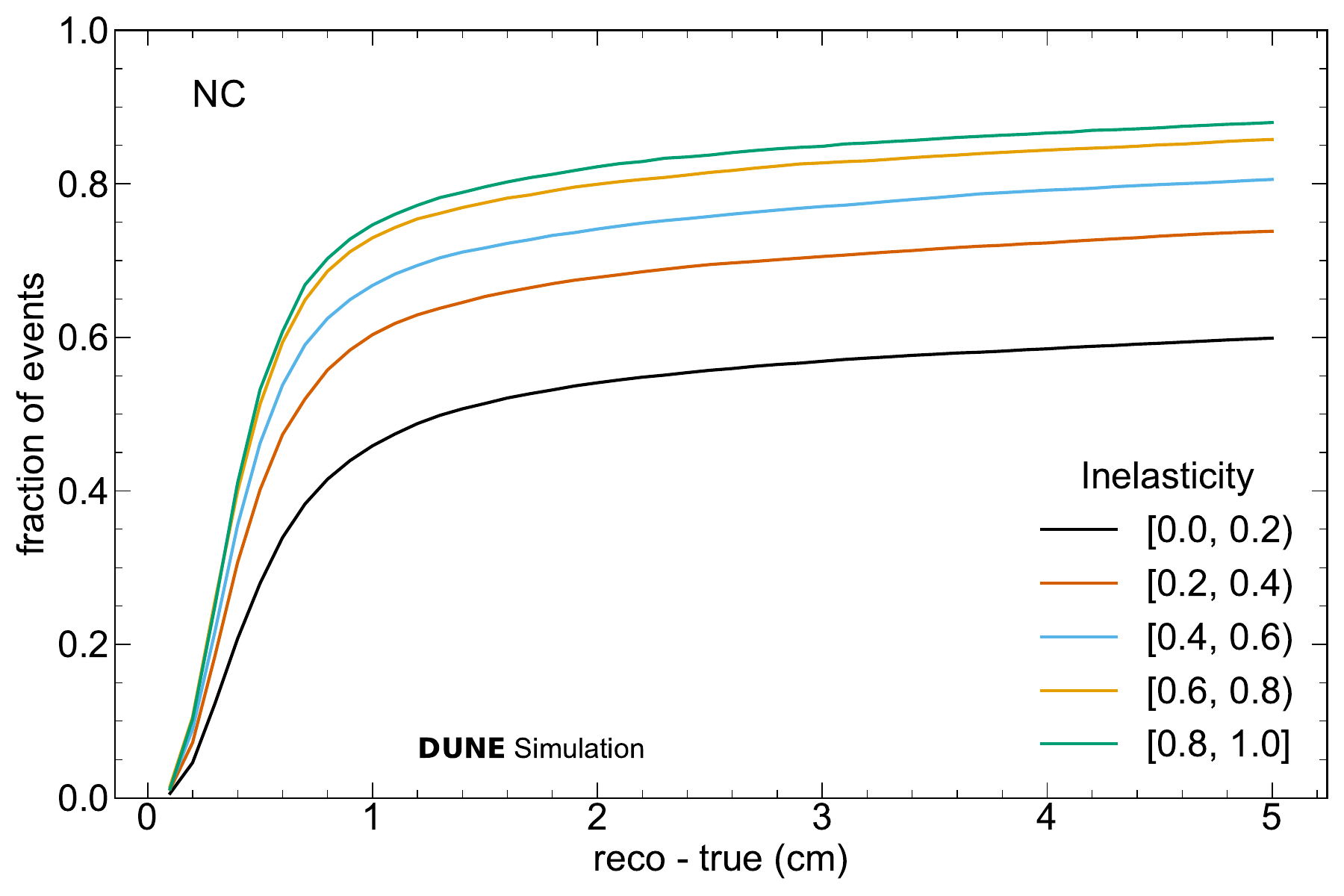}
        \caption{}
    \end{subfigure}
    \begin{subfigure}{0.45\textwidth}
        \centering
        \includegraphics[width=\textwidth]{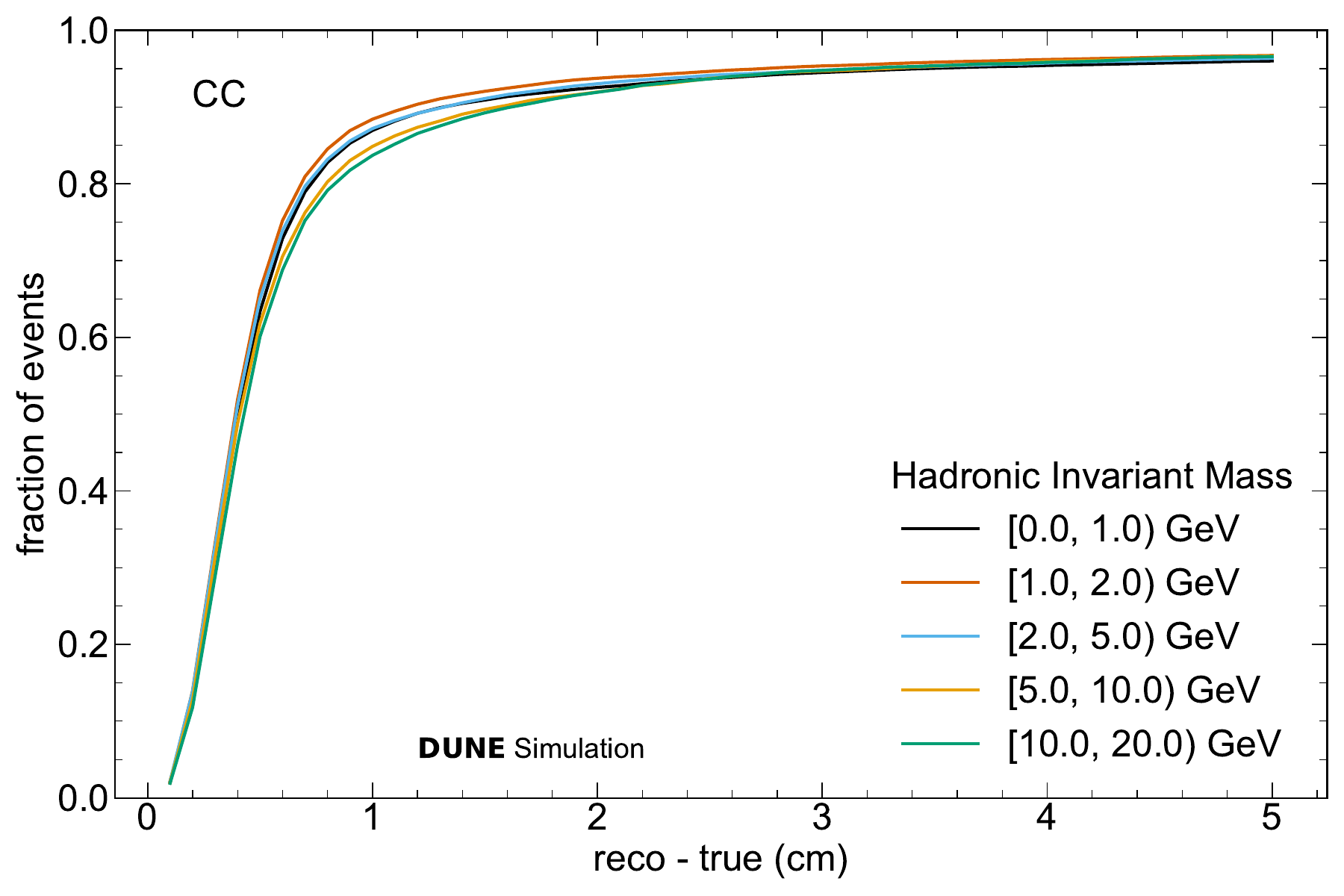}
        \caption{}
    \end{subfigure}\hfill
    \begin{subfigure}{0.45\textwidth}
        \centering
        \includegraphics[width=\textwidth]{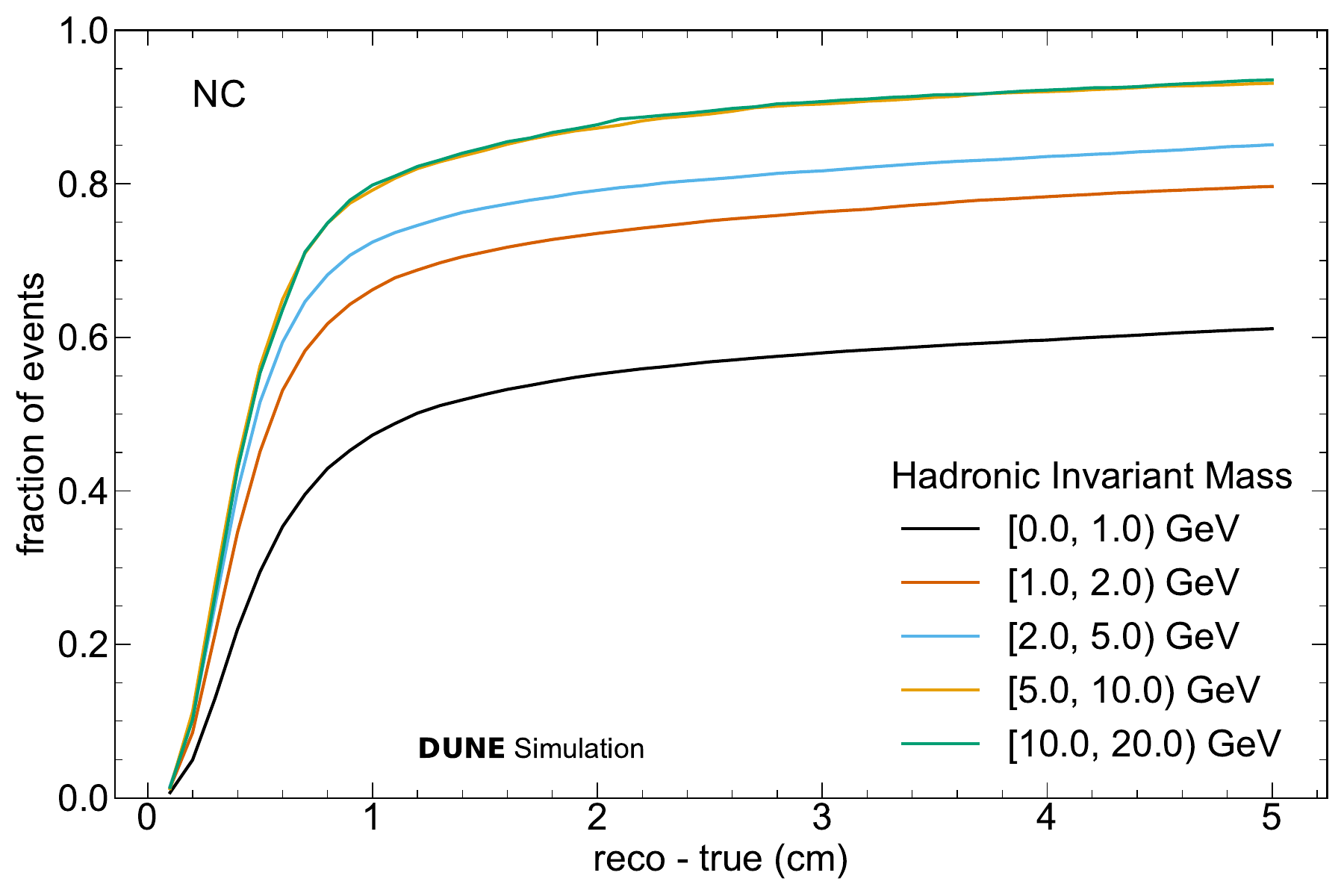}
        \caption{}
    \end{subfigure}
    \caption{Fraction of vertices reconstructed within a given distance of the true neutrino interaction (all flavours) vertex as a function of inelasticity for (a) CC and (b) NC events, and as a function of hadronic invariant mass for (c) CC and (d) NC events. The CC interactions show little sensitivity to inelasticity and hadronic invariant mass, while NC interactions approach CC performance only at higher inelasticity and hadronic invariant mass.}\label{fig:accel_hd_eff_dl_inel}
\end{figure*}

More generally, we can consider vertex reconstruction performance as a function of inelasticity (i.e. the fraction of incident neutrino energy not carried away by the outgoing lepton). Fig.~\ref{fig:accel_hd_eff_dl_inel} shows vertex reconstruction performance for different bins of inelasticity for (a) CC and (b) NC interactions. A few interesting features are evident in these figures. For the NC case, we see behaviour entirely consistent with that seen when considering the number of final state protons and pions. Without a leading lepton, the vertexing must instead depend upon protons emerging from the nucleus, or perhaps the charged particles produced in downstream interactions of neutral final state particles, which will provide imperfect pointing information by the nature of their production. As the contribution of the visible hadronic component increases, the vertex reconstruction performance improves, starting from a low baseline where much of the momentum is carried away by the neutrino, leaving little for the hadronic system and therefore little visible charge deposition. For CC interactions however, there is a subtle difference. Vertex reconstruction performance is high across bins of inelasticity, with small improvements as inelasticity increases at first, but then turns over at high inelasticity to yield slightly worse performance. The fraction of events reconstructed within 1\,cm of the true vertex is around 84\% below inelasticities of 0.8, but drops to around 82\% by inelasticities above this level. Initially, increases in the hadronic contribution usefully increase particle multiplicity and add to pointing information, but as the hadronic component begins to dominate, highly complex topologies with dense charge deposition and secondary interactions can form plausible primary interaction vertex candidates, which can lead to larger vertex misidentification rates. Fig.~\ref{fig:accel_hd_eff_dl_inel} also shows vertex reconstruction performance for different bins of hadronic invariant mass for (c) CC and (d) NC interactions. The fraction of events reconstructed within 1\,cm of the true vertex is around 85\% for hadronic invariant masses below 5\,GeV, but drops to around 82\% for hadronic invariant masses above this level. As for inelasticity, the effect on CC interactions is relatively small, with similar evidence for turnover as the hadronic component becomes very large and topologies become more complex. For NC interactions we see improved performance as the hadronic invariant mass increases, consistent with the picture for inelasticity, though highlighting that at lower energies, even if inelasticity is high, the reduced charge deposition associated with lower hadronic invariant mass still yields lower performance.

\begin{figure*}[tbh]
    \begin{subfigure}{0.45\textwidth}
        \centering
        \includegraphics[width=\textwidth]{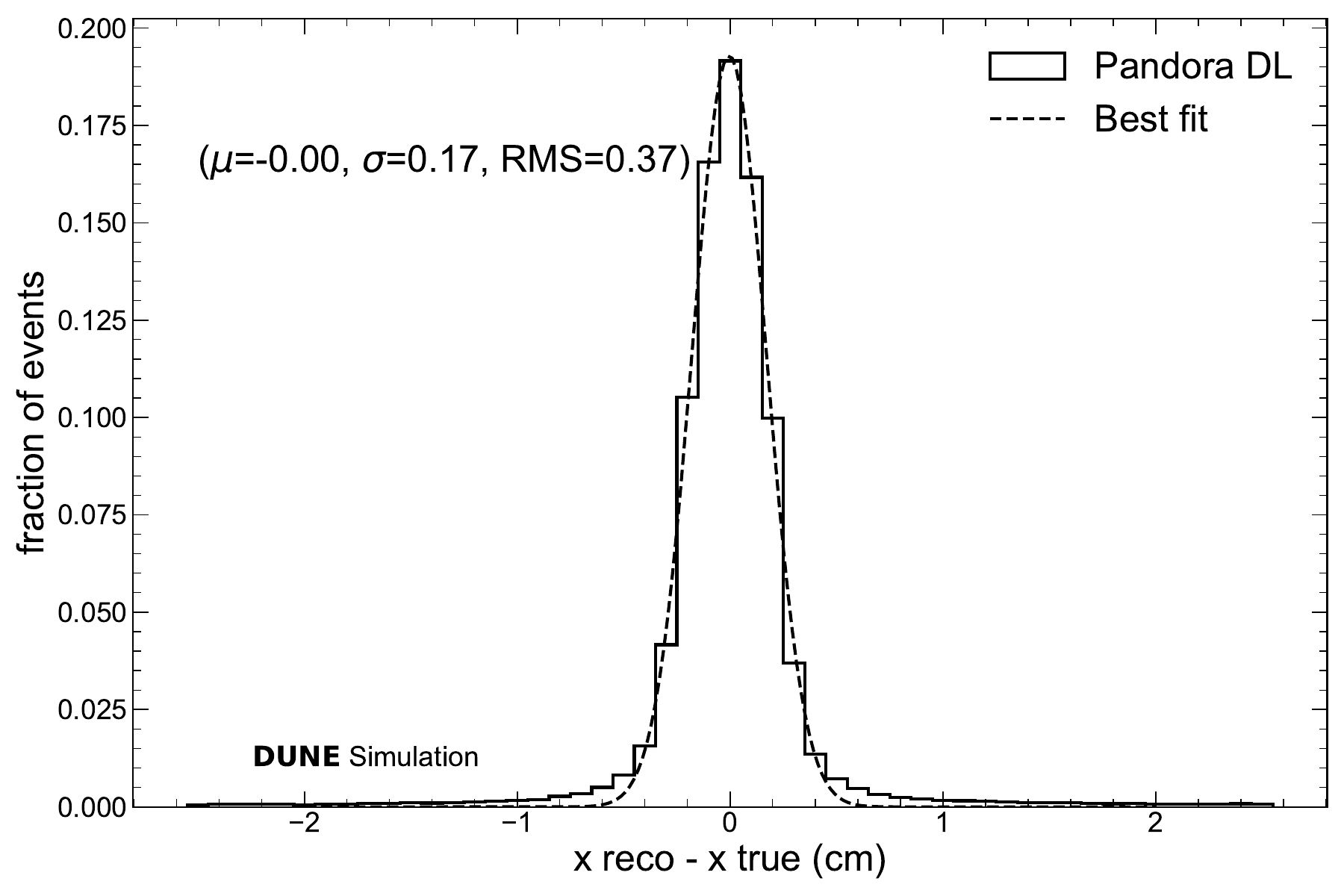}
    \end{subfigure}\hfill
    \begin{subfigure}{0.45\textwidth}
        \centering
        \includegraphics[width=\textwidth]{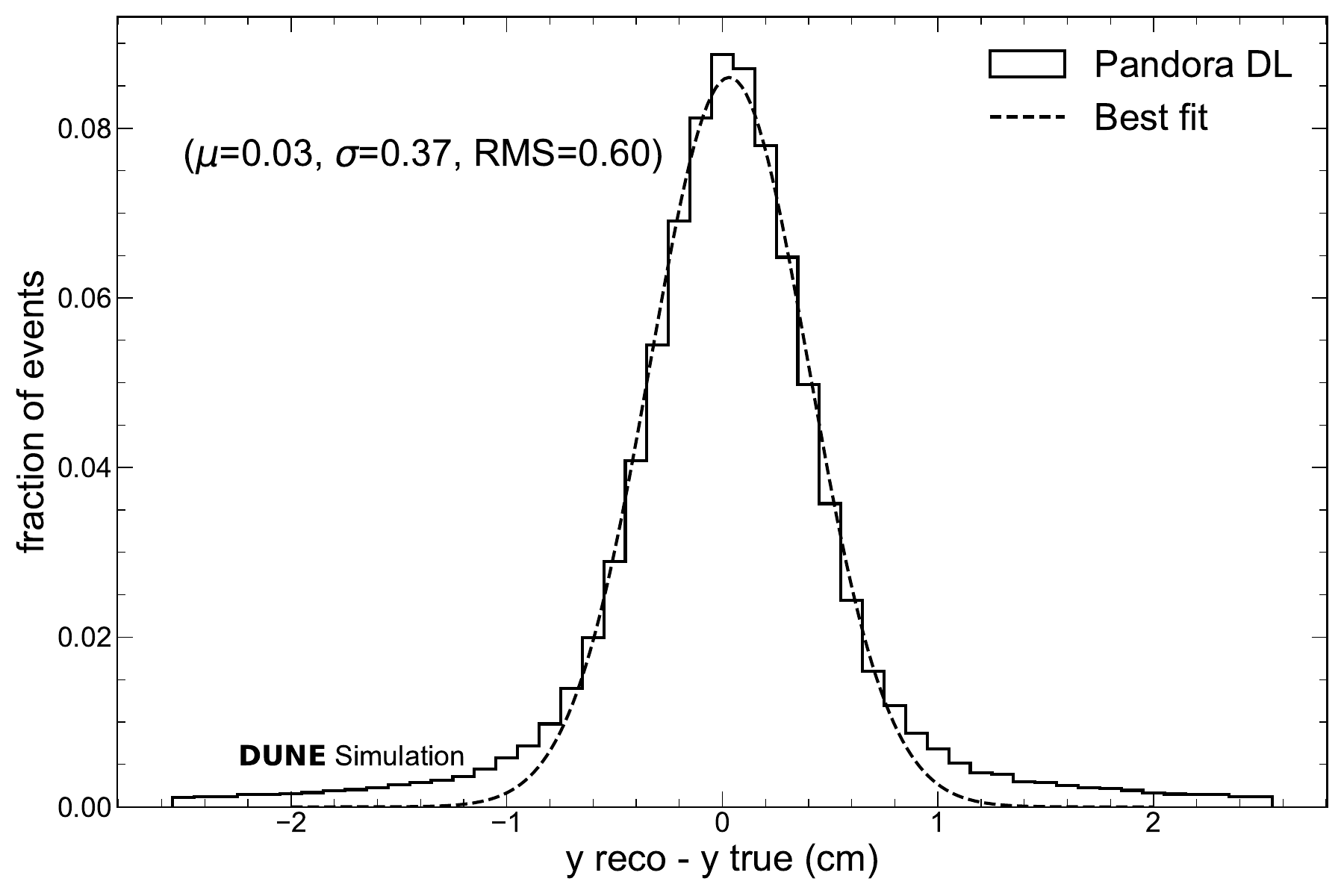}
    \end{subfigure}
    \begin{subfigure}{0.45\textwidth}
        \centering
        \includegraphics[width=\textwidth]{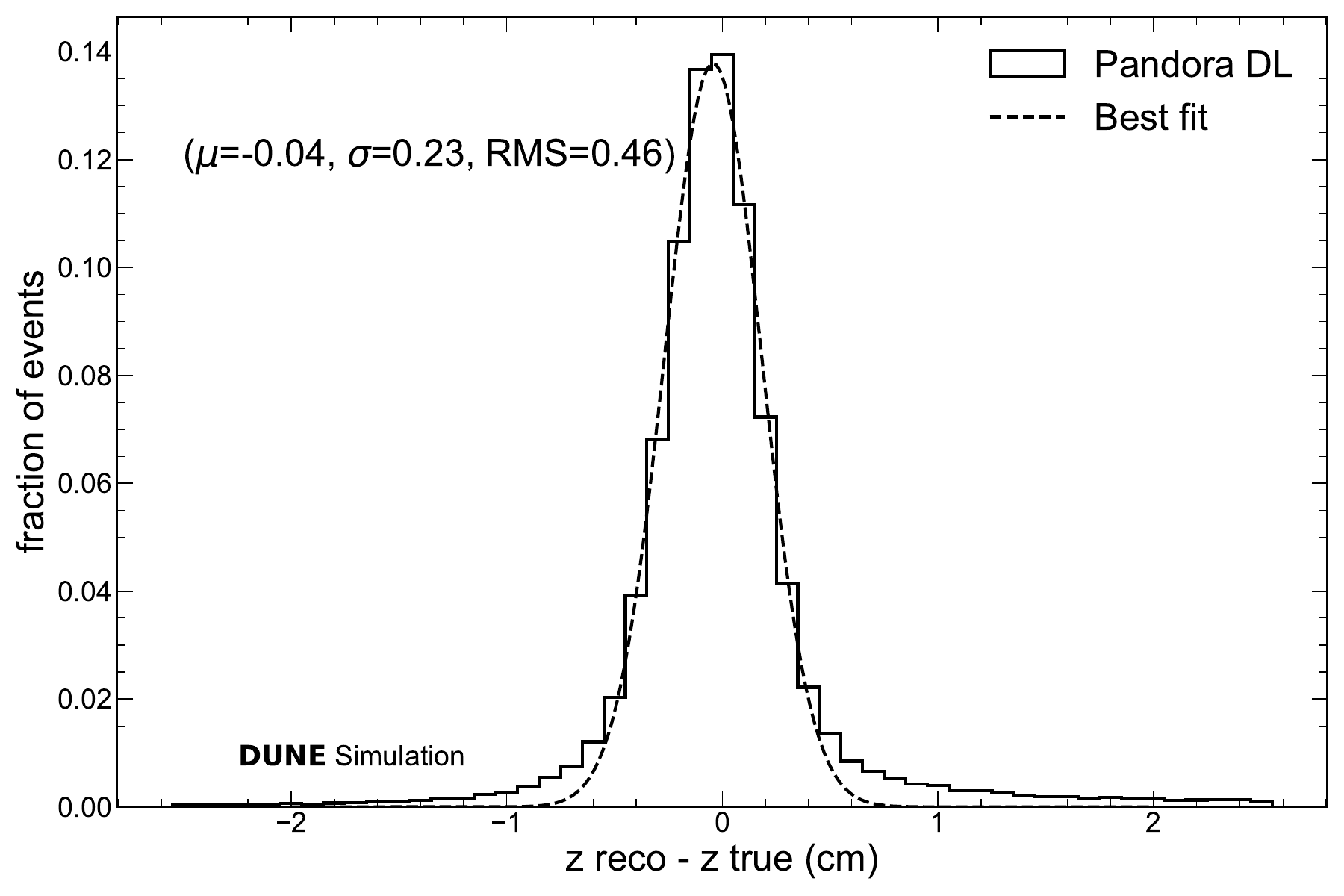}
    \end{subfigure}
    \caption{Vertex resolutions for each axis ($\mu$ and $\sigma$ are the mean and standard deviation of the fitted Gaussian, with RMS being the Root Mean Square of the distribution of reco - true values). All flavours and horn currents are combined here due to minimal differences in resolution between samples. The reduced resolution in Y comes from the need to infer this coordinate from the Z coordinates in more than one plane.}\label{fig:accel_hd_resolution}
\end{figure*}

Fig.~\ref{fig:accel_hd_resolution} shows the distribution of the difference between the three-dimensional positions of the reconstructed and true interaction vertex locations for each individual dimension, across all flavours, interaction types and horn currents. Of note in these distributions, in addition to the large fraction of events where the vertex is reconstructed to within 1\,cm of the true vertex, is the lack of bias or skew in the distributions. The distributions are centred on zero in all three dimensions and are as likely to be reconstructed upstream/left/below as downstream/right/above. The Y coordinate can only be inferred from the overlap of at least two readout channels (no channel provides a direct measurement in Y), which explains the reduced resolution, while the collection plane provides a direct measurement in Z (channels in the induction planes span a range of Z coordinates), and all three readout planes share a common X coordinate.

In summary, the vertex reconstruction performance exhibits an evolution whereby performance is lowest for those events where there is little information in the vicinity of the vertex, as expected. This is particularly acute for NC interactions with low inelasticity and low hadronic invariant mass, whereas these issues are largely offset in CC interactions by the presence of a leading lepton. As inelasticity or hadronic invariant mass increase, performance improves, as higher particle multiplicity yields more charge deposition leading back to the vertex, and naturally makes vertex identification easier. Eventually, however, increases in the number of secondary vertices, which act as additional candidates, and more overlapping particle trajectories smearing the paths back to the true interaction vertex, limit performance at the highest inelasticities and hadronic invariant masses in CC interactions.

\section{Robustness testing}\label{sec:model_dep}
The final state particles emerging from a simulated neutrino interaction depend upon the choice of generator and nuclear model. It is common to observe differences in the number of final state protons depending on the choices made. The particle multiplicity in the vicinity of the neutrino interaction vertex clearly affects the resolution of the vertex reconstruction. Therefore, in this article we investigate how changes to the number of final state protons impacts network performance and if any biases emerge. It is beyond the scope of this article to perform detailed generator and model comparisons, rather we seek to isolate particle multiplicity effects that might change vertex reconstruction efficiency and resolution. In particular, we compare the standard DUNE simulation described in Section~\ref{sec:simulated_data} to equivalent samples in which final state protons with momentum below 0.4\,GeV/c are suppressed (hereafter, for brevity, the 0p sample) and to equivalent samples in which final state neutrons are replaced by final state protons (hereafter the n$\rightarrow$p sample), altering the number of small tracks emerging from the neutrino interaction vertex. These are event-by-event modifications to the generator-level particle content, such that a given generated event has specific particles either suppressed, or exchanged for an alternative flavour, while the remaining particle content is unchanged. These changes are deliberately extreme to test that the network continues to produce sensible results in response to substantial changes in particle multiplicity around the vertex.

For each sample we generate 1000 $\nu_\mu$ events and 1000 $\nu_e$ events. The sample generation procedure is as described above, but with the following key alterations:
\begin{itemize}
    \item Randomisation: The generator step is seeded such that the same 2000 provisional events are generated for each of the standard simulation, the 0p sample and the n$\rightarrow$p sample. This allows for direct comparisons between otherwise equivalent events.
    \item Final state particle changes: Prior to the Geant4 stage, for the 0p sample, any final state proton with a momentum below 0.4\,GeV/c has its status set to zero, ensuring it is not propagated in Geant4. For the n$\rightarrow$p sample, final state neutrons are replaced by otherwise equivalent protons.
\end{itemize}

\begin{figure*}[tbh]
  \centering
  \includegraphics[width=0.5\textwidth]{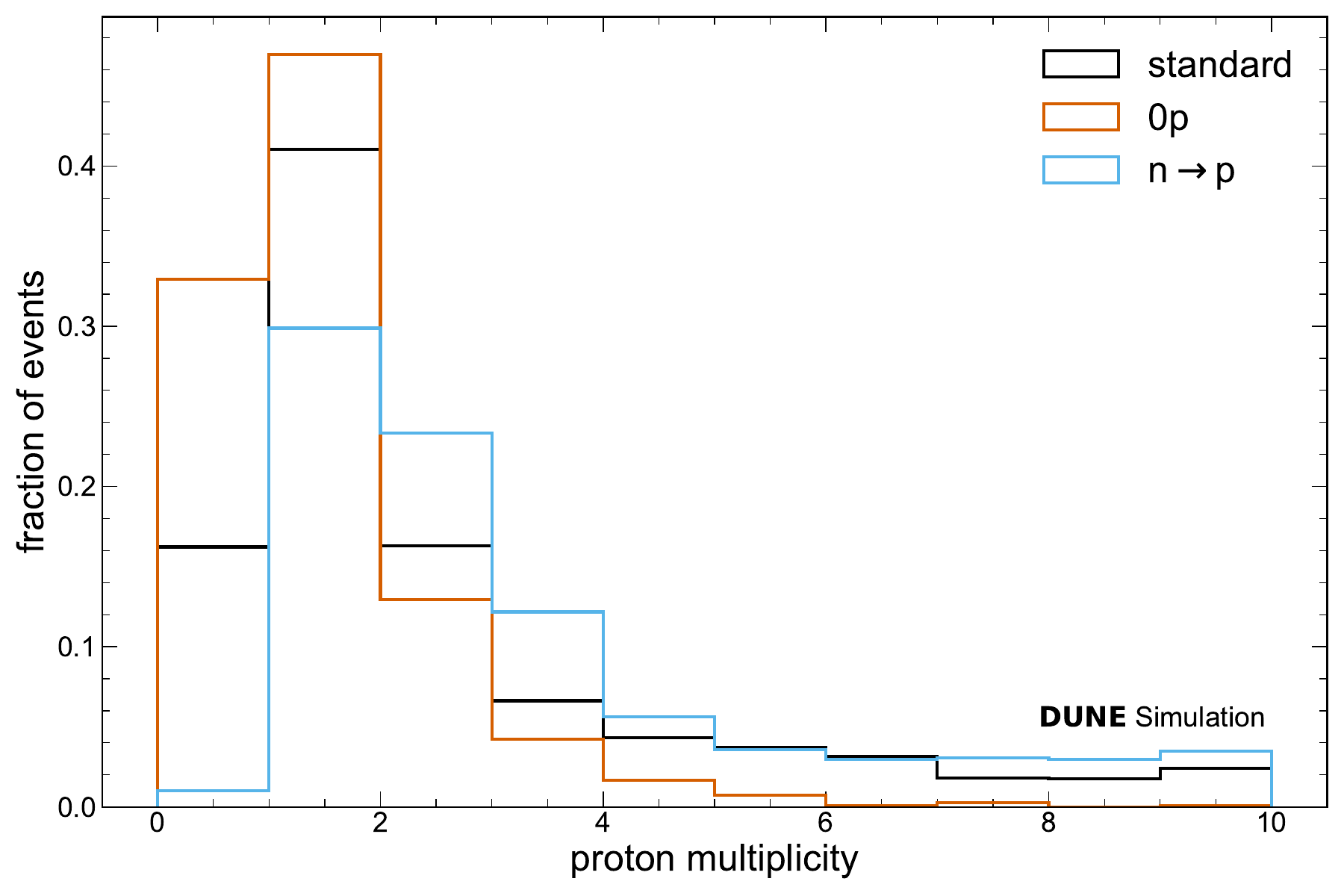}
  \caption{The proton multiplicity for the different samples used for robustness checks.}
  \label{fig:proton_multiplicity}
\end{figure*}

The resultant proton multiplicity, up to 10 final state protons, of the three samples is shown in Fig.~\ref{fig:proton_multiplicity}.

\begin{figure*}[tbh]
    \begin{subfigure}{0.45\textwidth}
        \centering
        \includegraphics[width=\textwidth]{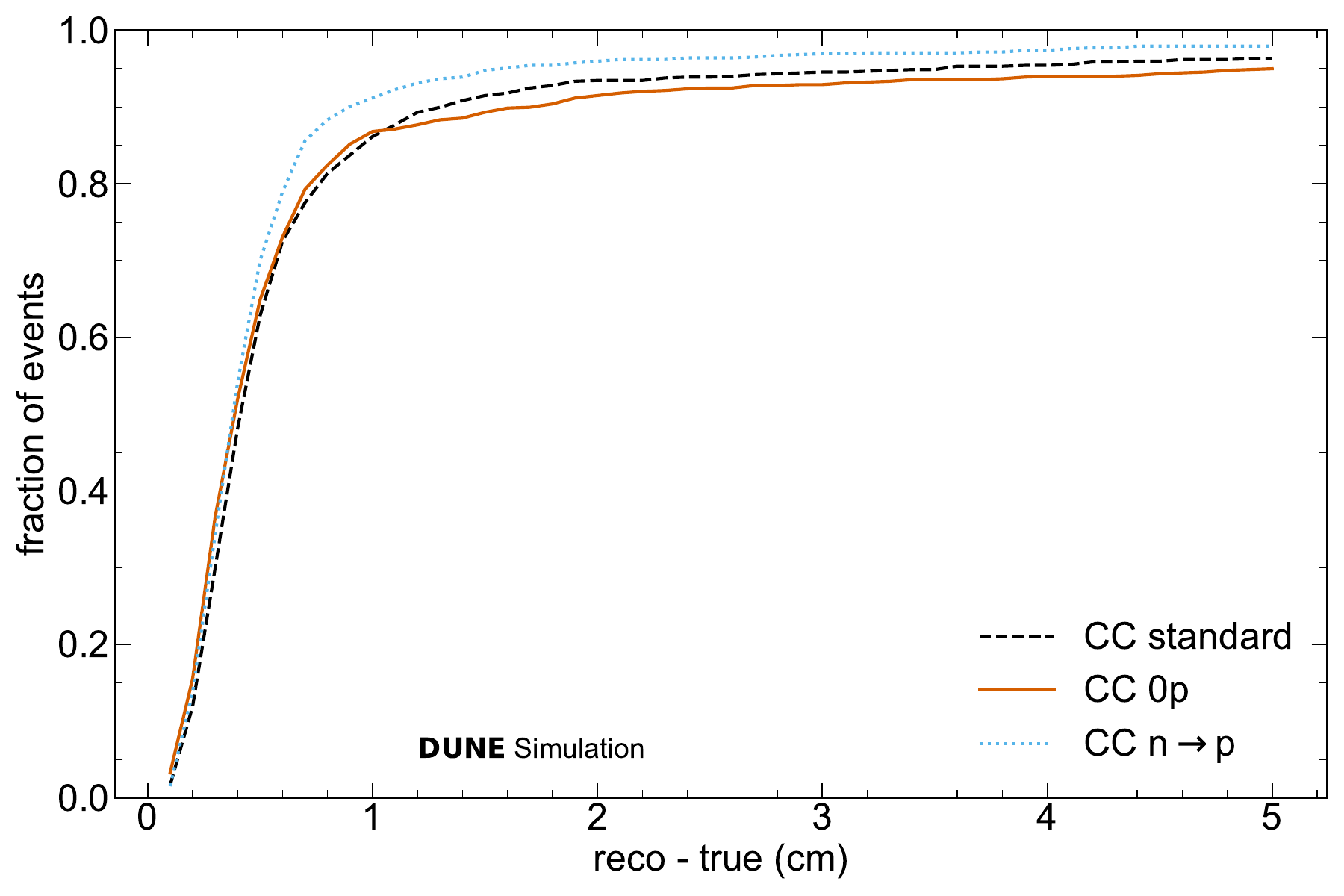}
    \end{subfigure}\hfill
    \begin{subfigure}{0.45\textwidth}
        \centering
        \includegraphics[width=\textwidth]{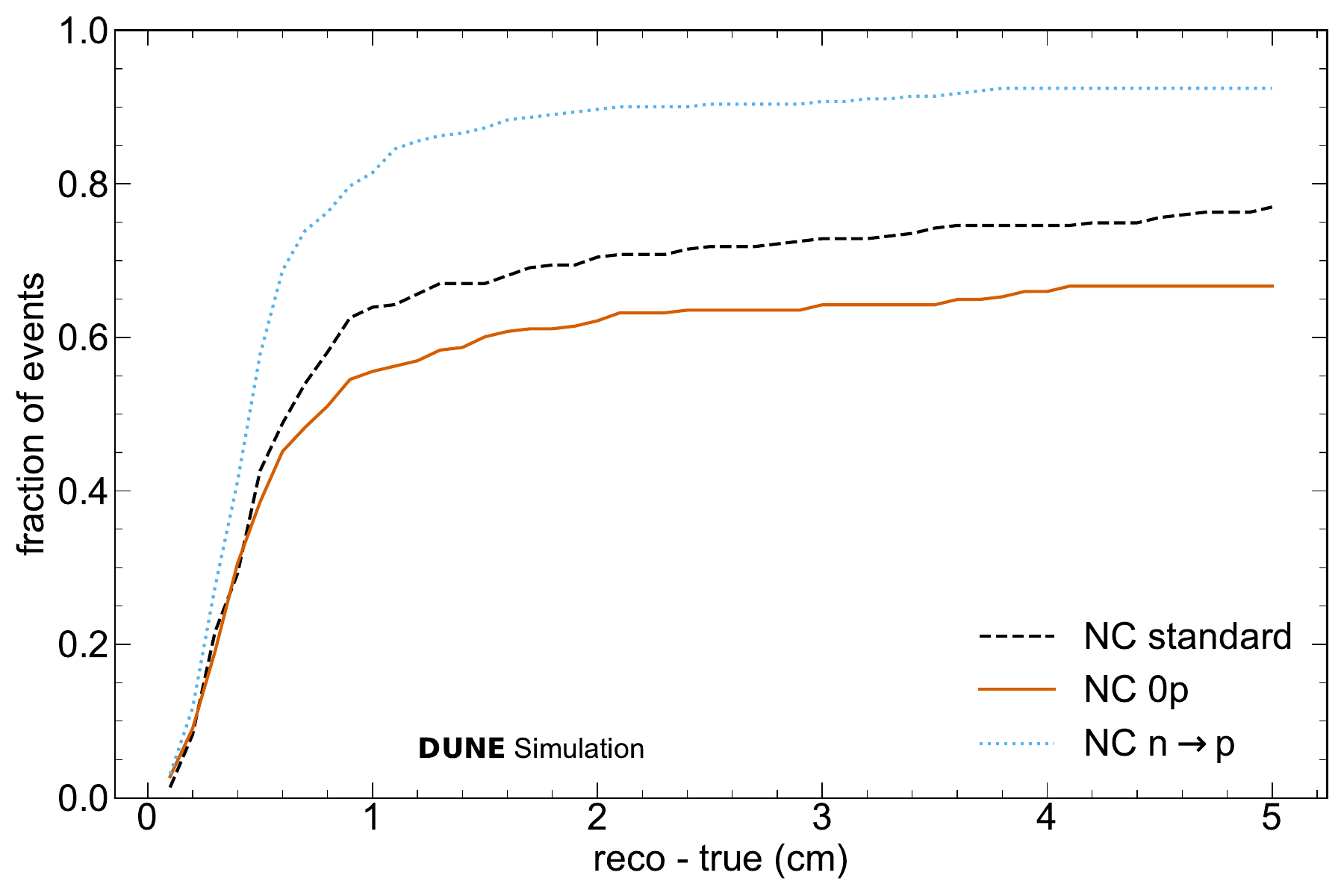}
    \end{subfigure}
    \caption{Fraction of reconstructed vertices as a function of distance to the true vertex for the standard, n$\rightarrow$p and 0p samples, split into (left) CC and (right) NC interactions.}\label{fig:accel_hd_model_dep}
\end{figure*}

\begin{figure*}[tbh]
    \begin{tabular}{c|c}
    \begin{subfigure}{0.45\textwidth}
        \centering
        \includegraphics[width=\textwidth]{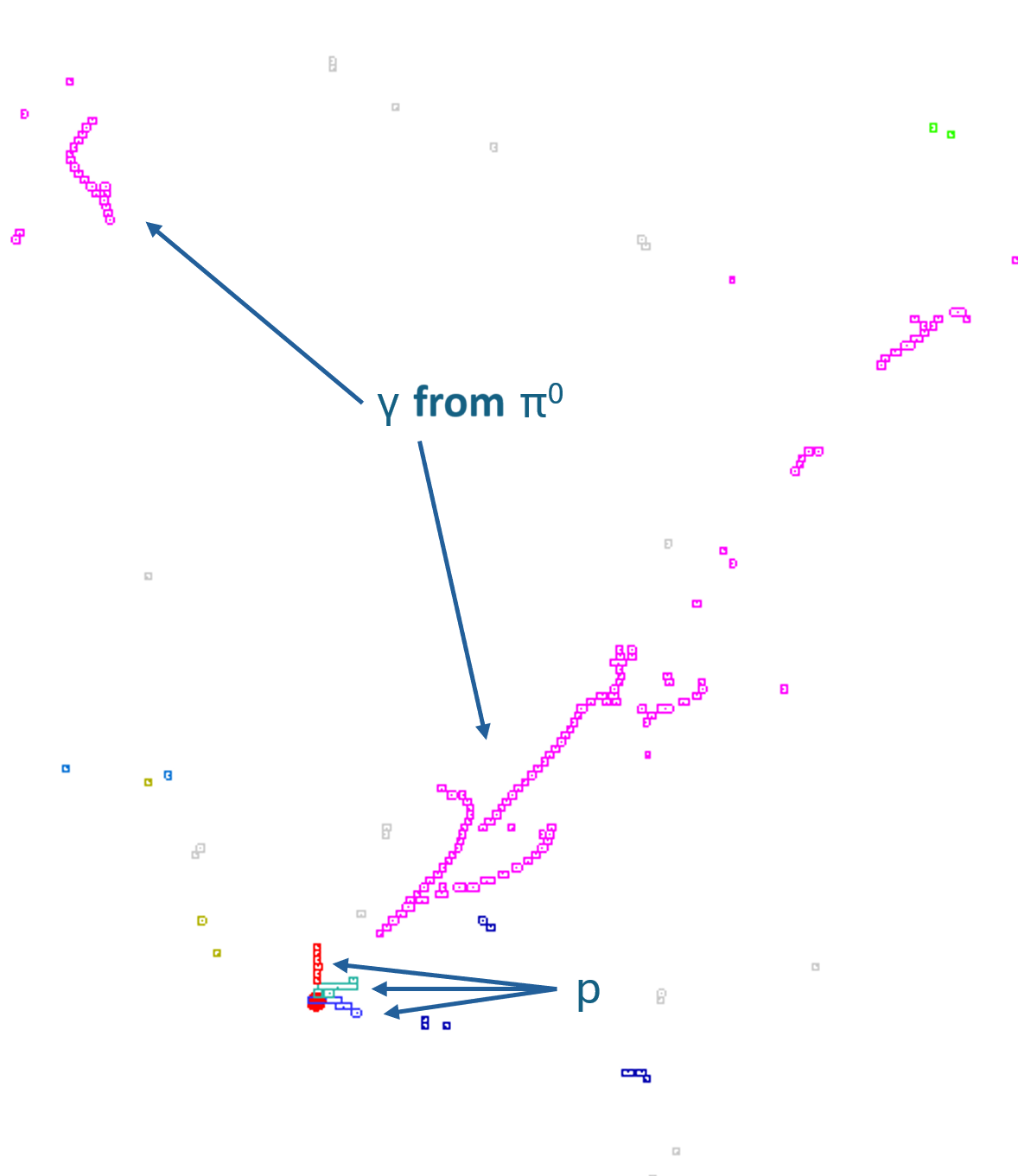}
    \end{subfigure}
    &
    \begin{subfigure}{0.45\textwidth}
        \centering
        \includegraphics[width=\textwidth]{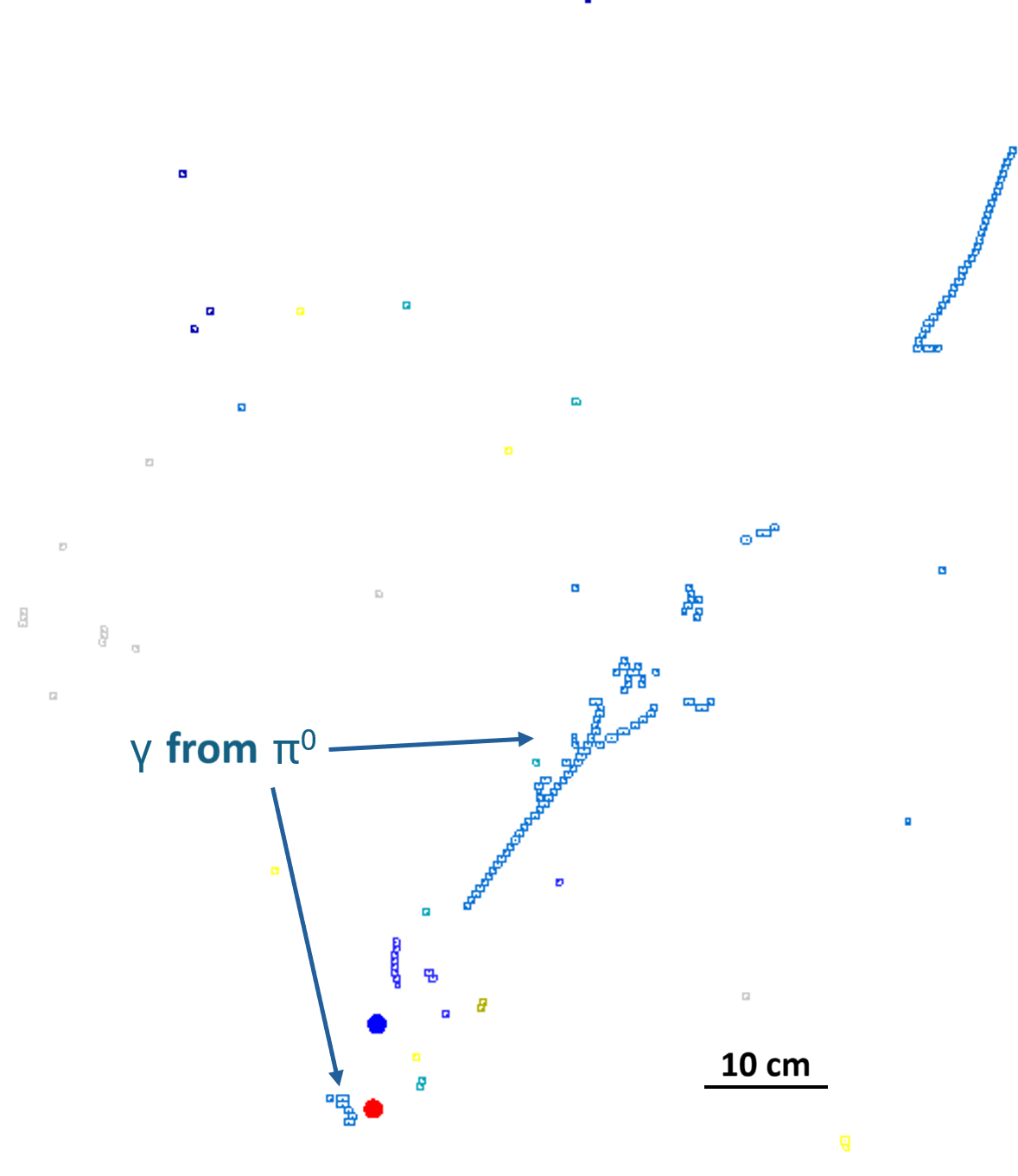}
    \end{subfigure}
    \end{tabular}
    \caption{1.6~GeV NC interaction with  a $\pi^0\rightarrow\gamma\gamma$, nine neutrons and either (left) three or (right) zero protons in the final state. The true interaction vertex is indicated by the blue circle (hidden below the reconstructed vertex in the left image), while the reconstructed interaction vertex is indicated by the red circle. Particle colours are arbitrary and not correlated between left and right.}\label{fig:accel_hd_model_dep_base_0p_nc}
\end{figure*}

\begin{figure*}[tbh]
    \begin{tabular}{c|c}
    \begin{subfigure}{0.45\textwidth}
        \centering
        \includegraphics[width=\textwidth]{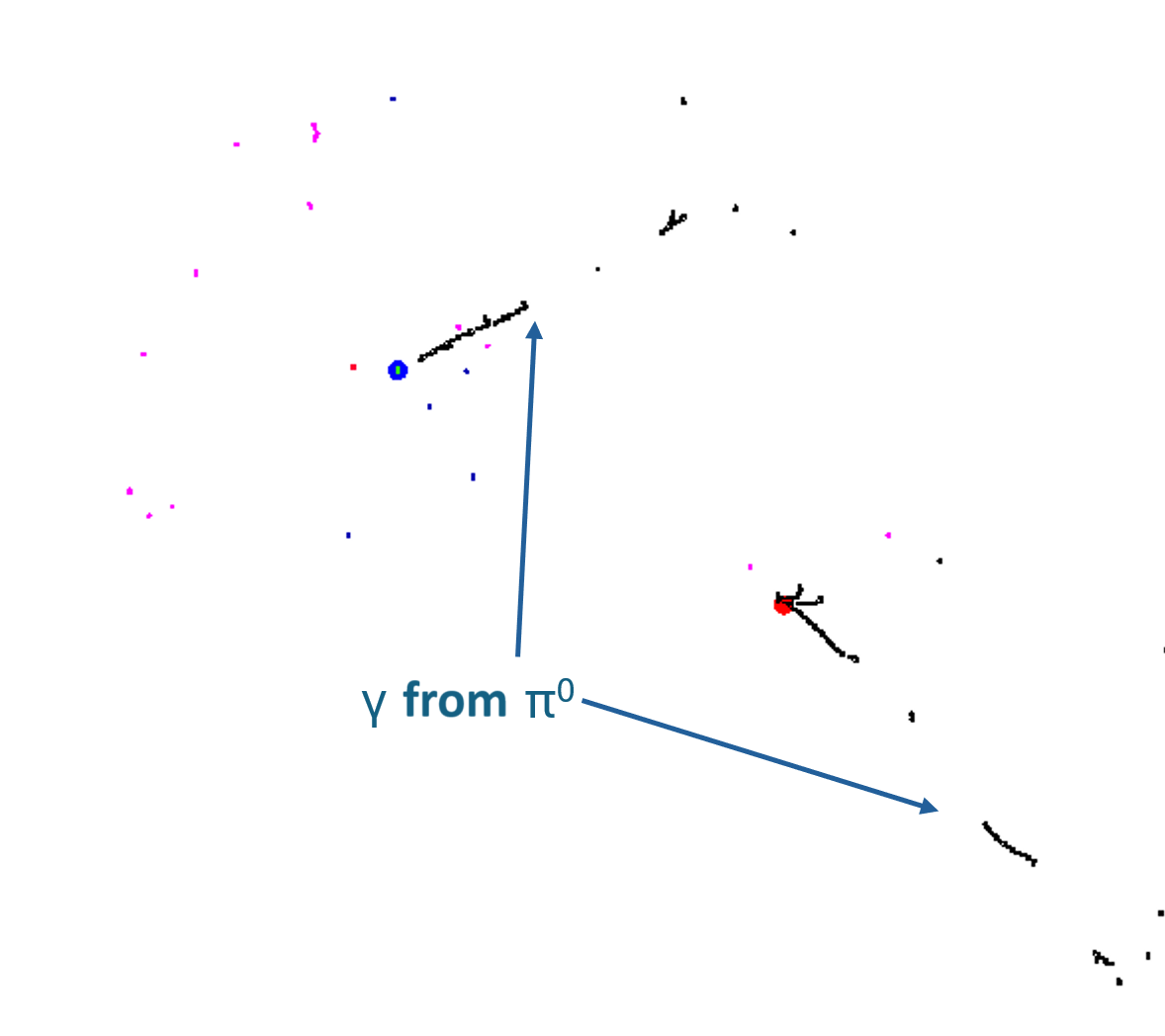}
    \end{subfigure}
    &
    \begin{subfigure}{0.45\textwidth}
        \centering
        \includegraphics[width=\textwidth]{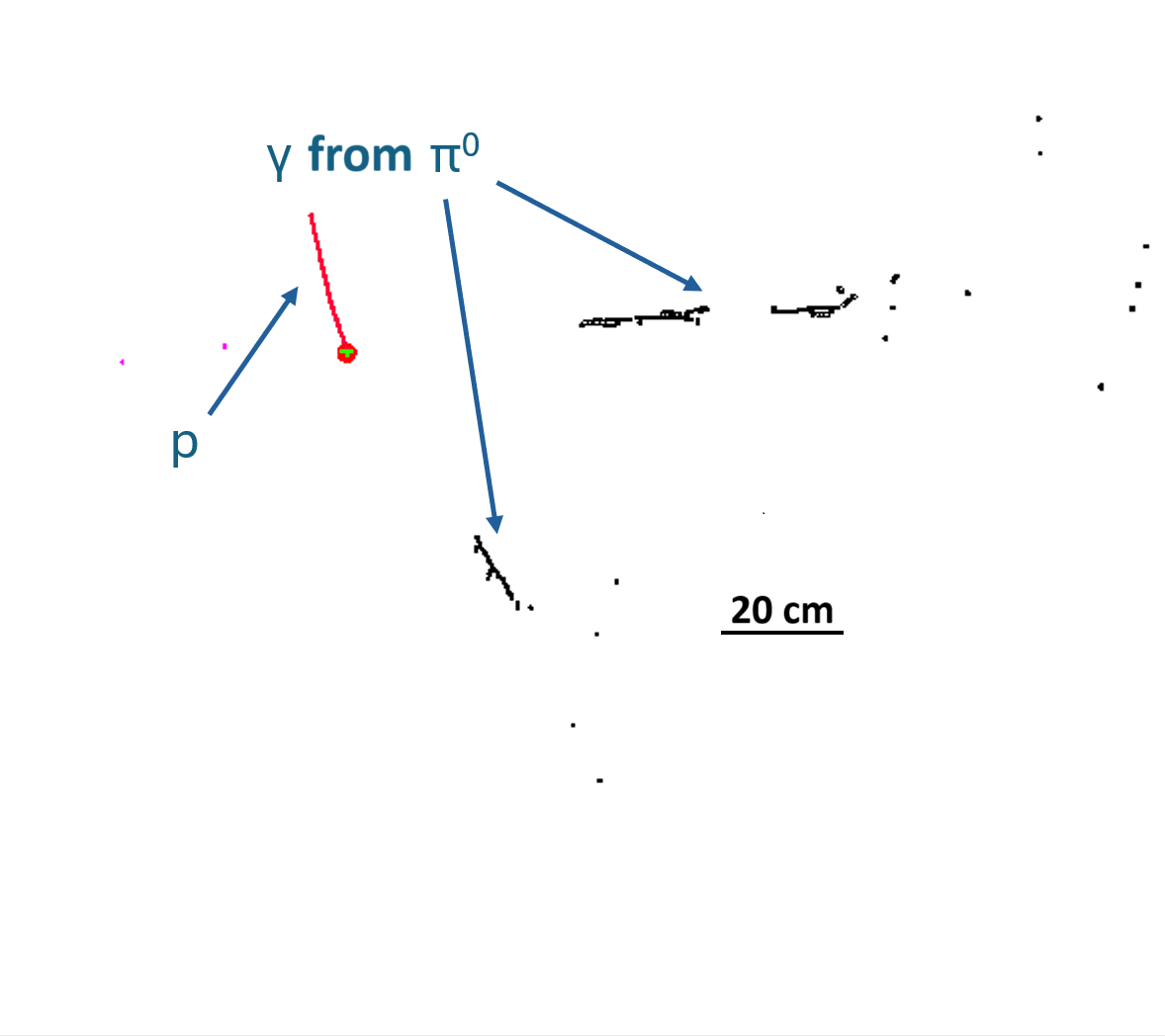}
    \end{subfigure}
    \end{tabular}
    \caption{2.8~GeV NC interaction with a $\pi^0\rightarrow\gamma\gamma$ and either (left) a neutron or (right) proton in the final state. The true interaction vertex is indicated by the blue circle, while the reconstructed interaction vertex is indicated by the red circle. Particle colours are arbitrary and not correlated between left and right.}\label{fig:accel_hd_model_dep_base_fs_nc}
\end{figure*}

Fig.~\ref{fig:accel_hd_model_dep} depicts the change in vertex reconstruction performance as the number of final state particles varies. For charged current interactions the differences are modest. Vertex resolution improves with the number of protons as we move from the proton-poor, through standard to proton-rich samples. Most notably, the proton-rich sample achieves a higher overall reconstruction efficiency below 5\,cm. The performance difference for neutral current interactions is much larger. Here, the 0p sample further reduces particle multiplicity around the vertex, where a leading lepton is already absent, yielding many more catastrophic failures (e.g. Fig.~\ref{fig:accel_hd_model_dep_base_0p_nc}). Conversely, the proton-rich sample is able to enhance pointing information in the region of the vertex and thereby offset the lack of a leading lepton to a large degree (e.g. Fig.~\ref{fig:accel_hd_model_dep_base_fs_nc}).

\begin{figure*}[tbh]
    \begin{subfigure}{0.45\textwidth}
        \centering
        \includegraphics[width=\textwidth]{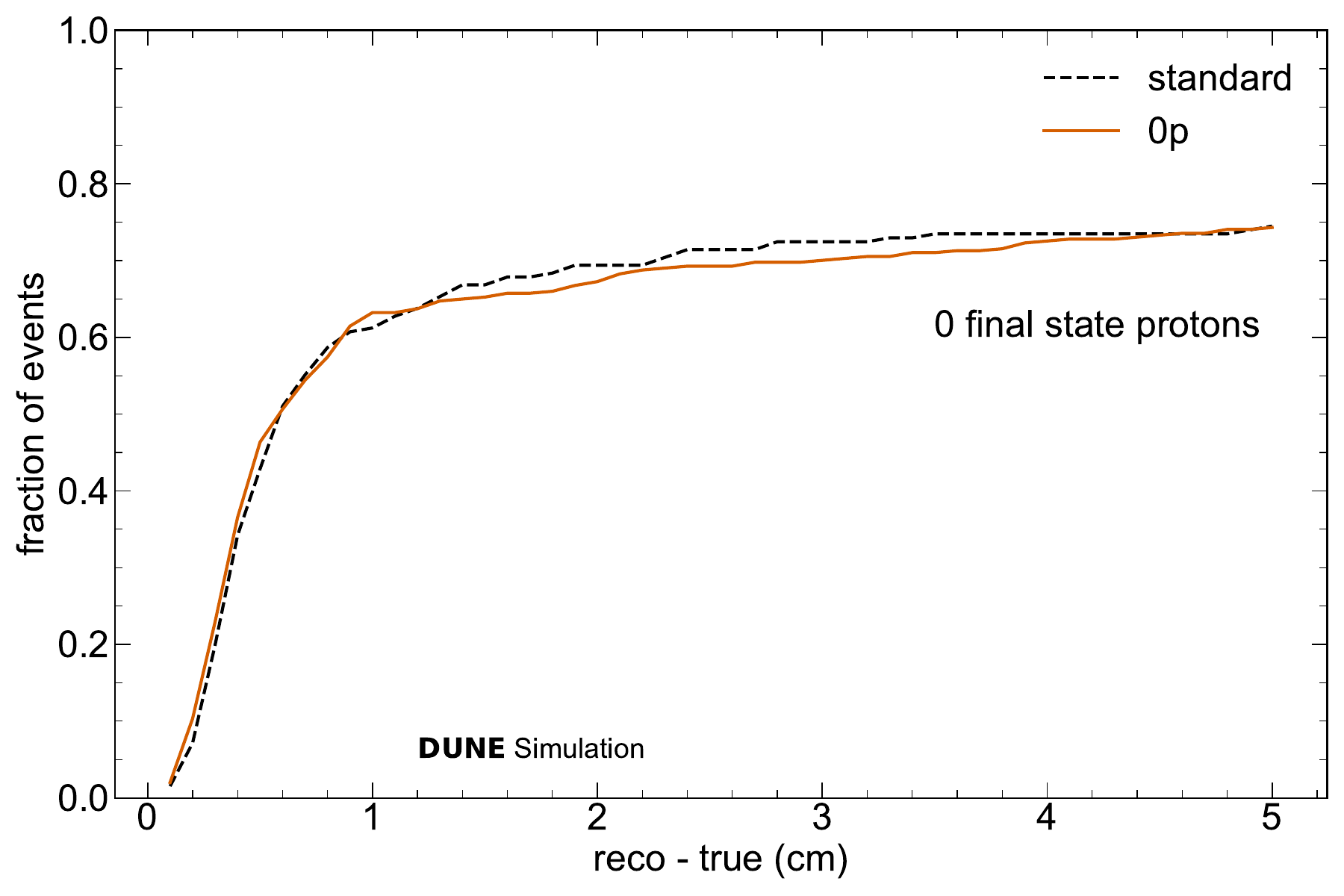}
    \end{subfigure}\hfill
    \begin{subfigure}{0.45\textwidth}
        \centering
        \includegraphics[width=\textwidth]{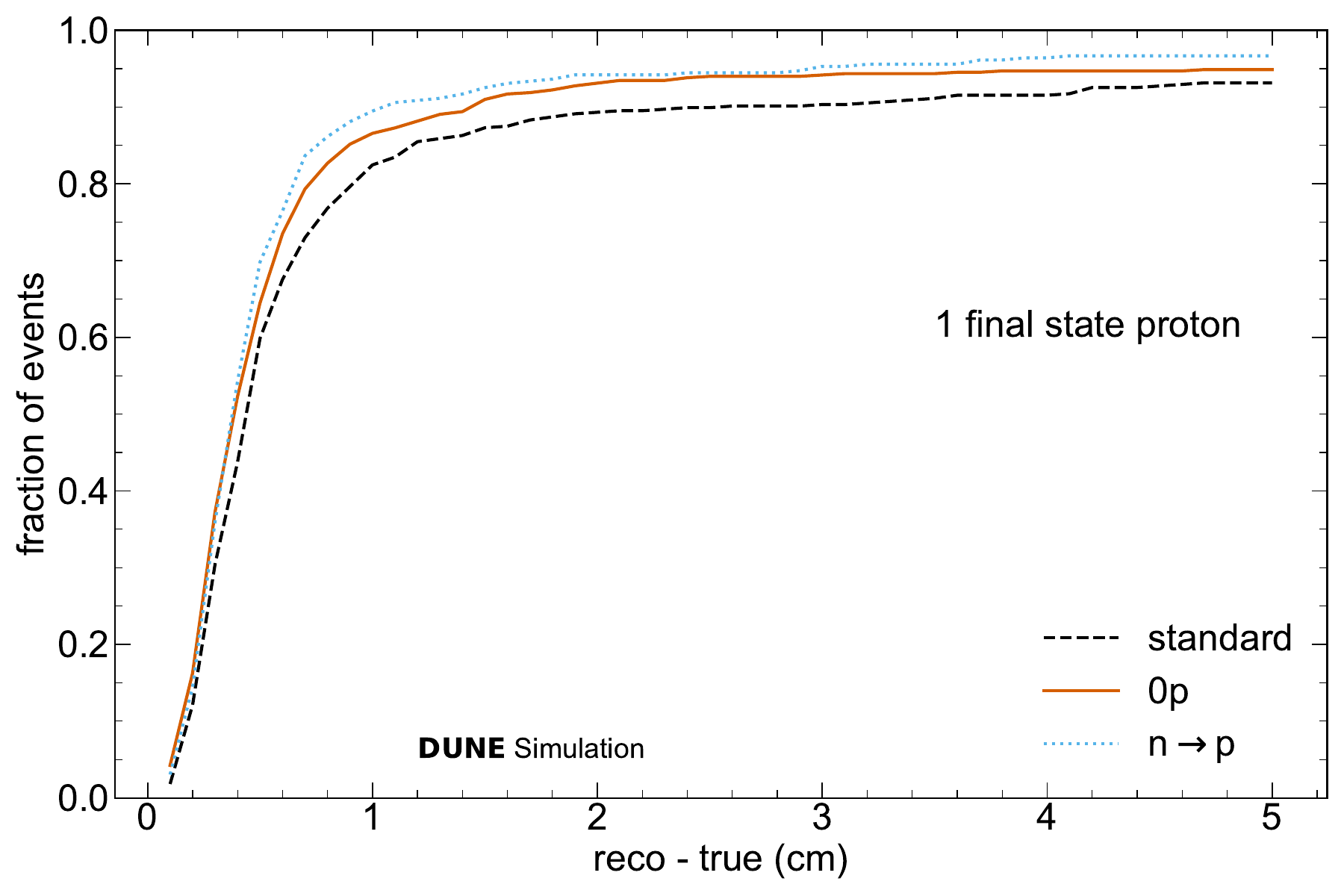}
    \end{subfigure}
    \begin{subfigure}{0.45\textwidth}
        \centering
        \includegraphics[width=\textwidth]{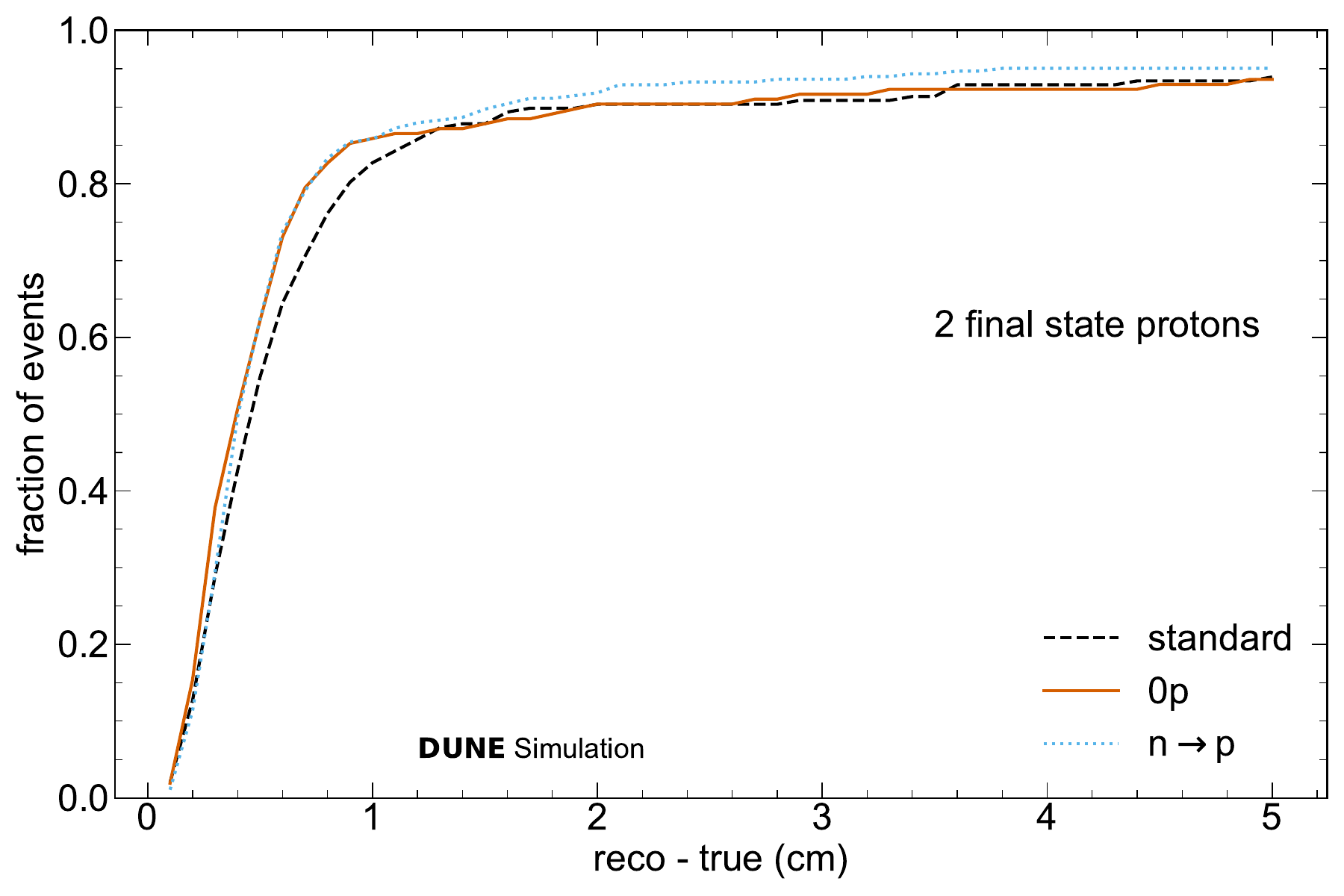}
    \end{subfigure}\hfill
    \begin{subfigure}{0.45\textwidth}
        \centering
        \includegraphics[width=\textwidth]{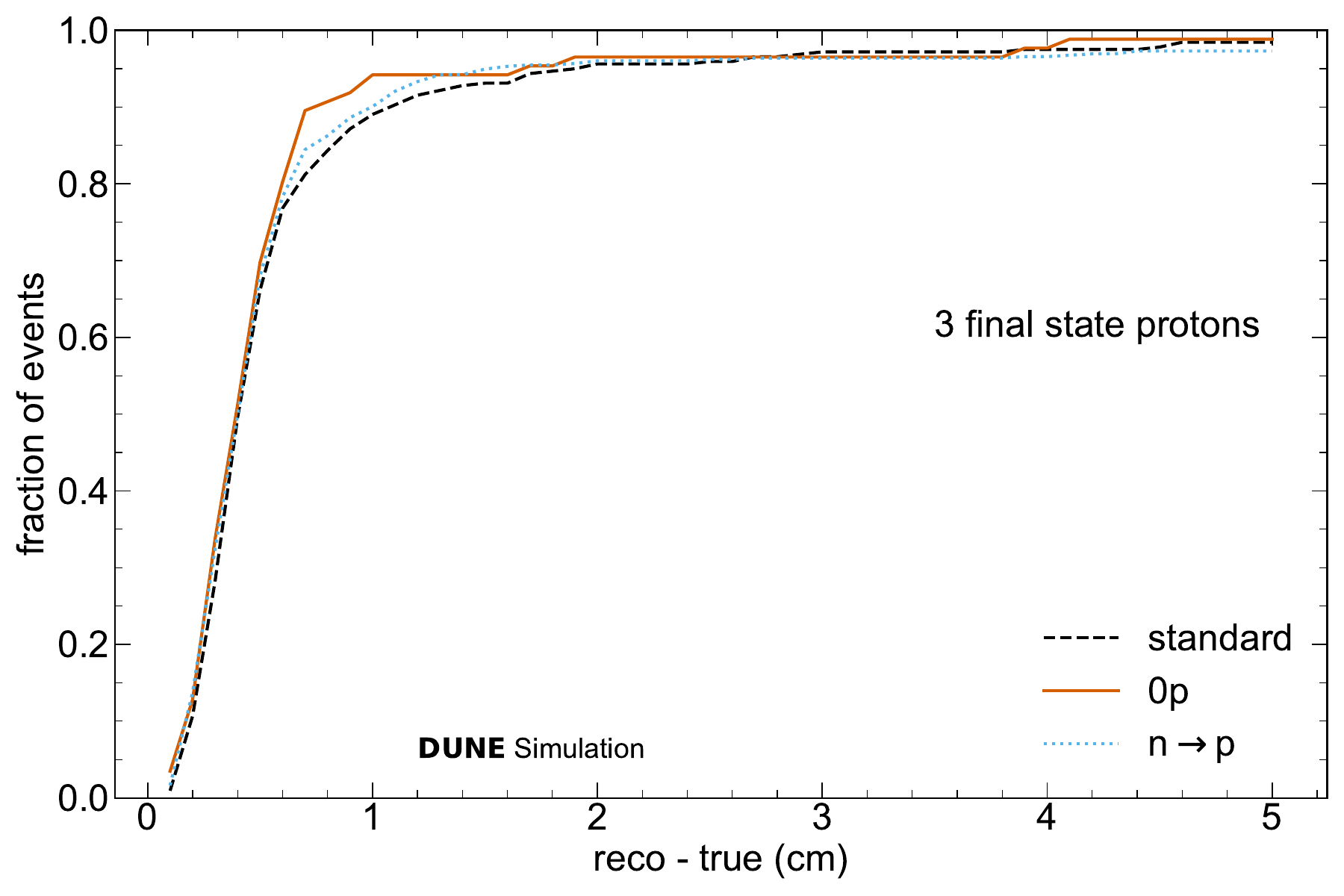}
    \end{subfigure}
    \caption{Fraction of reconstructed vertices as a function of distance to the true vertex broken down by the final state proton multiplicity (all momenta) of events. The n$\rightarrow$p sample is omitted from the 0 proton case due to a lack of events for a meaningful comparison.}\label{fig:accel_hd_model_dep_cf_by_mult}
\end{figure*}

\begin{figure*}[tbh]
    \begin{subfigure}{0.45\textwidth}
        \centering
        \includegraphics[width=\textwidth]{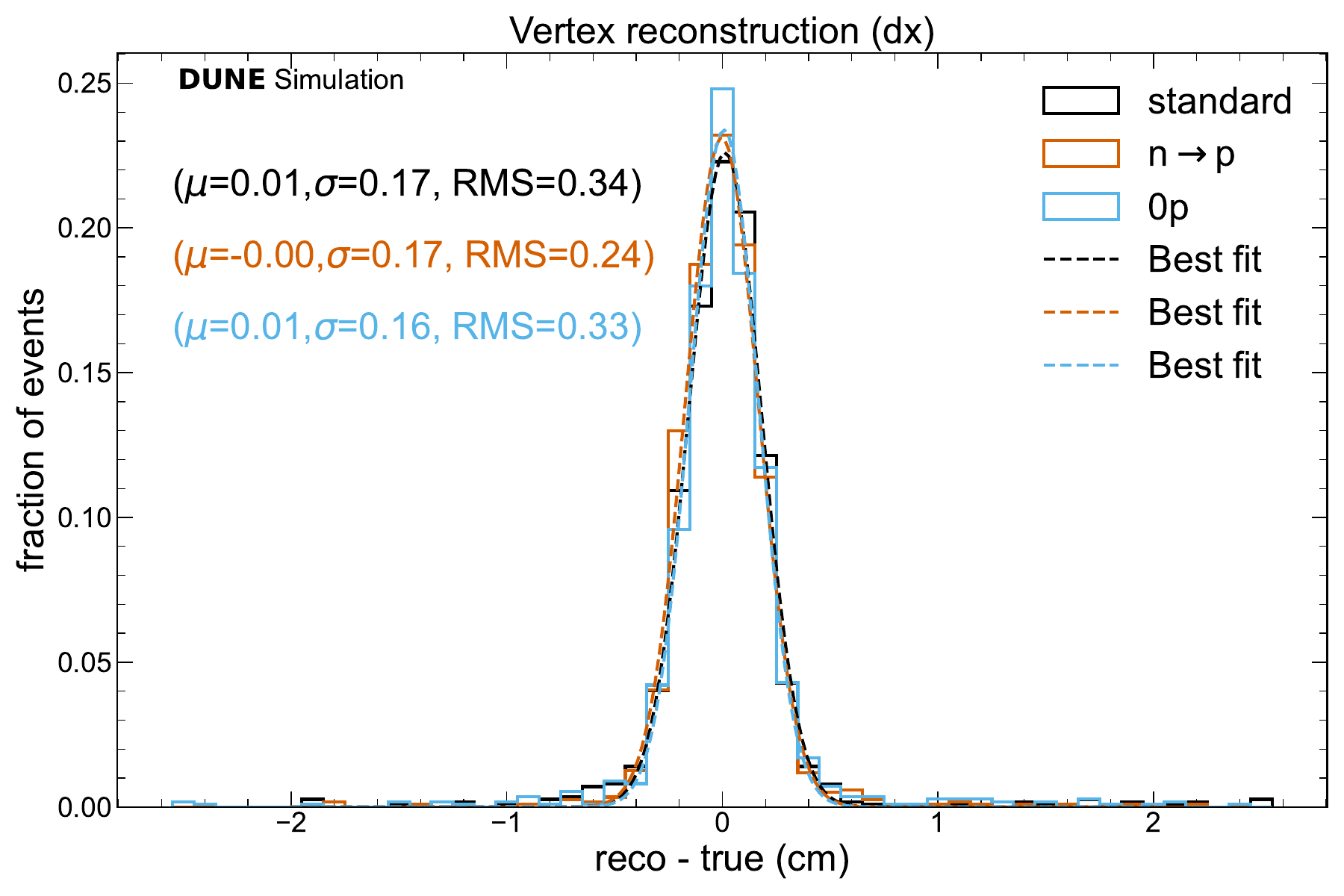}
    \end{subfigure}\hfill
    \begin{subfigure}{0.45\textwidth}
        \centering
        \includegraphics[width=\textwidth]{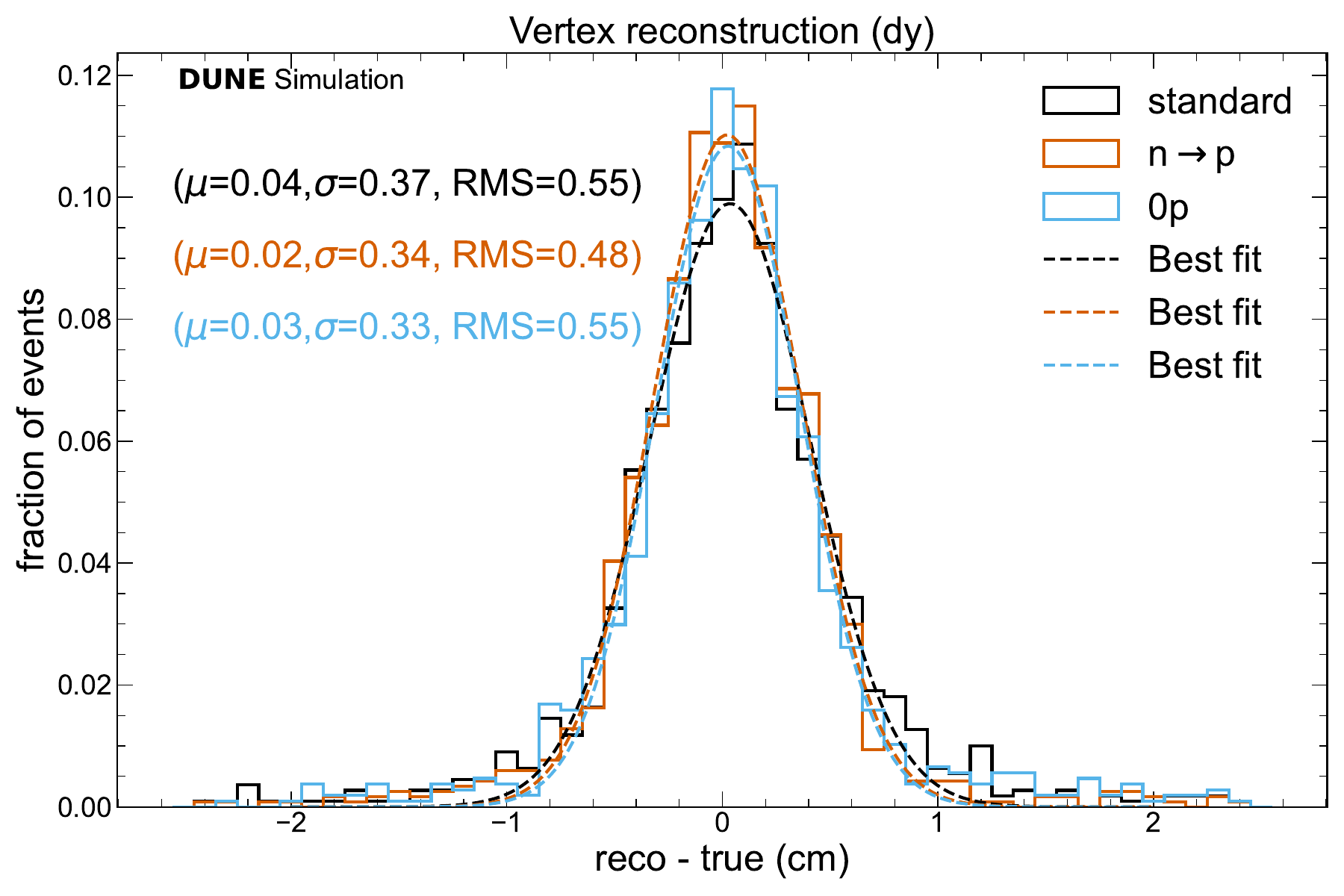}
    \end{subfigure}
    \begin{subfigure}{0.45\textwidth}
        \centering
        \includegraphics[width=\textwidth]{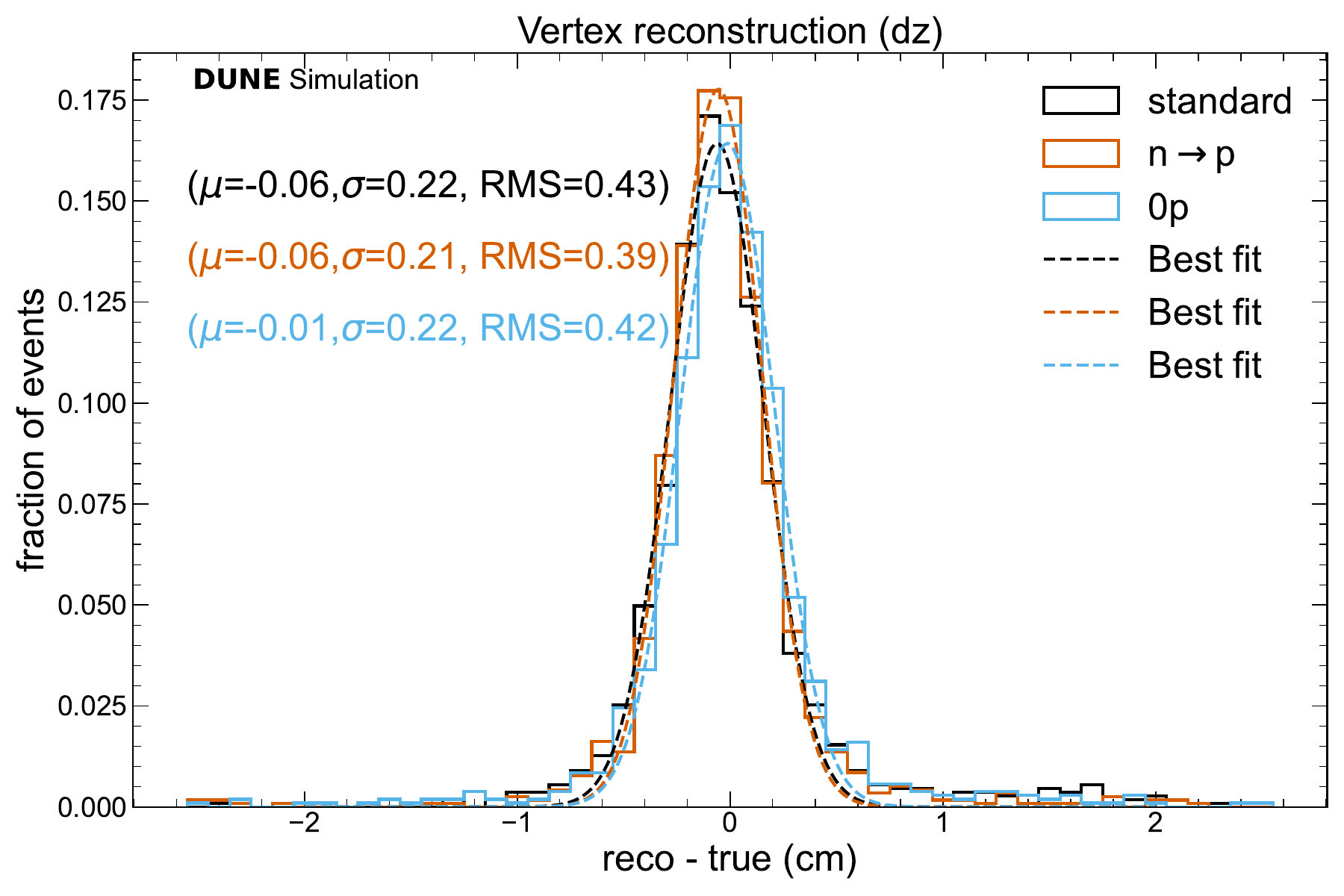}
    \end{subfigure}
    \caption{Vertex resolutions for the standard, n$\rightarrow$p, and 0p samples for each axis ($\mu$ and $\sigma$ are the mean and standard deviation of the fitted Gaussian, with RMS being the Root Mean Square of the distribution of reco - true values).}\label{fig:model_resolution}
\end{figure*}

The performance as a function of proton multiplicity is depicted in Fig.~\ref{fig:accel_hd_model_dep_cf_by_mult}. It can be seen that the performance in the different samples is consistent for equivalent proton multiplicity and therefore the difference in performance between the samples is driven by the changes in the distribution of proton multiplicity over the whole sample. Furthermore, it can be seen in Fig.~\ref{fig:model_resolution} that while the number of final state protons affects vertex resolution, there is no evidence that the number of protons in the final state biases vertex reconstruction in any particular direction, with differences in $\mu$ covered by one tenth of one channel spacing.

The source of the changes in vertex resolution can be illustrated by comparing a few events. Fig.~\ref{fig:accel_hd_model_dep_base_fs_nc} shows a 2.8~GeV neutrino undergoing a neutral current interaction to produce a $\pi^0$ along with a neutron (left) in the final state, and the corresponding event with the neutron replaced by a proton (right). In the former case, one photon from the $\pi^0$ decay produces charge deposition close to the true vertex, while the second photon produces charge deposition farther away, and this is chosen as the reconstructed vertex. In the case where the neutron is replaced by a proton there is an additional anchor point for the network to use in inferring the vertex location. In addition, two photons also point towards this location, yielding an accurate vertex reconstruction.

\begin{figure*}[tbh]
    \begin{tabular}{c|c}
    \begin{subfigure}{0.45\textwidth}
        \centering
        \includegraphics[width=\textwidth]{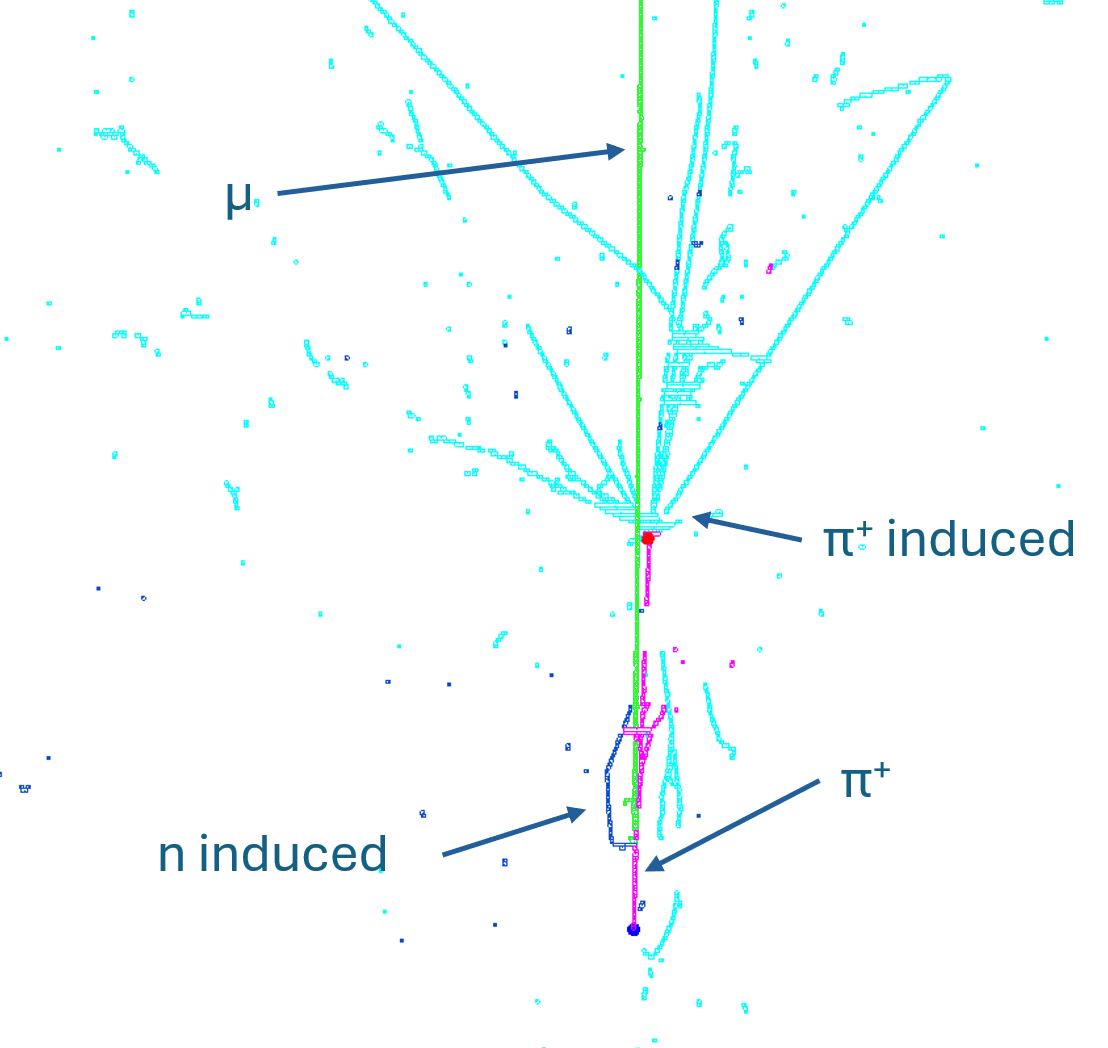}
    \end{subfigure}
    &
    \begin{subfigure}{0.45\textwidth}
        \centering
        \includegraphics[width=\textwidth]{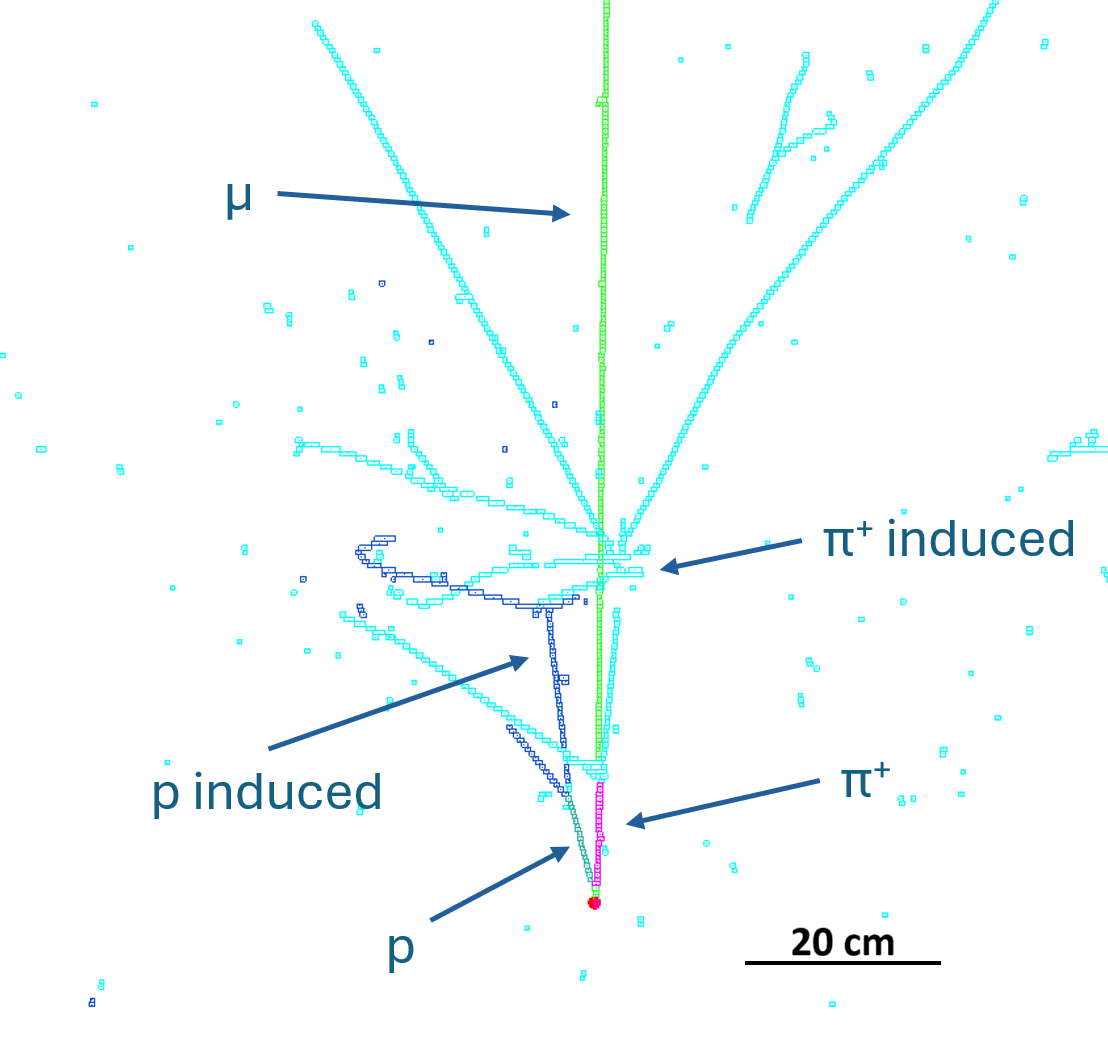}
    \end{subfigure}
    \end{tabular}
    \caption{24.9~GeV CC interaction with  a $\mu$, a $\pi^+$ and either (left) a neutron or (right) proton in the final state. The true interaction vertex is indicated by the blue circle, while the reconstructed interaction vertex is indicated by the red circle. Particle colours are arbitrary.}\label{fig:accel_hd_model_dep_base_fs_cc}
\end{figure*}

Fig.~\ref{fig:accel_hd_model_dep_base_fs_cc} shows a 24.9~GeV neutrino undergoing a charged current interaction to produce a $\mu$, a $\pi^+$ and a neutron (left) in the final state, and the corresponding event with the neutron replaced by a proton (right). Here, the high-energy (5.9~GeV) $\pi^+$ (magenta) interacts to produce considerable subsequent activity. The initial colinearity of the muon and pion results in a single track-like deposition emerging from the vertex and therefore the network picks the subsequent pion interaction vertex as the likely vertex candidate. With the neutron replaced by the proton (dark green), the additional track-like deposition emerging from the true interaction vertex yields a correct reconstruction.

If we instead consider removal of protons from an interaction, Fig.~\ref{fig:accel_hd_model_dep_base_0p_nc} depicts a 1.6~GeV neutral current interaction with a neutral pion, nine neutrons and either three (left), or zero (right) protons in the final state. The three short proton tracks emerging from the vertex (left) provide a clear handle for accurate vertex reconstruction, whereas the separation between the true vertex and the photons (right) from the neutral pion decay, along with diffuse neutron-induced activity results in an offset between the true and reconstructed vertex location.

In summary, when changing the proton multiplicity we see overall efficiency changes as a function of particle multiplicity. In NC interactions we observe more catastrophic failures where the vertex is shifted far from the correct region, which is to be expected, and a feature that would not be unique to machine learning methods, given that identifying a neutrino interaction vertex requires charge deposition in the detector that can be traced back to that location. However, we do not observe biases for those events that remain in the correct region --- high particle multiplicity at the vertex does not appear to smear out the vertex resolution, or shift the reconstructed vertex position along the beam direction.

\section{Future extensions}\label{sec:extensions}
There are many avenues for optimising and extending the vertex finding concept discussed in this article. In the first instance there are relatively simple, if time intensive optimisations that can be considered. For example, the width of the rings used to define the distance between a given hit and the vertex. The number of classes, and their width could be optimised to increase the resolution of each ring - in principle this approach could be extended to per-pixel regression with a Gaussian error defining the width of the band rather than per-pixel classification with a discrete width. It is also expected that the number of networks per pass can be reduced from three to two, by leveraging symmetries in the induction planes to train a single network to process both induction planes, thereby reducing the computational requirements of training.
%

An additional technical enhancement would be to introduce sparse convolutions or switch to a graph network. At present the non-hit regions are processed alongside the hit regions and given the sparse nature of the inputs, the computational and memory overhead can expect to be improved by moving to sparse convolutions or a graph network. In addition, the contribution of non-hit regions to the outputs of convolutions and transpose convolutions might reasonably be expected to limit the performance of the semantic segmentation, with sparse convolutions ensuring that only active hit regions contribute to this process. Finally, the present need for two passes to provide a sufficiently high-resolution reconstructed vertex position may be eliminated by the ability to represent hits via unstructured input tensors as opposed to fixed size, two dimensional images.

A non-technical extension includes extending the technique to identify secondary vertices. When finding a single vertex, the truth definition is each hit's distance to that single vertex. This can be modified to encode the distance to the closest vertex, partitioning the plane as a Voronoi diagram \cite{Voronoi1908a, Voronoi1908b}.

Finally, this method will be applied to additional detector contexts, such as the vertical-drift far detector, and additional samples, such as atmospheric and supernova neutrino samples, and extended to include the full far detector geometry, rather than the workspace geometry.

\section{Conclusion}\label{sec:conclusion}
Support for deep neural networks has been integrated into the Pandora pattern recognition reconstruction workflow using LibTorch. A U-ResNet classifying hits with respect to their distance to the neutrino interaction vertex has been implemented in the context of a DUNE horizontal-drift far detector and acts as a performant vertex finding technique that substantially outperforms the previous BDT implementation, with an increase in the efficiency of vertex reconstruction within 1\,cm of the true vertex of more than 20\% in all flavours. Given the transferability of Pandora's algorithms to other detector contexts, it is expected that this approach will also be effective in a vertical-drift far detector, though perhaps with a small reduction in resolution reflecting the larger induction channel spacing in the vertical-drift detector. Charged current interactions yield highly performant vertex finding, while neutral current performance is reduced by the absence of pointing information from a leading lepton (though still much improved relative to the BDT implementation). Interaction vertices are identified with equivalent efficiency in $\nu_e$ and $\nu_\mu$ samples, while $\nu_\tau$ performance is reduced due to a larger neutral current fraction. In general, precise vertex reconstruction for events with little charge deposition in the vicinity of the neutrino interaction vertex is very difficult, as expected, but vertex reconstruction performance rapidly improves as particle multiplicity in this region increases. This performance then plateaus and even over-turns for the most complex events with secondary and tertiary vertices acting as plausible alternative candidates, and overlapping particle trajectories smearing the path back to the true interaction vertex. Evaluation of robustness in terms of final state proton multiplicity shows no direction bias in the reconstructed vertex position. Though neutral current interactions yield lower vertex efficiency (more catastrophic failures) as the proton multiplicity goes to zero (which is to be expected given reduced pointing information), charged current interactions appear insensitive to proton multiplicity. These improvements in vertex reconstruction will facilitate improvements in hit clustering, particle characterisation and subsequent high-level reconstruction quantities, such as estimates of neutrino energy, by avoiding inappropriate splitting and merging of particles, and errors in parent-child relationships.

\backmatter

\section*{Data availability statement}
This manuscript has no associated data or the data will not be deposited. [Authors’ comment: The Pandora event reconstruction described in this paper operates on low-level data and does not directly produce physics results, hence there is no associated data release.]

\section*{Code availability statement}
This manuscript has associated code/software in a data repository. [Authors’ comment: The Pandora event reconstruction software can be found at \url{https://github.com/PandoraPFA}.]

\bmhead{Acknowledgements}

This document was prepared by the DUNE collaboration using the
resources of the Fermi National Accelerator Laboratory 
(Fermilab), a U.S. Department of Energy, Office of Science, 
HEP User Facility. Fermilab is managed by Fermi Research Alliance, 
LLC (FRA), acting under Contract No. DE-AC02-07CH11359.
%
%
This work was supported by
CNPq,
FAPERJ,
FAPEG and 
FAPESP,                         Brazil;
CFI, 
IPP and 
NSERC,                          Canada;
CERN;
M\v{S}MT,                       Czech Republic;
ERDF, 
Horizon Europe, 
MSCA and NextGenerationEU,      European Union;
CNRS/IN2P3 and
CEA,                            France;
INFN,                           Italy;
FCT,                            Portugal;
NRF,                            South Korea;
Generalitat Valenciana, 
Junta de Andaluc\'ia-FEDER, 
MICINN, and 
Xunta de Galicia,               Spain;
SERI and 
SNSF,                           Switzerland;
T\"UB\.ITAK,                    Turkey;
The Royal Society and 
UKRI/STFC,                      United Kingdom;
DOE and 
NSF,                            United States of America.

\bigskip
\bibliographystyle{ieeetr}
\bibliography{sn-article}

\end{document}